\RequirePackage{fix-cm}
\documentclass[a4,twocolumn,epjc3,comma,sort&compress,natbib]{svjour3}
\bibpunct{[}{]}{,}{n}{}{,} 

\journalname{Eur. Phys. J. C}

\usepackage{latexsym}
\usepackage{amsmath}
\usepackage{amssymb}
\usepackage{amsfonts}
\usepackage{dsfont}
\usepackage{slashed}
\usepackage{bm}

\usepackage[mathscr,scaled=1.15]{urwchancal}
\DeclareFontFamily{OT1}{pzc}{}
\DeclareFontShape{OT1}{pzc}{m}{it}%
{<-> s * [1.15] pzcmi7t}{}
\DeclareMathAlphabet{\mathpzc}{OT1}{pzc}{m}{it}

\usepackage{color}
\definecolor{purple}{rgb}{0.5,0,0.5}
\definecolor{blue}{rgb}{0.0,0,0.9}
\definecolor{prdblue}{rgb}{0.133,0.118,0.498}
\usepackage[colorlinks=true, pdfstartview=FitV, linkcolor=prdblue, citecolor= prdblue, urlcolor=prdblue]{hyperref}

\usepackage{enumitem}
\usepackage{multirow}
\usepackage{subfigure}
\usepackage{textcomp}

\usepackage{supertabular}
\usepackage{placeins}
\usepackage{epsfig}
\usepackage{graphicx}

\usepackage{dcolumn,booktabs}
\newcolumntype{d}[1]{D{.}{.}{#1}}

\begin{document}

\title{$\,$\\[-6ex]\hspace*{\fill}{\normalsize{\sf\emph{Preprint no}.\ NJU-INP 037/21}}\\[1ex]
Masses of positive- and negative-parity hadron ground-states,\\ including those with heavy quarks}

\author{Pei-Lin Yin\thanksref{ePLY,NJUPT}
        \and
       Zhu-Fang Cui\thanksref{eZFC,NJU,INP} 
       \and
       Craig D.~Roberts\thanksref{eCDR,NJU,INP}
       \and
       Jorge Segovia\thanksref{eJS,USe,INP}
}

\thankstext{ePLY}{yinpl@njupt.edu.cn}
\thankstext{eZFC}{phycui@nju.edu.cn}
\thankstext{eCDR}{cdroberts@nju.edu.cn}
\thankstext{eJS}{jsegovia@upo.es}

\authorrunning{Pei-Lin Yin \emph{et al}.} 

\institute{School of Science, Nanjing University of Posts and Telecommunications, Nanjing 210023, China \label{NJUPT}
            \and
            School of Physics, Nanjing University, Nanjing, Jiangsu 210093, China \label{NJU}
           \and
           Institute for Nonperturbative Physics, Nanjing University, Nanjing, Jiangsu 210093, China \label{INP}
           \and
           Dpto.\ Sistemas F\'isicos, Qu\'imicos y Naturales, Univ.\ Pablo de Olavide, E-41013 Sevilla, Spain \label{USe}
            }

\date{2021 February 24}

\maketitle

\begin{abstract}
A symmetry-preserving treatment of a vec-tor$\times$vector contact interaction is used to compute spectra of ground-state $J^P = 0^\pm, 1^\pm$ $(f\bar g)$ mesons, their partner diquark correlations, and $J^P=1/2^\pm, 3/2^\pm$ $(fg$ $h)$ baryons, where $f,g,h \in \{u,d,s,c,b\}$.  Results for the leptonic decay constants of all mesons are also obtained, including scalar and pseudovector states involving heavy quarks.  The spectrum of baryons produced by this chiefly algebraic approach reproduces the 64 masses known empirically or computed using lattice-regularised quantum chromodynamics with an accuracy of 1.4(1.2)\%.  It also has the richness of states typical of constituent-quark models and predicts many baryon states that have not yet been observed.  The study indicates that dynamical, nonpointlike diquark correlations play an important role in all baryons; and, typically, the lightest allowed diquark is the most important component of a baryon's Faddeev amplitude.
\end{abstract}



\maketitle


\section{Introduction}
\label{Sec:Introduction}
The empirical hadron spectrum is rich \cite{Zyla:2020zbs}, even without including the array of recently discovered exotic states \cite{Richard:2016eis, Esposito:2016noz, Liu:2019zoy, Barabanov:2020jvn, Eichmann:2020oqt}.  Yet, contemporary theory still predicts more states than have been observed.  This is especially true when one goes beyond systems comprised of light $\{u,d\}$-quarks and includes hadrons seeded by all flavours of valence quarks and/or antiquarks whose lifetime is long enough to produce measurable bound states, \emph{i.e}.\ also includes systems containing $\{s, c, b\}$ quarks and/or related antiquarks.  The empirically ``missing'' states are being sought in a worldwide effort  \cite{Galanti:2018bnp, KETZER2020103755, Carman:2020qmb, Brodsky:2020vco, NStar2020}.

Regarding calculations of hadron spectra, numerical simulations of lattice-regularised quantum chromodynamics (lQCD) provide a direct connection with the Standard Model Lagrangian.  Many collaborations are tackling the problem.  There are successes and challenges \cite{Bazavov:2009bb, Fodor:2012gf, Briceno:2017max, Hansen:2019nir}; and spectrum calculations are described, \emph{e.g}.\ in Refs.\,\cite{Durr:2008zz, Edwards:2011jj, Brown:2014ena, Lubicz:2017asp, Bali:2017pdv, Mathur:2018epb, Bahtiyar:2020uuj}.  Notably, few results are available on negative parity states and hadron radial excitations.

Explaining the mass and structure of parity partners in hadron spectra is crucial to understanding strong interactions because they would be degenerate if chiral symmetry were not dynamically broken, as discussed for the $a_1$- and $\rho$-mesons in Ref.\,\cite{Weinberg:1967kj}.  This dynamical chiral symmetry breaking (DCSB) is a corollary of emergent hadronic mass (EHM), the mechanism responsible for the $m_p \sim 1\,$GeV scale characteristic of visible matter \cite{Roberts:2020udq, Roberts:2020hiw, Roberts:2021xnz, Roberts:2021nhw}.  Current-quark masses enter into quantum chromodynamics (QCD) via the Higgs mechanism of explicit chiral symmetry breaking; so studying the spectra of states built from $\{u,d,s, c, b\}$ quarks and/or related antiquarks opens the way to exploring constructive interference between Nature's two distinct mass generating mechanisms.

Quark models have also been widely employed in the calculation of hadron spectra; see, \emph{e.g}.\ 
Refs.\,\cite{Aznauryan:2012ba, Crede:2013sze, Segovia:2013wma, Giannini:2015zia, Fernandez:2021zjq}.
Owing to complications introduced by the need to preserve chiral symmetry and its pattern of breaking in QCD, \emph{i.e}.\ to faithfully express DCSB, quark models are most reliable for baryons and mesons containing heavy quarks.  In such systems, Higgs-induced current-quark masses play the dominant role and differences between current- and constituent-quarks are least noticeable; see, \emph{e.g}.\ Ref.\,\cite[Fig.\,2.5]{Roberts:2021nhw}.

During the past decade, continuum Schwinger function methods, particularly Dyson-Schwinger equations, have increasingly been used in spectrum calculations following improvements in both \cite{Horn:2016rip, Eichmann:2016yit, Burkert:2017djo, Fischer:2018sdj, Barabanov:2020jvn, Qin:2020rad}: (\emph{i}) understanding the capacities and limitations of the approach; and (\emph{ii}) the quality and range of the description of hadron properties.
Recently, predictions for meson and baryon spectra in some of the low-lying flavour-SU$(N_f=5)$ multiplets were delivered \cite{Qin:2018dqp, Qin:2019hgk}.  They were made using the rainbow-ladder (RL) truncation of the continuum bound-state equations.  This is the leading-order in a systematic scheme \cite{Binosi:2016rxz, El-Bennich:2016qmb, Binosi:2016wcx, Fu:2017azw, Williams:2018adr, Xu:2018cor, Oliveira:2018fkj, Qin:2020jig}, which enables both: (\emph{i}) a unified, symmetry-preserving description of mesons and baryons, accounting properly for DCSB; and (\emph{ii}) a traceable connection to quantum chromodynamics (QCD).  A next challenge here is to advance beyond the leading-order truncation.  That effort will likely benefit from the use of high-performance computing.

Complementing such studies, Refs.\,\cite{Yin:2019bxe, Gutierrez-Guerrero:2019uwa} followed a different, largely algebraic path to calculating the spectra of ground-state pseudoscalar and vector $(f\bar g)$ mesons and $J^P=1/2^+, 3/2^+$ $(fgh)$ baryons, where $f,g,h \in \{u,d,s,c,b\}$.  Exploiting the fact that a had-ron's mass is a volume-integrated (long-wavelength) \linebreak quantity, so not very sensitive to details of the system's wave function, Refs.\,\cite{Yin:2019bxe, Gutierrez-Guerrero:2019uwa} used a symmetry-preserving treatment of a vector$\times$vector contact interaction (CI) \cite{Roberts:2011wy} to deliver insights into features of these systems that can be obscured in approaches that rely heavily on computer resources.  For instance, the analysis found diquark correlations to be important in all baryons studied; and owing to the dynamical character of the diquarks, the lightest allowed diquark correlation is typically that which defines the dominant component of a baryon's Faddeev amplitude.

Herein, we extend Refs.\,\cite{Yin:2019bxe, Gutierrez-Guerrero:2019uwa} and explore the capacity of the contact interaction to also explain the spectra of scalar and pseudovector mesons and $J^P=1/2^-$, $3/2^-$ baryons.
The meson spectrum is canvassed in Sec.\,\ref{Sec:Meson}, along with a discussion of leptonic decay constants, including those of scalar and pseudovector states.  The masses and decay constants of $\pi, K, \eta_c, \eta_b$ mesons are used to determine the current-masses of the $\{u=d,s,c,b\}$ quarks and basic CI parameters.  Predictions are then made for another sixty-six quantities.

Connections with diquark masses and correlation amplitudes are drawn in Sec.\,\ref{SecDiquarks}.  They are important because soft diquarks seemingly play an important role in hadron structure \cite{Barabanov:2020jvn}; hence, serve usefully in developing a quark+diquark approximation to the baryon three-body problem \cite{Cahill:1988dx, Burden:1988dt, Reinhardt:1989rw, Efimov:1990uz}.

Results for the spectra of ground-state $J=1/2^\pm$, $3/2^\pm$ baryons are discussed in Sec.\,\ref{Sec:Baryons}.   Following Refs.\,\cite{Roberts:2011cf, Eichmann:2016hgl, Lu:2017cln}, eight parameters are introduced to complete the CI definition in the baryon sector, with four serving to correct for omission of resonant (meson cloud) contributions in the quark+diquark Faddeev equation.  Masses for eighty baryons subsequently emerge as predictions along with their CI spin-flavour Faddeev amplitudes.

A summary and perspective is provided in Sec.\,\ref{Sec:Epilogue}.

Appendices are included to sketch the CI, \ref{Sec:Contact}; and detail the structure of flavour-SU$(5)$ baryon Faddeev amplitudes, \ref{AppFaddeev}.

\section{Meson Spectrum}
\label{Sec:Meson}
Ref.\,\cite{Yin:2019bxe} calculated the masses and leptonic decay constants of ground-state pseudoscalar and vector $(f\bar g)$ me-sons, $f,g \in \{u,d,s,c,b\}$.  Herein, we expand the coverage to include kindred scalar and pseudovector mesons, \emph{viz}.\ their parity partners.  All CI bound-states satisfy a Bethe-Salpeter equation of the form in Eq.\,\eqref{LBSEI}.  The pseudoscalar solution amplitude is given in Eq.\,\eqref{PSBSA} and the other systems are described by the following amplitudes:
\begin{subequations}
\label{fDecayConstant}
\begin{align}
\Gamma_\mu^{1^{-}}(Q) &= \gamma_\mu^\perp E_{1^{-}}(Q)\,,\\
\Gamma^{0^{+}}(Q) &= \textrm{I}_{D}E_{0^{+}}(Q)\,,\\
\Gamma_\mu^{1^{+}}(Q) &= \gamma_5 \gamma_\mu^\perp E_{1^{+}}(Q)\,,
\end{align}
\end{subequations}
where $Q_\mu \gamma_\mu^\perp = 0$ and $\textrm{I}_{D}$ is the $4\times 4$ identity matrix in spinor space.  When a momentum-dependent interaction is used, the amplitudes have a richer structure \cite{LlewellynSmith:1969az, Krassnigg:2009zh}; but the terms in Eqs.\,\eqref{fDecayConstant} are dominant \cite{Maris:2000ig, Chang:2011ei, Qin:2011xq}.

In terms of the amplitudes in Eqs.\,\eqref{fDecayConstant}, canonically normalised via Eq.\,\eqref{normcan} and its analogues, the leptonic decay constants are calculated using $(t_+ = t+Q)$
\begin{subequations}
\begin{align}
f_{1^{-}}m_{1^{-}} &= \frac{N_{c}}{3}\int\frac{\textrm{d}^{4}t}{(2\pi)^{4}}\textrm{tr}\big[\gamma_{\mu}S_{f}(t_+)
\Gamma_{\mu}^{1^{-}}(Q)S_{g}(t)\big]\,, \\
f_{0^{+}}P_{\mu}&= N_{c}\int\frac{\textrm{d}^{4}t}{(2\pi)^{4}}\textrm{tr}\big[i\gamma_{\mu}S_{f}(t_+)\Gamma^{0^{+}}(Q)
S_{g}(t)\big]\,,\\
f_{1^{+}}m_{1^{+}}&=\frac{N_{c}}{3}\int\frac{\textrm{d}^{4}t}{(2\pi)^{4}} \textrm{tr}\big[\gamma_5\gamma_{\mu}S_{f}(t_+)
\Gamma_{\mu}^{1^{+}}(Q)S_{g}(t)\big]\,,
\end{align}
\end{subequations}
with $N_c=3$, and $f_{0^-}$ given in Eq.\,\eqref{ffg}.  In $f=g$ systems \cite{Bhagwat:2006py}: $f_{0^{+}}\equiv 0$.

\begin{table}[t]
\caption{\label{Tab:MesonSpectrum}
Mesons: computed masses, Bethe-Salpeter amplitudes, and decay constants.
Empirical masses from Ref.\,\cite{Zyla:2020zbs};
entries marked by ``$\ast$'',
from Ref.\,\cite{Mathur:2018epb}.
Empirically unknown decay constants quoted from lQCD \cite{Becirevic:1998ua, Chiu:2007bc, Davies:2010ip, McNeile:2012qf, Donald:2012ga, Colquhoun:2015oha, Bali:2017pdv}.
%
%
%
(Dimensioned quantities in GeV.
Underlined entries from Table~\ref{Tab:DressedQuarks}.
``---'' indicates no empirical/lQCD results available for comparison.)
}
\begin{tabular*}
{\hsize}
{c@{\extracolsep{0ptplus1fil}}
l@{\extracolsep{0ptplus1fil}}||
l@{\extracolsep{0ptplus1fil}}
l@{\extracolsep{0ptplus1fil}}|
l@{\extracolsep{0ptplus1fil}}
l@{\extracolsep{0ptplus1fil}}|
l@{\extracolsep{0ptplus1fil}}
l@{\extracolsep{0ptplus1fil}}
}
\hline\hline
$J^{P}$ &      Meson               &     $m^{\rm CI}$     &        $m^{\rm e/l}$         &         $E$      &      $F$         &       $f^{\rm CI}$         &       $f^{\rm e/l}$   \\
\hline
$0^{-}$ & $\pi(u\bar{d})$          &      \underline{0.14}       &          0.14            &        3.59     &     0.47$\ $        &         \underline{0.10}          &           0.092         \\
        & $K(u\bar{s})$            &      \underline{0.50}       &          0.50            &        3.82     &     0.59$\ $        &         \underline{0.11}          &           0.11         \\

        & $D(u\bar{c})$            &      1.93       &          1.87            &        3.09     &     0.36$\ $        &         0.16          &           0.15         \\
        & $D_{s}(s\bar{c})$        &      2.01       &          1.97            &        3.23     &     0.48$\ $        &         0.17          &           0.18         \\
        & $\eta_{c}(c\bar{c})$     &      \underline{2.98}       &          2.98            &        3.25     &     0.72$\ $        &         \underline{0.24}          &           0.24         \\
        & $B(u\bar{b})$            &      5.42       &          5.28            &        1.67     &     0.095$\ $        &         0.17          &           0.14         \\
        & $B_{s}(s\bar{b})$        &      5.50       &          5.37            &        1.79     &     0.14$\ $        &         0.18          &           0.16         \\
        & $B_{c}(c\bar{b})$        &      6.28       &          6.28            &        3.38     &     0.61$\ $        &         0.27          &           0.35         \\
        & $\eta_{b}(b\bar{b})$     &     \underline{9.40}       &          9.40            &        3.18     &     0.81$\ $        &         \underline{0.41}          &           0.41         \\
\hline
$1^{-}$ & $\rho(u\bar{d})$         &      0.93       &          0.78            &        1.53     &              &         0.13          &           0.15         \\
        & $K^{\ast}(u\bar{s})$     &      1.03       &          0.89            &        1.63     &               &         0.12          &           0.16         \\
        & $\phi(s\bar{s})$         &      1.13       &          1.02            &        1.74     &               &         0.12          &           0.17         \\
        & $D^{\ast}(u\bar{c})$     &      2.14       &          2.01            &        1.20     &               &         0.15          &           0.17         \\
        & $D_{s}^{\ast}(s\bar{c})$ &      2.23       &          2.11            &        1.22     &              &         0.16          &           0.19         \\
        & $J/\psi(c\bar{c})$       &      3.19       &          3.10            &        1.19     &               &         0.20          &           0.29         \\
        & $B^{\ast}(u\bar{b})$     &      5.47       &          5.33            &        0.70     &               &         0.16          &           0.12         \\
        & $B_{s}^{\ast}(s\bar{b})$ &      5.56       &          5.42            &        0.71     &               &         0.16          &           0.15         \\
        & $B_{c}^{\ast}(c\bar{b})$ &      6.39       &          6.33$^\ast$     &        1.37     &              &         0.23          &           0.30         \\
        & $\Upsilon(b\bar{b})$     &      9.49       &          9.46            &        1.48     &              &         0.38          &           0.46         \\
\hline
$0^{+}$ & $\sigma(u\bar{d})$       &      1.22       &           ---             &        0.66     &              &         0          &            ---          \\
        & $\kappa(u\bar{s})$       &      1.34       &           ---             &        0.65     &               &         0.023          &            ---          \\
        & $D_{0}^{*}(u\bar{c})$    &      2.39       &          2.30            &        0.37     &              &         0.11          &            ---          \\
        & $D_{s0}^{*}(s\bar{c})$   &      2.51       &          2.32            &        0.35     &              &         0.083          &            $0.081(6)$          \\
        & $\chi_{c0}(c\bar{c})$    &      3.46       &          3.42            &        0.23     &               &         0         &            ---          \\
        & $B_{0}^{\ast}(u\bar{b})$    &      5.64       &           ---             &        0.22     &               &         0.19          &            ---          \\
        & $B_{s0}^{\ast}(s\bar{b})$&      5.74       &           ---             &        0.21     &               &         0.17          &            ---          \\
        & $B_{c0}^{\ast}(c\bar{b})$&      6.63       &          6.71$^\ast$           &        0.18     &               &         0.068          &            ---          \\
        & $\chi_{b0}(b\bar{b})$    &      9.78       &          9.86            &        0.10     &               &         0          &            ---          \\
\hline
$1^{+}$ & $a_{1}(u\bar{d})$        &      1.37       &          1.23            &        0.32     &              &         0.29          &            ---          \\
        & $K_{1}(u\bar{s})$        &      1.48       &          1.25            &        0.32     &              &         0.27          &            ---          \\
        & $f_{1}(s\bar{s})$        &      1.59       &          1.43            &        0.32     &               &         0.25          &            ---          \\
        & $D_{1}(u\bar{c})$        &      2.48       &          2.42            &        0.19     &               &         0.32          &            ---          \\
        & $D_{s1}(s\bar{c})$       &      2.59       &          2.46            &        0.18     &               &         0.31          &            $0.14(1)$          \\
        & $\chi_{c1}(c\bar{c})$    &      3.51       &          3.51            &        0.12     &               &         0.28          &            ---          \\
        & $B_{1}(u\bar{b})$        &      5.70       &          5.73            &        0.11     &              &         0.39          &            ---          \\
        & $B_{s1}(s\bar{b})$       &      5.79       &          5.83            &        0.11     &               &         0.37          &            ---          \\
        & $B_{c1}(c\bar{b})$&      6.66       &          6.74$^\ast$           &        0.096     &               &         0.24          &            ---          \\
        & $\chi_{b1}(b\bar{b})$    &      9.79       &          9.89            &        0.055     &               &         0.22          &            ---          \\
\hline\hline
\end{tabular*}
\end{table}

The explicit CI form for the vector-meson Bethe-Salpeter equation may be obtained simply by generalising Eqs.\,(18)\,--\,(20) in Ref.\,\cite{Chen:2012qr}; and those for the scalar and pseudovector channels from Ref.\,\cite[Appendices\,B.3, B.4]{Chen:2012qr}.
The latter, however, deserve additional attention.

It has long been known that RL truncation has some defects in the $0^+$, $1^+$ channels because these systems possess significant orbital angular momentum and RL fails to generate sufficient spin-orbit repulsion \cite{Chang:2009zb, Fischer:2009jm, Chang:2011ei, Qin:2020jig}.  Nonperturbatively generated corrections to RL truncation, such as those originating in EHM-induced dressed-quark anomalous chromomagnetic moments \cite{Chang:2010hb}, can remedy this defect and generate the empirical splitting between parity partners \cite{Chang:2011ei} and radial excitations \cite{Qin:2020jig}.  (Such corrections largely cancel in $0^-$ and $1^-$ channels.)  An ameliorating expedient was introduced in \linebreak Ref.\,\cite{Roberts:2011cf}, and refined in Refs.\,\cite{Eichmann:2016hgl, Lu:2017cln}.  Namely, one includes a multiplicative factor in the Bethe-Salpeter kernel for each of these channels, $g_{\rm SO}^{0^+}$, $g_{\rm SO}^{1^+}$, with $g_{\rm SO}^{0^+}$ chosen to produce a mass difference of approximately $0.3\,$GeV between the quark-core of the $0_{u\bar d}^+$ and that of the $\rho$-meson (as obtained with beyond-RL kernels \cite{Chang:2011ei}) and $g_{\rm SO}^{1^+}$ fixed to obtain the empirical size of the $a_1$-$\rho$ mass-splitting:
\begin{equation}
\label{SOfactors}
g_{\rm SO}^{0^+}=0.32\,, \quad  g_{\rm SO}^{1^+} = 0.25\,.
\end{equation}
(\emph{N.B}.\ In a Poincar\'e-covariant treatment, no bound state is purely $S$- or $P$-wave; and $g_{\rm SO}=1$ indicates no additional interaction beyond that generated by the RL kernel.)

At this point, all current-quark masses are determined; the coupling and cutoff in a given flavour channel are fixed using Eqs.\,\eqref{alphaLambda}, \eqref{alphaIRMass}, with the argument being the empirical pseudoscalar meson mass in this channel; and the spin-orbit correction factors are defined in Eq.\,\eqref{SOfactors}.
Consequently, apart from the four underlined entries, every CI result in Table~\ref{Tab:MesonSpectrum} is a prediction.  Recalling Ref.\,\cite{Yin:2019bxe}, the $F_{0^-}$ (pseudovector) component of each pseudoscalar meson is nonzero.  On average, it is 15(6)\% of the $E_{0^-}$ (pseudoscalar) piece; so, the pseudovector component is quantitatively important in all cases.

\begin{figure*}[t]
\vspace*{2ex}

\leftline{\hspace*{0.5em}{\large{\textsf{A}}}}
\vspace*{-3ex}
\includegraphics[width=0.95\textwidth]{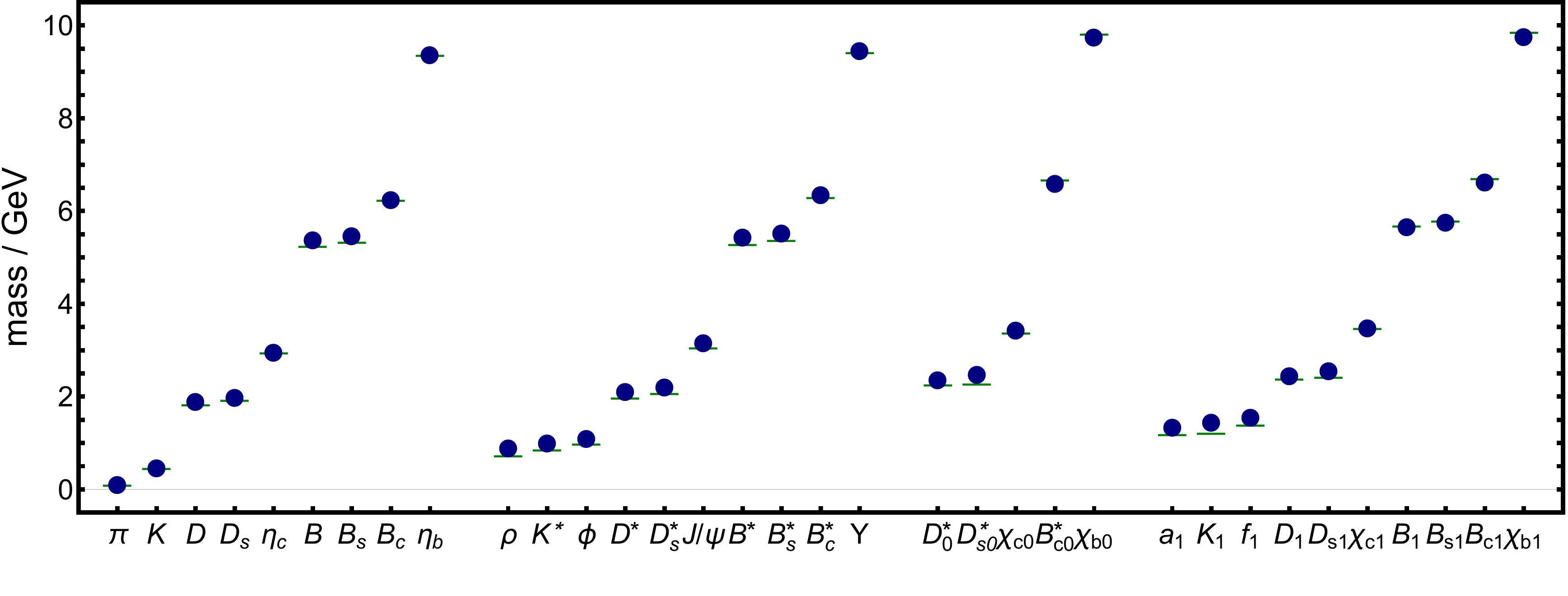}
\vspace*{1ex}

\leftline{\hspace*{0.5em}{\large{\textsf{B}}}}
\vspace*{-3ex}
\includegraphics[width=0.95\textwidth]{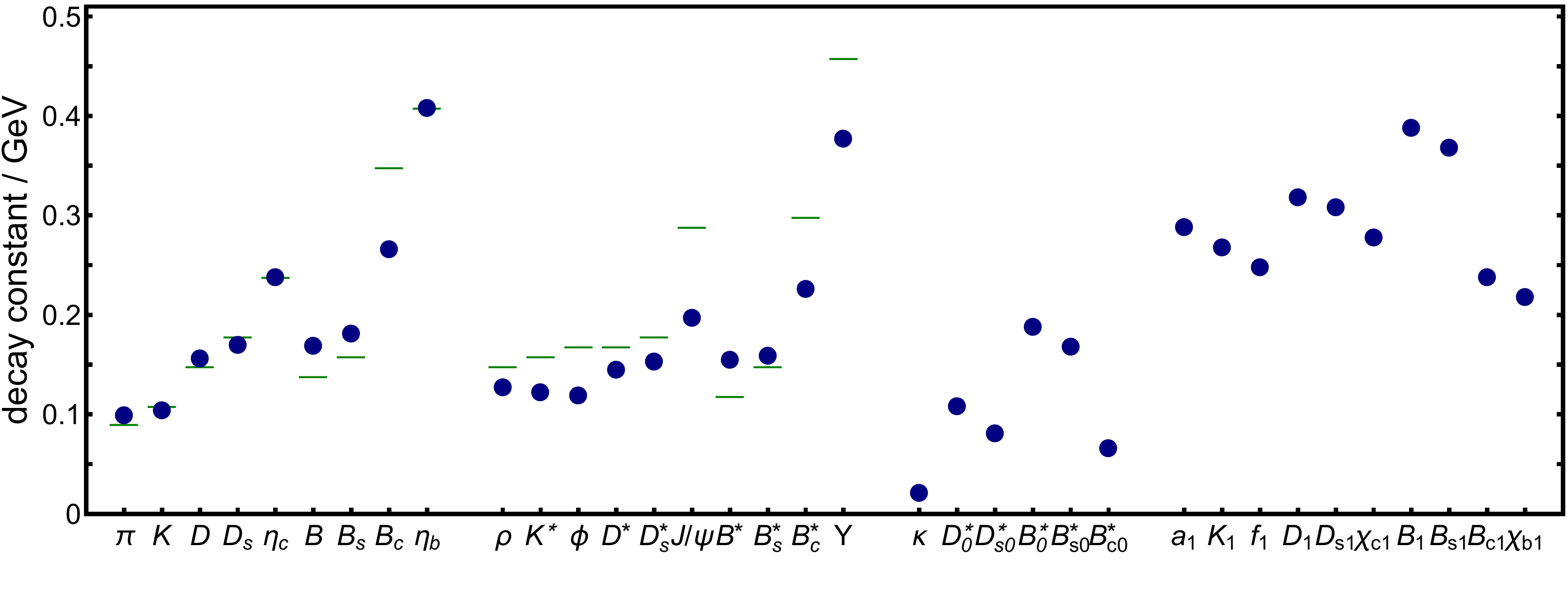}

\caption{\label{FigMeson}
\emph{Upper panel}\,--\,{\sf A}.
Comparison between CI predictions for meson masses and available experiment \cite{Zyla:2020zbs} and $m_{B^\ast_c}$, $m_{B_{c0}^{\ast}}$, $m_{B_{c1}}$ from Ref.\,\cite{Mathur:2018epb}.
\emph{Lower panel}\,--\,{\sf B}.
CI predictions for meson leptonic decay constants compared with experiment \cite{Zyla:2020zbs}, where known, and lQCD otherwise \cite{Becirevic:1998ua, Chiu:2007bc, Davies:2010ip, McNeile:2012qf, Donald:2012ga, Colquhoun:2015oha}.
In both panels, contact-interaction predictions are depicted as (blue) circles and comparison values by (green) bars.
(Pictorial representation of results in Table~\ref{Tab:MesonSpectrum}.)
}
\end{figure*}

To assist with understanding the comparisons in Table~\ref{Tab:MesonSpectrum}, we also represent them in Fig.\,\ref{FigMeson}.
Considering Fig.\,\ref{FigMeson}A, one sees that the CI delivers good estimates for the masses of both positive and negative parity ground-state flavour-SU$(5)$ mesons.  Aspects of its symmetry-preserving formulation skew results toward overestimates in most cases \cite{Roberts:2010rn}, but the mean relative-difference between theory and experiment/lQCD is only 5(6)\%.  (This value is obtained by comparing columns~1 and 2 in Table~\ref{Tab:MesonSpectrum}, omitting the underlined entries in column~1.)

Furthermore, as found with pseudoscalar and vector mesons in Ref.\,\cite{Yin:2019bxe}, the computed masses of scalar and pseudovector mesons neatly follow a pattern prescribed by equal spacing rules (ESRs) \cite{Okubo:1961jc, GellMann:1962xb, Qin:2018dqp, Chen:2019fzn, Qin:2019hgk}, \emph{e.g}.\
%
\begin{subequations}
\label{ESRexamples}
\begin{align}
m_{K_1} & \approx (m_{f_1}+m_{a_1})/2 \,,\\
m_{B_{s0}^{\ast}} - m_{B_{0}^\ast} & \approx m_{D_{s0}^\ast} - m_{D_{0}^\ast} \nonumber \\
& \approx m_{B_{s1}} - m_{B_{1}}
 \approx m_{D_{s1}} - m_{D_{1}} \,,\\
m_{B_{c1}}-m_{B_{s1}} & \approx m_{B_{c0}^{\ast}}-m_{B_{s0}^{\ast}}\,,\\
(m_{\chi_{b1}} - m_{\chi_{c1}})/2  & \approx (m_{\chi_{b0}} - m_{\chi_{c0}})/2 \nonumber \\
& \approx m_{B_{s1}} - m_{D_{s1}}
\!\!\approx m_{B_{s0}^{\ast}} - m_{D_{s0}^\ast} .  \label{MbMc}
\end{align}
\end{subequations}
These relations expose the scales that may be identified with the splittings between the spectrum-generating constituent-like quark masses, \emph{e.g}.\ Eqs.\,\eqref{MbMc} reveal a $b-c$ spectrum mass difference ${\mathsf d}_M^{bc} = 3.18(4)\,$GeV.  This is a fair match with $M_b-M_c = 3.23\,$GeV from Table~\ref{Tab:DressedQuarks}.  As evident in this table, the Higgs mechanism of mass generation is the origin of such splittings.

Interesting patterns are also found by comparing parity partners, \emph{e.g}.\
{\allowdisplaybreaks
\begin{subequations}
\begin{align}
0.44(1)\,{\rm GeV:} & \ m_{a_1} - m_{\rho} \approx m_{K_1} - m_{K^\ast} \nonumber \label{a1}\\
& \quad \approx m_{f_1}-m_\phi\,, \\
0.34(2)\,{\rm GeV:} & \ m_{\chi_{c1}} - m_{J/\psi} \approx m_{D_{s1}} - m_{D_s^\ast}  \nonumber \\
& \quad \approx m_{D_1} - m_{D^\ast} \,, \label{Xc1}\\
%
0.48(2)\,{\rm GeV:} & \ m_{\chi_{c0}} - m_{\eta_c} \approx m_{D_{s0}^\ast} - m_{D_s} \nonumber \\
& \quad \approx m_{D_{0}^\ast} - m_{D} \,, \label{Xc0}\\
0.25(4)\,{\rm GeV:} & \ m_{\chi_{b1}} - m_{\Upsilon} \approx m_{B_{c1}}-m_{B_c^\ast} \nonumber  \\ & \quad \approx m_{B_{s1}}-m_{B_s^\ast} \approx m_{B_1}-m_{B^\ast} \,, \label{Xb1}\\
0.30(8)\,{\rm GeV:} & \ m_{\chi_{b0}} - m_{\eta_b} \approx m_{B_{c0}^{\ast}}-m_{B_c} \nonumber  \\ & \quad \approx m_{B_{s0}^{\ast}}-m_{B_s} \approx m_{B_{0}^{\ast}}-m_{B}\,. \label{Xb0}
\end{align}
\end{subequations}}
\hspace*{-0.4\parindent}These results reveal the scale of DCSB in meson spectra.  Overall, parity partners are split by $0.35(10)\,$GeV.
However, comparing the impact of Higgs couplings into QCD, the expression of DCSB becomes weaker with increasing current-quark mass: Eq.\,\eqref{a1} \emph{cf}.\ \eqref{Xc1} \emph{cf}.\ \eqref{Xb1}; and Eq.\,\eqref{Xc0} \emph{cf}.\ \eqref{Xb0}.
Moreover, DCSB generates larger scalar-pseudoscalar splittings than pseudovector-vector: Eq.\,\eqref{Xb0} cf.\ \eqref{Xb1}; and Eq.\,\eqref{Xc0} \emph{cf}.\ \eqref{Xc1}.

%

CI results for meson leptonic decay constants are also listed in Table~\ref{Tab:MesonSpectrum} and drawn in Fig.\,\ref{FigMeson}B.  These quantities describe quark+antiquark annihilation at a single spacetime point, so they are sensitive to ultraviolet physics.  In QCD, this is expressed through the appearance of a logarithmic ultraviolet divergence in integrals like those in Eqs.\,\eqref{fDecayConstant}, which is compensated by the dressed-quark wave-function renormalisation constant \cite{Maris:1997hd, Ivanov:1998ms}.  It is therefore unsurprising that the cutoff-regularised CI supplies a poorer description of meson decay constants than it does of their masses.  It would be worse if Eqs.\,\eqref{alphaLambda}, \eqref{alphaIRMass} were not implemented.  Nevertheless, in comparison with known empirical or lQCD values, the picture is fair: trends are generally reproduced and the mean-absolute-relative-difference between the entries in columns~5 and 6 of Table~\ref{Tab:MesonSpectrum} is 15(11)\%.
(Given that very little is known about the decay constants of scalar and pseudovector mesons, they are excluded from this comparison.)
Peculiarities of the CI's formulation mean that the description is better for pseudoscalars than it is for vector mesons \cite{Roberts:2010rn}.
Further improvement can be achieved by additional tuning of Eq.\,\eqref{alphaLambda}, including its extension into the light-quark sector and consideration of the particular properties of heavy+light mesons.

Notwithstanding these observations, as with pseudoscalar and vector mesons, analogues of Eqs.\,\eqref{ESRexamples} are applicable, \emph{e.g}.\
{\allowdisplaybreaks
\begin{subequations}
\label{ESRfexamples}
\begin{align}
f_{K_1} & \approx [ f_{a_1} + f_{f_1} ]/2 \,,\\
f_{B_{c1}} & \approx [ f_{\chi_{c1}} + f_{\chi_{b1}} ]/2 \,,\\
%
f_{B_{0}^\ast}-f_{B_{s0}^{\ast}} & \approx f_{D_{0}^\ast}- f_{D_{s0}^\ast}  \nonumber \\
& \approx f_{B_{1}}- f_{B_{s1}}
 \approx f_{D_{1}}-f_{D_{s1}}  \,,\\
%
f_{B_{s1}}-f_{B_{c1}} & \approx f_{B_{s0}^{\ast}}-f_{B_{c0}^{\ast}}\,,\\
%
f_{B_{s1}} - f_{D_{s1}} &  \approx f_{B_{s0}^{\ast}} - f_{D_{s0}^\ast} \,.
\end{align}
\end{subequations}}
Such relations are not surprising when one recalls that such decay constants are order parameters for chiral symmetry breaking \cite{Bender:1996bm}.

Regarding Fig.\,\ref{FigMeson}B, one finds that using the CI the decay constants of vector mesons and those of pseudovector mesons have the opposite tendency with increasing current-quark mass, \emph{e.g}.\
\begin{align}
& f_{B^\ast} < f_{B_s^\ast}< f_{B_c^\ast} < f_\Upsilon \nonumber \\
\mbox{\emph{cf}.}\ & f_{B_1} > f_{B_{s1}}> f_{B_{c1}} > f_{\chi_{b1}}.
\end{align}
It is currently unclear whether this is a CI artefact.  However, it does follow the trend set by the meson Bethe-Salpeter amplitudes in Table~\ref{Tab:MesonSpectrum} and match qualitatively with the behaviour of scalar-meson decay constants.  The question could be answered by using a mo-mentum-dependent quark+antiquark scattering kernel \cite{Qin:2011xq}.

\section{Diquark Mass Spectrum}
\label{SecDiquarks}
\begin{figure}[t]
\centerline{%
\includegraphics[clip, width=0.45\textwidth]{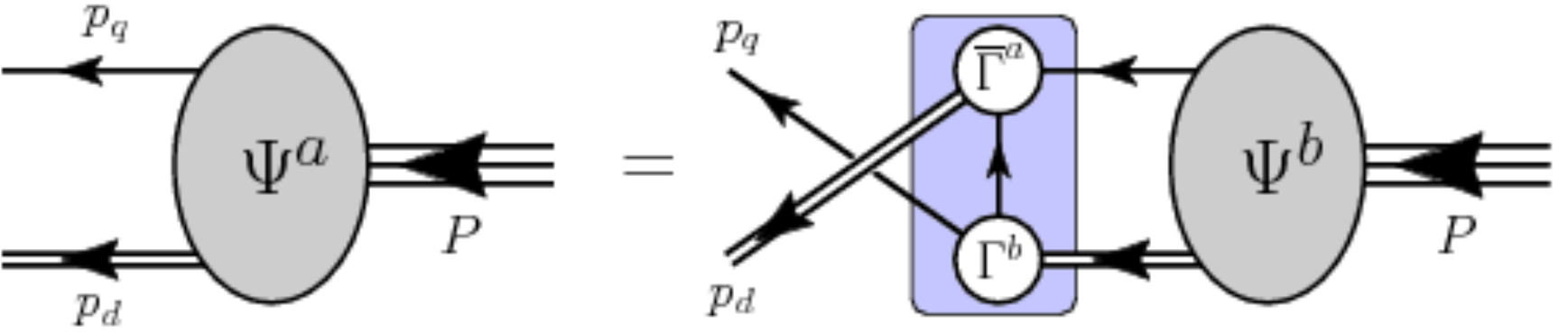}}
\caption{\label{figFaddeev}
Covariant Faddeev equation: linear integral equation satisfied by the matrix-valued Faddeev amplitude, $\Psi$, for a baryon of total momentum $P= p_d+p_q$, which expresses the relative-momentum correlation between the quarks and nonpointlike-diquarks within the baryon.  The shaded rectangle is the Faddeev equation kernel:
\emph{single line}, dressed-quark propagator (Appendix~\ref{Sec:Contact}); $\Gamma$,  diquark correlation amplitude (Sec.\,\ref{SecDiquarks}); and \emph{double line}, diquark propagator (Eqs.\,\eqref{diquarkprops}.) }
\end{figure}

In approaching the problem of calculating the spectrum of flavour-SU$(5)$ ground-state baryons with $J^P = 1/2^\pm, 3/2^\pm$, we use a quark+diquark approximation to the Faddeev equation introduced elsewhere \cite{Cahill:1988dx, Burden:1988dt, Reinhardt:1989rw, Efimov:1990uz}.  As apparent in Fig.\,\ref{figFaddeev}, here the diquarks are fully dynamical, appearing in a Faddeev kernel that requires their continual breakup and reformation.  In consequence, they are vastly different from the static, pointlike degrees-of-freedom considered in early models of baryons \cite{Anselmino:1992vg}.

To proceed, therefore, one must calculate the masses and amplitudes for all diquark correlations that can exist in $J^P = 1/2^\pm, 3/2^\pm$ baryons.  Using the CI, this means the following colour-$\bar 3$ correlations: flavour-$\overline{10}$ scalar; flavour-${15}$ pseudovector; flavour-$\overline{10}$ pseudoscalar; and flavour-$\overline{10}$ vector.\footnote{A $\bar 3_c$ flavour-$15$ vector-diquark is generally possible, but is not supported by a RL-like treatment of the CI \cite{Lu:2017cln}.   Using momentum-dependent interactions, flavour-symmetric vector correlations remain strongly suppressed \cite{Chen:2017pse}.}
This is a straightforward exercise because the RL Bethe-Salpeter equation for a $J^P$ diquark is generated from that for a $J^{-P}$ meson by simply multiplying the meson kernel by a factor of $1/2$ \cite{Cahill:1987qr}.  For instance, working from Eq.\,\eqref{LBSEI}, the mass and amplitude for a flavour-$\overline{10}$ scalar diquark is obtained from the following equation:
\begin{equation}
\Gamma^C_{[fg]}(Q) =  -  \frac{8 \pi}{3} \frac{\alpha_{\rm IR}}{m_G^2}
\int \!\! \frac{d^4t}{(2\pi)^4} \! \gamma_\mu S_f(t_+) \Gamma^C_{[fg]}(Q) S_g(t) \gamma_\mu \,,
\label{LBSEqq}
\end{equation}
where the correlation amplitude is $\Gamma_{[fg]}(Q)$ and
\begin{subequations}
\begin{align}
\Gamma^C_{[fg]}(Q) & := \Gamma_{[fg]}(Q)C^\dagger\\
& = \gamma_5 \left[ i E_{[fg]}(Q) + \frac{1}{2 M_{f g}} \gamma\cdot Q F_{[fg]}(Q)\right]\,,
\end{align}
\end{subequations}
with $C=\gamma_2\gamma_4$ being the charge-conjugation matrix.  The canonical normalisation conditions are likewise \linebreak amended, with the multiplicative factor being $2/3$ in this case.

In a study of flavour-SU$(3)$ baryon spectra \cite{Lu:2017cln}, an additional step was found to be necessary in order to arrive at realistic results.  Namely, considering that valence-quarks within a diquark are less tightly correlated than the valence-quark and -antiquark in a bound-state meson, then spin-orbit repulsion in diquarks should be weaker than it is in mesons.  This was effected by writing
\begin{equation}
\label{eqsSO}
g_{\rm SO}^{1^-_{qq},0^-_{qq}} = g_{\rm SO}^{1^+_{q\bar q},0^+_{q\bar q}} \times {\mathpzc s}_{\rm SO} \,,\;
{\mathpzc s}_{\rm SO}=1.8\,,
\end{equation}
so that the modification factor in the RL-like diquark Bethe-Salpeter equations is nearer unity; hence, generates less repulsion.  The value of ${\mathpzc s}_{\rm SO}$ was chosen in concert with
$g_{DB}^{\pi_\Psi \pi_d}$, Eq.\,\eqref{gDBval}, in order to ensure that the dressed-quark core mass of the nucleon's parity partner is not lower than that of its first radial excitation.

\begin{table}[t]
\caption{\label{diquarkspectrum}
Diquark correlations: computed masses (GeV) and Bethe-Salpeter amplitudes of scalar, axial-vector, pseudoscalar and vector diquarks.  \emph{N.B}.\ We work in the isospin-symmetry limit; so, e.g.\ $[ds]_{0^{+}}=[us]_{0^{+}}$.}
\begin{center}
\begin{tabular*}
{\hsize}
{l@{\extracolsep{0ptplus1fil}}
l@{\extracolsep{0ptplus1fil}}|
l@{\extracolsep{0ptplus1fil}}
l@{\extracolsep{0ptplus1fil}}|
l@{\extracolsep{0ptplus1fil}}
l@{\extracolsep{0ptplus1fil}}
}
\hline\hline
$J^{P}$     &        Diquark$\ $         &   $m_{qq}^{CI}$  & $m_{q\bar{q}}^{CI}$ $\ $ &      $E_{qq}$      &      $F_{qq}$   \\
\hline
$0^{+}$     &     $[ud]_{0^{+}}$        &      0.78       &      0.14$\ $          &     2.74     &     0.31  \\
            &     $[us]_{0^{+}}$        &      0.93       &      0.50$\ $          &     2.88     &     0.39  \\

            &     $[uc]_{0^{+}}$        &      2.15       &      1.93$\ $          &     1.97     &     0.22  \\
            &     $[sc]_{0^{+}}$        &      2.26       &      2.01$\ $          &     1.99     &     0.29  \\
            &     $[ub]_{0^{+}}$        &      5.51       &      5.42$\ $          &     1.05     &     0.059  \\
            &     $[sb]_{0^{+}}$        &      5.60       &      5.50$\ $          &     1.05     &     0.083  \\
            &     $[cb]_{0^{+}}$        &      6.48       &      6.28$\ $          &     1.42     &     0.25  \\
\hline
$1^{+}$     &     $\{uu\}_{1^{+}}$      &      1.06       &      0.93$\ $          &     1.31     &     \\
            &     $\{us\}_{1^{+}}$      &      1.16       &      1.03$\ $          &     1.36     &         \\
            &     $\{ss\}_{1^{+}}$      &      1.26       &      1.13$\ $          &     1.43     &          \\
            &     $\{uc\}_{1^{+}}$      &      2.24       &      2.14$\ $          &     0.89     &          \\
            &     $\{sc\}_{1^{+}}$      &      2.34       &      2.23$\ $          &     0.87     &          \\
            &     $\{cc\}_{1^{+}}$      &      3.30       &      3.19$\ $          &     0.69     &          \\
            &     $\{ub\}_{1^{+}}$      &      5.53       &      5.47$\ $          &     0.51     &          \\
            &     $\{sb\}_{1^{+}}$      &      5.62       &      5.56$\ $          &     0.50     &          \\
            &     $\{cb\}_{1^{+}}$      &      6.50       &      6.39$\ $          &     0.62     &          \\
            &     $\{bb\}_{1^{+}}$      &      9.68       &      9.49$\ $          &     0.48     &          \\
\hline
$0^{-}$     &     $[ud]_{0^{-}}$        &      1.15       &      1.22$\ $          &     1.07     &          \\
            &     $[us]_{0^{-}}$        &      1.28       &      1.34$\ $          &     1.05     &          \\

            &     $[uc]_{0^{-}}$        &      2.35       &      2.39$\ $          &     0.62     &          \\
            &     $[sc]_{0^{-}}$        &      2.48       &      2.51$\ $          &     0.56     &          \\
            &     $[ub]_{0^{-}}$        &      5.61       &      5.64$\ $          &     0.36     &          \\
            &     $[sb]_{0^{-}}$        &      5.72       &      5.74$\ $          &     0.33     &          \\
            &     $[cb]_{0^{-}}$        &      6.62       &      6.63$\ $          &     0.28     &          \\
\hline
$1^{-}$     &     $[ud]_{1^{-}}$        &      1.33       &      1.37$\ $          &     0.51     &          \\
            &     $[us]_{1^{-}}$        &      1.44       &      1.48$\ $          &     0.51     &          \\
            &     $[uc]_{1^{-}}$        &      2.45       &      2.48$\ $          &     0.30     &          \\
            &     $[sc]_{1^{-}}$        &      2.56       &      2.59$\ $          &     0.28     &          \\
            &     $[ub]_{1^{-}}$        &      5.67       &      5.70$\ $          &     0.18     &          \\
            &     $[sb]_{1^{-}}$        &      5.77       &      5.79$\ $          &     0.17     &          \\
            &     $[cb]_{1^{-}}$        &      6.65       &      6.66$\ $          &     0.15     &          \\
\hline\hline
\end{tabular*}
\end{center}
\end{table}

\begin{figure*}[t]
\includegraphics[width=0.95\textwidth]{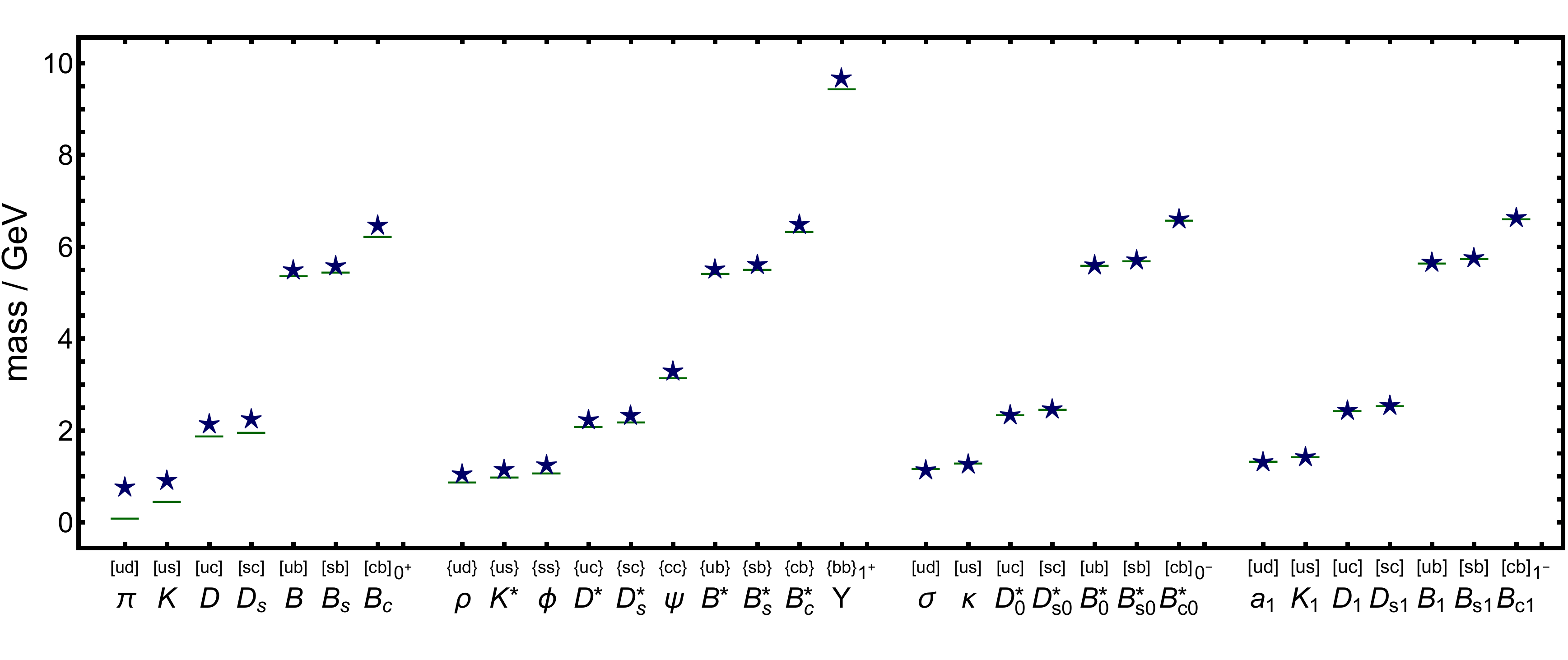}
\caption{\label{Fig:Diquark}
Computed masses of diquark correlations and their symmetry-related meson counterparts: diquarks -- (blue) stars and mesons -- (green) bars.  (Pictorial representation of Table~\ref{diquarkspectrum}.)
}
\end{figure*}

Following this approach, using parameters determined as described here, in Sec.\,\ref{Sec:Meson} and Appendix~\ref{Sec:Contact}, one obtains the diquark masses and amplitudes listed in Table~\ref{diquarkspectrum}.  Consistent with RL studies using realistic interactions \cite{Maris:2002yu, Maris:2004bp}, in the Bethe-Salpeter equation for a given $J^P$ diquark correlation we use the values of $\alpha_{\rm IR}$, $\Lambda_{\rm uv}$ associated with its $J^{-P}$ meson partner.  A $\pm 10$\% change in ${\mathpzc s}_{\rm SO}$ alters the $0^-$, $1^-$ values by $\mp 2$\%.
Evidently, the antisymmetric Dirac-scalar combination of any two quark flavours (scalar diquark) is always lighter than the symmetric $J^P=1^+$ combination (pseudovector diquark, denoted $\{fg\}_{1^+}$), followed by the pseudoscalar and then the pseudovector diquarks.  Just like mesons, the pattern of masses can be understood in terms of ESRs.


It is worth reiterating that diquark correlations are coloured and it is only in connection with partnering coloured objects (quark, another diquark, etc.) that a colour singlet system is obtained.  Hence, diquarks are confined.  That is not true in RL truncation \cite{Maris:2002yu}; but corrections to this leading-order analysis have been studied using an infrared-dominant interaction \cite{Munczek:1983dx}.  In fully self-consistent symmetry-preser\-ving treatments, such corrections eliminate diquark bound-state poles from the quark+quark scattering matrix, whilst preserving the strong correlations \cite{Bhagwat:2004hn}.  In these cases, one still has a physical interpretation of the RL masses; to wit, the quantity $\ell_{(fg)^{\!J^P}}:=1/m_{(fg)^{\!J^P}}$ is a spacetime distance that the diquark correlation can propagate before fragmentation.

Fig.\,\ref{Fig:Diquark} compares calculated diquark masses with those of their partner mesons.  The level ordering of diquark correlations is precisely the same as that for mesons.  Scalar and pseudovector diquarks are heavier than their partner mesons.  Owing to the introduction of ${\mathpzc s}_{\rm SO}$ in Eq.\,\eqref{eqsSO}, this ordering is reversed for pseudoscalar and vector diquark correlations.  Notwithstanding this, omitting the $[ud]_{0^+}$ and $[us]_{0^+}$ correlations, the mass of a diquark's partner meson is a reasonable guide to the diquark's mass: the mean difference in absolute value is $0.08(7)\,$GeV.

Owing to DCSB, the light-quark scalar diquark channels are atypical because their partner mesons are the almost-Nambu-Goldstone modes, $\pi$ and $K$.  In two-colour chiral-limit QCD, scalar diquarks are also Nambu-Gold-stone modes \cite{Bloch:1999vk}.   A hint of this remains in the three-colour theory, expressed in the relatively low masses of $[ud]_{0^+}$ and $[us]_{0^+}$, even though they are split widely from the $\pi$ and $K$.

It is the canonically normalised diquark Bethe-Sal-peter amplitudes that appear in baryon Faddeev equations.  Listed in Table~\ref{diquarkspectrum}, they exhibit a clear ordering:  in any given flavour sector, scalar diquarks have the largest amplitudes $\leftrightarrow$ strongest couplings.  This means that the lighter flavour-$\overline{10}$ scalar diquark correlations are typically favoured in all $J=1/2$ baryon amplitudes because the Faddeev equations involve the diquark-amp-litude-squared.  As will become apparent, under certain circumstances, \emph{e.g}.\ in baryons whose valence-quarks have widely different masses, this preference can be overcome.

Notably, the flavour-exchange symmetries of $J^P=3/2^\pm$ baryons mean that such CI states may only contain flavour-$15$ pseudovector diquark correlations, \emph{viz}.\ positive-parity diquarks.  Using momentum-dependent interactions, flavour-$15$ vector diquarks are possible.  \linebreak However, these correlations are weak; hence, even then, are unlikely to play a significant role.  Nevertheless, this expectation should be checked.


\section{Baryon Spectrum}
\label{Sec:Baryons}
Following Ref.\,\cite{Yin:2019bxe}, we use the Faddeev equation drawn in Fig.\,\ref{figFaddeev} to calculate the spectrum of ground-state fla-vour-SU$(5)$ $J^P=1/2^\pm, 3/2^\pm$ baryons.  In the isospin-symmetry limit, $88$ distinct states are supported.  Details are presented in \ref{AppFaddeev}.

\begin{table*}[t]
\caption{\label{ResultsHalf}
Computed mass and Faddeev amplitude for each ground-state flavour-SU$(5)$ $J^P=1/2^+$ baryon.
The last column highlights the baryon's dominant spin-flavour correlation.  The arrays of possibilities are given in Eqs.\,\eqref{halfnucleon}, \eqref{half2q1Q}, \eqref{HalfOneTwo}, \eqref{OneHalfQQQ}.
Empirical mass values, $M^{e}$, are taken from Ref.\,\cite{Zyla:2020zbs}; and lQCD results, $M^{l}$, from Refs.\,\cite{Durr:2008zz, Brown:2014ena, Mathur:2018epb}.
 The results under heading $M^3$ are the three-body Faddeev equation predictions in Ref.\,\cite{Qin:2019hgk}.  This analysis did not include SU$(3)$-flavour-symmetry breaking effects.
 (Masses in GeV.)
 }
\begin{center}
\begin{tabular*}
{\hsize}
{l@{\extracolsep{0ptplus1fil}}|
r@{\extracolsep{0ptplus1fil}}
r@{\extracolsep{0ptplus1fil}}
r@{\extracolsep{0ptplus1fil}}
r@{\extracolsep{0ptplus1fil}}|
r@{\extracolsep{0ptplus1fil}}
r@{\extracolsep{0ptplus1fil}}
r@{\extracolsep{0ptplus1fil}}
r@{\extracolsep{0ptplus1fil}}
r@{\extracolsep{0ptplus1fil}}
r@{\extracolsep{0ptplus1fil}}
r@{\extracolsep{0ptplus1fil}}
r@{\extracolsep{0ptplus1fil}}
r@{\extracolsep{0ptplus1fil}}
r@{\extracolsep{0ptplus1fil}}
r@{\extracolsep{0ptplus1fil}}
r@{\extracolsep{0ptplus1fil}}
r@{\extracolsep{0ptplus1fil}}
r@{\extracolsep{0ptplus1fil}}|
c@{\extracolsep{0ptplus1fil}}}
\hline\hline
   & $M^{\rm CI}$ & $M^{e}$ & $M^{l}$  & $M^{3}$ & $s^{r_{1}}$ & $s^{r_{2}}$ & $a_{1}^{r_{2}}$ & $a_{2}^{r_{2}}$ & $a_{1}^{r_{3}}$ & $a_{2}^{r_{3}}$ & $p^{r_{4}}$ & $p^{r_{5}}$ & $v_{1}^{r_{5}}$ & $v_{2}^{r_{5}}$ & $v_{1}^{r_{6}}$ & $v_{2}^{r_{6}}$ & $v_{1}^{r_{7}}$ & $v_{2}^{r_{7}}$ & dom.\ corr.\                  \\
\hline
$N$                &  0.98    &  0.94      &   0.94      &   0.95\;    &  0.88       &             & -0.38           & -0.06           &  0.27           &  0.05           & -0.02       &             & -0.02           &  0              &                 &                 &                 &                 & $u[ud]_{0^{+}}$                \\
$\Lambda$          &  1.10    &  1.12      &   1.11      &   1.11\;    &  0.66       &  0.59       &                 &                 & -0.45           & -0.08           & -0.02       & -0.03       &                 &                 & -0.01           &  0              & -0.01           &  0\;              & $s[ud]_{0^{+}}$                \\
$\Sigma$           &  1.20    &  1.19      &   1.17      &   1.11\;    &  0.84       &             & -0.45           &  0.13           &  0.26           &  0.01           & -0.01       &             & -0.02           &  0              &                 &                 &                 &                 & $u[us]_{0^{+}}$                \\
$\Xi$              &  1.27    &  1.32      &   1.32      &   1.28\;    &  0.89       &             & -0.31           &  0.04           &  0.33           &  0.05           & -0.03       &             & -0.02           &  0.01           &                 &                 &                 &                 & $s[us]_{0^{+}}$                \\
\hline
$\Lambda_{c}$      &  2.40    &  2.29      &   2.25      &   2.18\;    &  0.18       &  0.86       &                 &                 & -0.35           & -0.33           & -0.02       & -0.05       &                 &                 & -0.02           &  0.03           & -0.02           &  0.01\;           & $d[uc]_{0^{+}}-u[dc]_{0^{+}}$  \\
$\Sigma_{c}$       &  2.45    &  2.45      &   2.47      &   2.18\;    &  0.46       &             & -0.20           &  0.86           &  0.09           &  0.06           &  0.02       &             &  0              &  0              &                 &                 &                 &                 & $c\{uu\}_{1^{+}}$              \\
$\Xi_{c}$          &  2.55    &  2.47      &   2.43      &   2.35\;    &  0.18       &  0.84       &                 &                 & -0.36           & -0.35           & -0.02       & -0.06       &                 &                 & -0.02           &  0.03           & -0.02           &  0.01\;           & $s[uc]_{0^{+}}-u[sc]_{0^{+}}$  \\
$\Xi_{c}^{\prime}$ &  2.59    &  2.58      &   2.57      &   2.35\;    &  0.48       &             & -0.21           &  0.85           &  0.09           &  0.06           &  0.02       &             &  0              &  0              &                 &                 &                 &                 & $c\{us\}_{1^{+}}$              \\
$\Omega_{c}$       &  2.73    &  2.70      &   2.68      &   2.52\;    &  0.50       &             & -0.21           &  0.83           &  0.09           &  0.06           &  0.01       &             &  0              &  0              &                 &                 &                 &                 & $c\{ss\}_{1^{+}}$              \\
\hline
$\Lambda_{b}$      &  5.62    &  5.62      &   5.63      &   5.39\;    &  0.10       &  0.93       &                 &                 & -0.31           & -0.13           &  0          &  0.03       &                 &                 & -0.01           &  0.04           & -0.01           & -0.01\;           & $d[ub]_{0^{+}}-u[db]_{0^{+}}$  \\
$\Sigma_{b}$       &  5.75    &  5.81      &   5.86      &   5.39\;    &  0.33       &             & -0.12           &  0.93           &  0.04           &  0.10           &  0.01       &             &  0              &  0              &                 &                 &                 &                 & $b\{uu\}_{1^{+}}$              \\
$\Xi_{b}$          &  5.75    &  5.79      &   5.77      &   5.56\;    &  0.11       &  0.94       &                 &                 & -0.31           & -0.13           &  0          & -0.03       &                 &                 & -0.01           &  0.03           & -0.01           &  0\;              & $s[ub]_{0^{+}}-u[sb]_{0^{+}}$  \\
$\Xi_{b}^{\prime}$ &  5.88    &  5.94      &   5.93      &   5.56\;   &  0.37       &             & -0.11           &  0.92           &  0.05           &  0.12           &  0.01       &             &  0              &  0              &                 &                 &                 &                 & $b\{us\}_{1^{+}}$              \\
$\Omega_{b}$       &  6.00    &  6.05      &   6.06      &   5.73\;    &  0.42       &             & -0.11           &  0.89           &  0.05           &  0.14           &  0.02       &             &  0              & -0.01           &                 &                 &                 &                 & $b\{ss\}_{1^{+}}$              \\
\hline
$\Xi_{cc}$         &  3.64    &  3.62      &   3.61      &   3.42\;    &  0.88       &             & -0.34           &  0.30           &  0.10           &  0.06           & -0.05       &             & -0.03           &  0.03           &                 &                 &                 &                 & $c[uc]_{0^{+}}$                \\
$\Xi_{cb}$         &  6.97    &            &   6.94      &   6.63\;    &  0.07       & -0.77       &                 &                 &  0.62           & -0.01           &  0.12       &  0.01       &                 &                 &  0.02           & -0.01           &  0.01           &  0\;              & $u[cb]_{0^{+}}$                \\
$\Xi_{cb}^{\prime}$&  6.89    &            &   $6.96$      &   6.63\;    &  0.89       &             & -0.33           & -0.14           &  0.13           &  0.25           & -0.06       &             & -0.03           &  0.02           &                 &                 &                 &                 & $b[uc]_{0^{+}}+c[ub]_{0^{+}}$  \\
$\Xi_{bb}$         &  10.22   &            &   10.14     &   9.84\;    &  0.85       &             & -0.36           &  0.37           &  0.04           &  0.06           & -0.06       &             & -0.05           &  0.04           &                 &                 &                 &                 & $b[ub]_{0^{+}}$                \\
\hline
$\Omega_{cc}$      &  3.79    &            &   3.74      &   3.59\;    &  0.86       &             & -0.35           &  0.33           &  0.15           &  0.07           & -0.05       &             & -0.03           &  0.03           &                 &                 &                 &                 & $c[sc]_{0^{+}}$                \\
$\Omega_{cb}$      &  7.08    &            &   7.00      &   6.80\;    &  0.05       & -0.84       &                 &                 &  0.50           & -0.19           &  0.08       &  0.01       &                 &                 &  0.01           &  0              &  0.01           &  0\;              & $s[cb]_{0^{+}}$                \\
$\Omega_{cb}^{\prime}$ &  7.01    &            &   7.03      &   6.80\;    &  0.83       &             & -0.32           & -0.21           &  0.18           &  0.36           & -0.06       &             & -0.03           &  0.02           &                 &                 &                 &                 & $b[sc]_{0^{+}}+c[sb]_{0^{+}}$  \\
$\Omega_{bb}$      &  10.33   &            &   $\rule{-0.5em}{0ex}$10.27     &   10.01\;   &  0.83       &             & -0.38           &  0.39           &  0.07           &  0.09           & -0.07       &             & -0.05           &  0.04           &                 &                 &                 &                 & $b[sb]_{0^{+}}$                \\
\hline
$\Omega_{ccb}\ $     &  8.04    &            &   8.01      &   7.87\;    &  0.40       &             & -0.17           &  0.89           &  0.07           &  0.08           &  0.01       &             &  0              &  0              &                 &                 &                 &                 & $b\{cc\}_{1^{+}}$              \\
$\Omega_{cbb}\ $     &  11.22   &            &   11.20     &   11.08\;   &  0.80       &             & -0.39           &  0.40           &  0.16           &  0.14           & -0.05       &             & -0.03           &  0.03           &                 &                 &                 &                 & $b[cb]_{0^{+}}$                \\
\hline\hline
\end{tabular*}
\end{center}
\end{table*}

\subsection{Completing the Faddeev kernels}
As highlighted above, the kernel in Fig.\,\ref{figFaddeev} introduces binding through diquark breakup and reformation via exchange of a dressed-quark.  Again following Ref.\,\cite{Yin:2019bxe}, we exploit an oft used simplification, \emph{viz}.\ in the Faddeev equation for a baryon of type $B$, the quark exchanged between the diquarks, Eq.\,\eqref{AppendixFE}, is represented as
\begin{equation}
S_g^{\rm T}(k) \to \frac{g_B^2}{M_g}\,,
\label{staticexchange}
\end{equation}
where $g=l,s,c,b$ is the quark's flavour and $g_B$ is a coupling constant.  This is a variation on the  ``static approximation'' introduced in Ref.\,\cite{Buck:1992wz}.  It makes the Faddeev amplitudes momentum-independent, just like the diquark Bethe-Salpeter amplitudes.  Calculations reveal that it has little impact on the calculated masses \cite{Xu:2015kta}.
The couplings $g_{N}$, $g_{\Delta}$ are treated as parameters, with values chosen to obtain desired masses for the nucleon and $\Delta$-baryon, $m_N$, $m_\Delta$.

\begin{figure*}[t]
\vspace*{2ex}

\leftline{\hspace*{0.5em}{\large{\textsf{A}}}}
\vspace*{-3ex}
\includegraphics[width=0.99\textwidth]{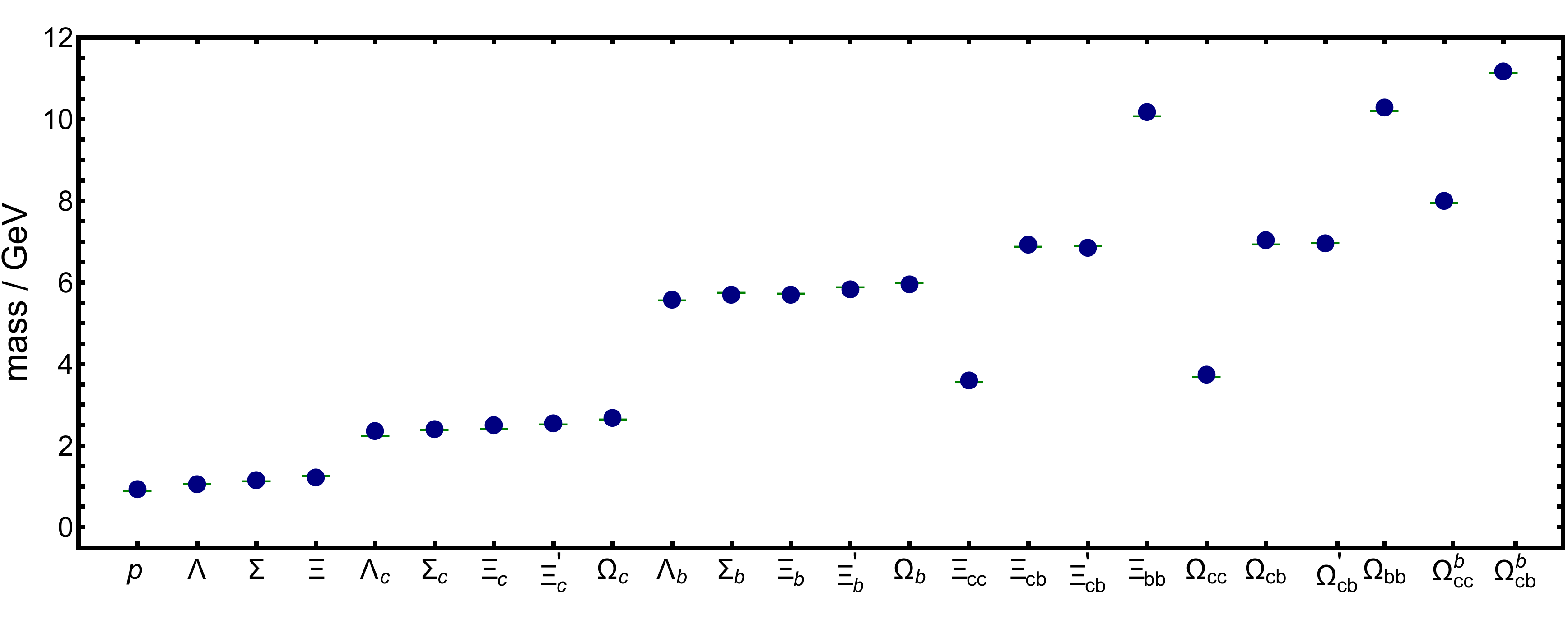}
\vspace*{1ex}

\leftline{\hspace*{0.5em}{\large{\textsf{B}}}}
\vspace*{-3ex}
\includegraphics[width=0.99\textwidth]{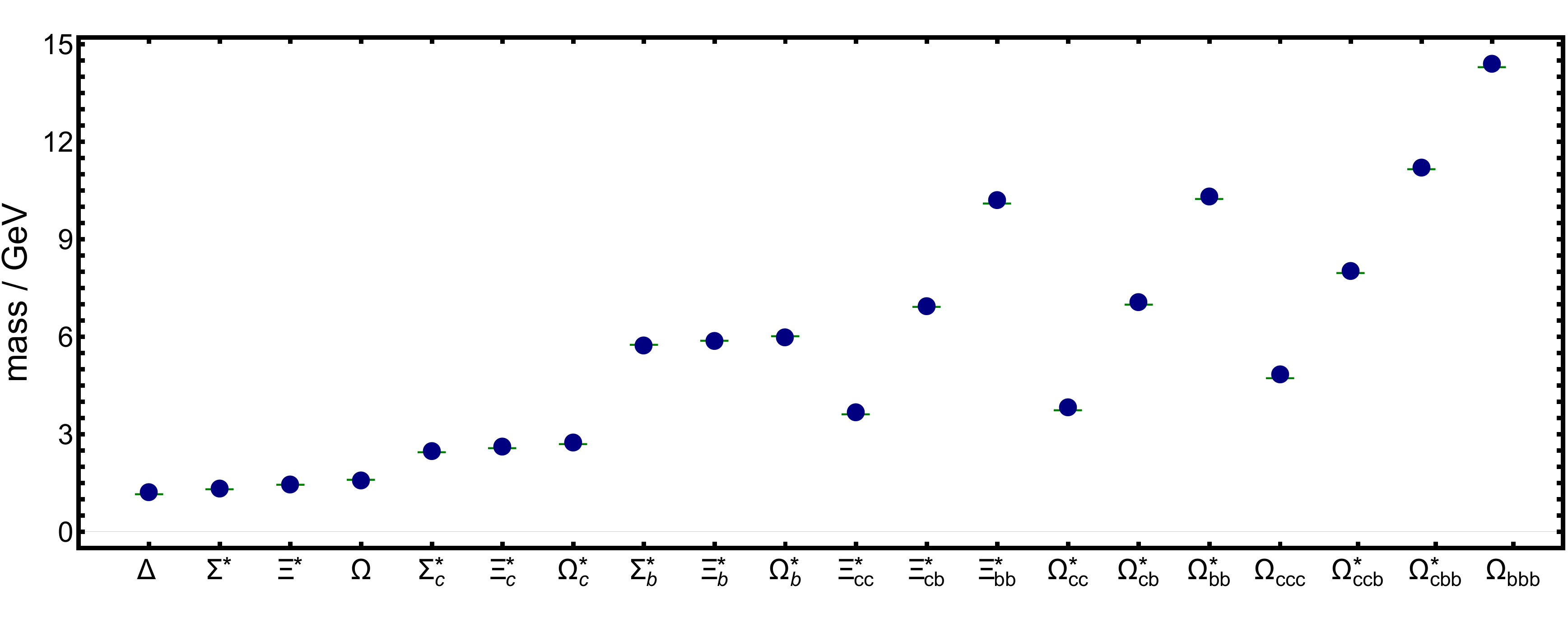}
\caption{\label{figResultsHalf}
\emph{Upper panel}\,--\,{\sf A}.
CI computed masses (in GeV) of ground-state flavour-SU$(5)$ $J^P=1/2^+$ baryons in Table~\ref{ResultsHalf} compared with either experiment  (first 15) \cite{Zyla:2020zbs} or lQCD (last 9) \cite{Brown:2014ena, Mathur:2018epb}.
\emph{Lower panel}\,--\,{\sf B}.
Similar comparison for ground-state flavour-SU$(5)$ $J^P=3/2^+$ baryons in Table~\ref{ResultsThreeHalf}: experiment  (first 9) \cite{Zyla:2020zbs}; or lQCD (last 11) \cite{Brown:2014ena, Mathur:2018epb}.
Both panels: CI results -- (blue) circles; reference values  -- (green) bars.
}
\end{figure*}

The Faddeev equation, Fig.\,\ref{figFaddeev}, generates what may be called a baryon's dressed-quark core \cite{Hecht:2002ej}: it omits what are typically described as meson-cloud contributions to the baryon masses, which work to reduce $m_N$, $m_\Delta$.  The sizes of such corrections has been estimated: for the nucleon, the reduction is roughly $0.2\,$GeV and for the $\Delta$ it is $0.16\,$GeV.  The choices \cite{Roberts:2011cf}
\begin{equation}
\label{gNgD}
g_N = 1.18\,,\; g_\Delta=1.56\,,
\end{equation}
produce $m_N=1.14\,$GeV and $m_\Delta=1.39\,$GeV, \emph{i.e}.\ inflated masses that leave room for correction by meson cloud effects.  (\emph{N.B}.\ The values in Eq.\,\eqref{gNgD} are deliberately different from those used in Ref.\,\cite{Yin:2019bxe}.)

Using the framework outlined above, Ref.\,\cite{Lu:2017cln} studied flavour-SU$(3)$ baryons, finding that the ground-state positive-parity octet baryons are primarily constituted from like-parity diquarks, with negligible contributions from negative-parity correlations.  Somewhat surprisingly, it found that their parity partners are also dominated by positive-parity diquark correlations; hence, too light.  As with mesons and diquarks, the missing element was identified as too little spin-orbit repulsion generated by RL-like kernels.  This was remedied by inserting an additional parameter into the Faddeev equation for $J^P=(1/2)^P$ baryons, \emph{i.e}.\ $g_{DB}^{\pi_\Psi \pi_d}$, a linear multiplicative factor, expressing EHM-induced enhancement of spin-orbit repulsion, attached to each diquark amplitude in the baryon's Faddeev equation kernel, Eq.\,\eqref{AppendixFE}:
\begin{equation}
\label{gDBval}
g_{DB}^{\pi_\Psi \pi_d} =
\left\{
\begin{array}{cl}
1.0 & \pi_\Psi = \pi_d \\
0.1 & \pi_\Psi = - \pi_d
\end{array}\right.\,,
\end{equation}
\emph{i.e}.\ nothing is done when the parity of the diquark correlation,  $\pi_d $, matches that of the host baryon, $\pi_\Psi$, but suppression is introduced when the parity is different.  The magnitude of the effect was chosen so that, in concert with Eqs.\,\eqref{eqsSO}, the dressed-quark core mass of the nucleon's parity
partner is not lower than that of its first radial excitation \cite{Lu:2017cln}.

Every element in the Faddeev kernel is now specified.  It remains only to select the channel, derive the explicit form for the associated algebraic equation, then solve the resulting matrix equation to obtain the baryon mass and amplitude.
Specific examples of the Faddeev equations for the various light-quark systems, along with their derivations, are presented in Ref.\,\cite{Chen:2012qr}; and equations for $\Xi_c^+$, $\Xi_c^{\prime +}$, $\Xi_c^{\ast +}$  are given in Ref.\,\cite{Yin:2019bxe}.  The equations solved herein are not any more complex, although they do involve larger kernel matrices because we also allow all baryons to contain pseudoscalar and vector diquarks.
As in Ref.\,\cite{Yin:2019bxe}, the ultraviolet cutoff in each baryon channel is identified with that of the lightest diquark in the system.  This is always the smallest value; hence, the dominant regularising influence.

\begin{table*}[t]
\caption{\label{ResultsThreeHalf}
Computed mass and Faddeev amplitude for each ground-state flavour-SU$(5)$ $J^P=3/2^+$ baryon.
The last column highlights the baryon's dominant spin-flavour correlation.  The possibilities are given in Eqs.\,\eqref{ThreeHalfDelta}, \eqref{ThreeHalfqQQ}, \eqref{ThreeHalfOneTwo}, \eqref{ThreeHalfQQQ}.
Empirical mass values, $M^{e}$, are taken from Ref.\,\cite{Zyla:2020zbs}; and lQCD results, $M^{l}$, from Refs.\,\cite{Durr:2008zz, Brown:2014ena, Mathur:2018epb}.
The results in column $M^3$ are the three-body Faddeev equation predictions in Ref.\,\cite{Qin:2019hgk}.
 (Masses in GeV.)
 }
\begin{center}
\begin{tabular*}
{\hsize}
{l@{\extracolsep{0ptplus1fil}}|
r@{\extracolsep{0ptplus1fil}}
r@{\extracolsep{0ptplus1fil}}
r@{\extracolsep{0ptplus1fil}}
r@{\extracolsep{0ptplus1fil}}|
r@{\extracolsep{0ptplus1fil}}
r@{\extracolsep{0ptplus1fil}}|
c@{\extracolsep{0ptplus1fil}}}
\hline\hline
 & $M^{\rm CI}$ & $M^{e}$ & $M^{l}$  & $M^{3}$\; & $a^{r_{1}}$ & $a^{r_{2}}$ & dom.\ corr.\ \\
\hline
$\Delta$           &   1.27   &   1.23     &   1.23      &   1.21\;    &  1       &            &  $u\{uu\}_{1^{+}}$                  \\
$\Sigma^{*}$       &   1.39   &   1.38     &   1.40      &   1.36\;    &  0.61       &  0.79\;       &  $u\{us\}_{1^{+}}$                  \\
$\Xi^{*}$          &   1.51   &   1.53     &   1.56      &   1.52\;    &  0.85       &  0.52\;       &  $s\{us\}_{1^{+}}$                  \\
$\Omega$           &   1.63   &   1.67     &   1.67      &   1.67\;    &  1       &            &  $s\{ss\}_{1^{+}}$                  \\
\hline
$\Sigma_{c}^{*}$   &   2.54   &   2.52     &   2.55      &   2.39\;    &  0.63       &  0.78\;       &  $u\{uc\}_{1^{+}}$                  \\
$\Xi_{c}^{*}$      &   2.67   &   2.65     &   2.65      &   2.55\;    &  0.61       &  0.79\;       &  $s\{uc\}_{1^{+}}+u\{sc\}_{1^{+}}$  \\
$\Omega_{c}^{*}$   &   2.80   &   2.77     &   2.76      &   2.70\;    &  0.59       &  0.81\;       &  $s\{sc\}_{1^{+}}$                  \\
\hline
$\Sigma_{b}^{*}$   &   5.79   &   5.83     &   5.88      &   5.60\;    &  0.66       &  0.75\;      &  $u\{ub\}_{1^{+}}$                  \\
$\Xi_{b}^{*}$      &   5.92   &   5.95     &   5.96      &   5.75\;    &  0.61       &  0.79\;       &  $s\{ub\}_{1^{+}}+u\{sb\}_{1^{+}}$  \\
$\Omega_{b}^{*}$   &   6.04   &            &   6.09      &   5.90\;    &  0.55       &  0.84\;       &  $s\{sb\}_{1^{+}}$                  \\
\hline
$\Xi_{cc}^{*}$     &   3.73   &            &   3.69      &   3.58\;    &  0.98       &  0.20\;      &  $c\{uc\}_{1^{+}}$                  \\
$\Xi_{cb}^{*}$     &   7.00   &            &   6.99      &   6.78\;    &  0.97       &  0.24\;       &  $b\{uc\}_{1^{+}}+c\{ub\}_{1^{+}}$  \\
$\Xi_{bb}^{*}$     &   10.26  &            &   10.18     &   9.98\;    &  0.99       &  0.08\;       &  $b\{ub\}_{1^{+}}$                  \\
\hline
$\Omega_{cc}^{*}$  &   3.88   &            &   3.82      &   3.73\;    &  0.96       &  0.28\;       &  $c\{sc\}_{1^{+}}$                  \\
$\Omega_{cb}^{*}$  &   7.12   &            &   7.06      &   6.93\;    &  0.94       &  0.35\;       &  $b\{sc\}_{1^{+}}+c\{sb\}_{1^{+}}$  \\
$\Omega_{bb}^{*}$  &   10.37  &            &   10.31     &   10.14\;   &  0.99       &  0.12\;       &  $b\{sb\}_{1^{+}}$                  \\
\hline
$\Omega_{ccc}$     &   4.90   &            &   4.80      &   4.76\;    &  1       &            &  $c\{cc\}_{1^{+}}$                  \\
$\Omega_{ccb}^{*}$ &   8.08   &            &   8.04      &   7.96\;    &  0.62       &  0.79\;       &  $c\{cb\}_{1^{+}}$                  \\
$\Omega_{cbb}^{*}$ &   11.26  &            &   11.23     &   11.17\;   &  0.96       &  0.28\;       &  $b\{cb\}_{1^{+}}$                  \\
$\Omega_{bbb}$     &   14.45  &            &   14.37     &   14.37\;   &  1       &            &  $b\{bb\}_{1^{+}}$                  \\
\hline\hline
\end{tabular*}
\end{center}
\end{table*}

\subsubsection{$J^P=1/2^+$}
\label{subsubHalf}
Our results for the masses and amplitudes of flavour-SU$(5)$ $J^P = 1/2^+$ baryons are listed in Table~\ref{ResultsHalf}.
As remarked in association with Eq.\,\eqref{gNgD}, the Faddeev kernel employed omits resonant contributions that serve to reduce baryon masses.  From each of the values in column~1 we have therefore subtracted
\begin{equation}
{\mathpzc z}_{N^+} = 0.16\,{\rm GeV}
\label{Nrescale}
\end{equation}
in order to express an empirically informed estimate of such corrections.  This value is the average difference between the unmodified CI result and experiment in the first four rows.   It matches the expectations described in connection with Eq.\,\eqref{gNgD}.  As indicated after Eq.\,\eqref{gNgD}, Ref.\,\cite{Yin:2019bxe} achieved a similar effect by inflating the values of $g_{N,\Delta}$, but we find that can produce unexpected (perhaps unphysical) modifications of the Faddeev amplitudes.

%
The computed masses in Table~\ref{ResultsHalf} are also compared with empirical or lQCD values in Fig.\,\ref{figResultsHalf}A: the mean absolute-relative-difference is $1.3(1.3)$\%.

Compared with the results in column~4, drawn from Ref.\,\cite{Qin:2019hgk}, the analogous difference is $5.2(2.8)$\%.  In that study, the Faddeev equations were solved in a fully-consistent RL truncation, eschewing a quark+diquark approximation.  Moreover, the results are \emph{ab initio} predictions, whereas we introduced $g_N$, ${\mathpzc z}_{N^+}$ -- Eqs.\,\eqref{staticexchange}, \eqref{Nrescale} -- to adjust the overall scale of the $J^P=1/2^+$ spectrum.
Notwithstanding these things, the level of agreement confirms the validity of the ESRs used to complete the spectrum calculations in Ref.\,\cite{Qin:2019hgk}; and the fact that our results exceed the accuracy of Ref.\,\cite{Qin:2019hgk} indicates that we have implemented a phenomenologically efficacious CI formulation.

The $1/2^+$ Faddeev amplitudes are represented in Table~\ref{ResultsHalf} by the strength of the coefficient that multiplies the flavour and Dirac structure specified by Eqs.\,\eqref{SAPD}, \eqref{halfnucleon}, \eqref{half2q1Q}, \eqref{HalfOneTwo}, \eqref{OneHalfQQQ}.  For instance, the $\Lambda$ entry in column~6 is ``$-0.45$'', indicating that this baryon contains a pseudovector diquark component $[d\{us\}_{1^{+}}-u\{ds\}_{1^{+}}]/\sqrt{2}$ with $i\gamma_5\gamma_\mu$ coefficient $a_1^+=-0.45$ in \linebreak Eq.\,\eqref{Aa1}.

The amplitudes in Table~\ref{ResultsHalf} confirm the conclusions in Ref.\,\cite{Yin:2019bxe}.  Namely, concerning ground-state flavour-$SU(5)$ $J^P=1/2^+$ baryons:
($\mathpzc a$) the lightest participating diquark correlation usually defines the most important component of a baryon's Faddeev amplitude and this is true even if a pseudovector diquark is the lightest channel available; and
($\mathpzc b$) light-diquark dominance may be overcome in flavour channels for which the bound-state's spin-flavour structure and the quark-exchange character of the kernel in Fig.\,\ref{figFaddeev} lead dynamically to a preference for mixed-flavour correlations.

\begin{table*}[t]
\caption{\label{ResultsHalfMinus}
Computed mass (in GeV) and Faddeev amplitude for each ground-state flavour-$SU(5)$ $J^P=1/2^-$ baryon: the last column identifies the baryon's dominant spin-flavour correlation.  The arrays of possibilities are given in Eqs.\,\eqref{halfnucleon}, \eqref{half2q1Q}, \eqref{HalfOneTwo}, \eqref{OneHalfQQQ}.
Empirical mass values, $M^{e}$, are taken from Ref.\,\cite{Zyla:2020zbs}; and lQCD results, $M^{l}$, from Ref.\,\cite{Bahtiyar:2020uuj}.
The results under heading $M^3$ are the three-body Faddeev equation predictions in Ref.\,\cite{Qin:2019hgk}, which did not include SU$(3)$-flavour-symmetry breaking effects.
}
\begin{center}
\begin{tabular*}
{\hsize}
{l@{\extracolsep{0ptplus1fil}}|
r@{\extracolsep{0ptplus1fil}}
r@{\extracolsep{0ptplus1fil}}
r@{\extracolsep{0ptplus1fil}}
r@{\extracolsep{0ptplus1fil}}|
r@{\extracolsep{0ptplus1fil}}
r@{\extracolsep{0ptplus1fil}}
r@{\extracolsep{0ptplus1fil}}
r@{\extracolsep{0ptplus1fil}}
r@{\extracolsep{0ptplus1fil}}
r@{\extracolsep{0ptplus1fil}}
r@{\extracolsep{0ptplus1fil}}
r@{\extracolsep{0ptplus1fil}}
r@{\extracolsep{0ptplus1fil}}
r@{\extracolsep{0ptplus1fil}}
r@{\extracolsep{0ptplus1fil}}
r@{\extracolsep{0ptplus1fil}}
r@{\extracolsep{0ptplus1fil}}
r@{\extracolsep{0ptplus1fil}}|
c@{\extracolsep{0ptplus1fil}}}
\hline\hline
& $M^{\rm CI}$ & $M^{e}$ & $M^{l}$  & $M^{3}$ & $s^{r_{1}}$ & $s^{r_{2}}$ & $a_{1}^{r_{2}}$ & $a_{2}^{r_{2}}$ & $a_{1}^{r_{3}}$ & $a_{2}^{r_{3}}$ & $p^{r_{4}}$ & $p^{r_{5}}$ & $v_{1}^{r_{5}}$ & $v_{2}^{r_{5}}$ & $v_{1}^{r_{6}}$ & $v_{2}^{r_{6}}$ & $v_{1}^{r_{7}}$ & $v_{2}^{r_{7}}$ & dom.\ corr.\  \\
\hline
$N$                &  1.55    &  1.54      &            &   1.54\;    &  0.35       &             &  0.04           &  0              & -0.03           &  0              & -0.92       &             & -0.06           &  0.18           &                 &                 &                 &                 & $u[ud]_{0^{-}}$                \\
$\Lambda$          &  1.65    &  1.67      &            &   1.58\;    &  0.27       &  0.24       &                 &                 &  0.03           &  0              & -0.66       & -0.62       &                 &                 & -0.06           &  0.14           & -0.05           &  0.13\;           & $s[ud]_{0^{-}}$                \\
$\Sigma$           &  1.70    &  1.68      &            &   1.58\;    &  0.46       &             &  0.17           & -0.13           & -0.08           &  0.06           & -0.85       &             &  0.06           &  0.13           &                 &                 &                 &                 & $u[us]_{0^{-}}$                \\
$\Xi$              &  1.78    &           &            &   1.62\;    &  0.39       &             &  0.04           & -0.01           & -0.04           &  0.02           & -0.90       &             & -0.05           &  0.19           &                 &                 &                 &                 & $s[us]_{0^{-}}$                \\
\hline
$\Lambda_{c}$      &  2.67    &  2.59      &   2.67      &   2.67\;    &  0.07       &  0.45       &                 &                 &  0.06           & -0.03           & -0.21       & -0.83       &                 &                 & -0.09           &  0.14           &  0.03           &  0.20\;           & $d[uc]_{0^{-}}-u[dc]_{0^{-}}$  \\
$\Sigma_{c}$       &  2.84    &           &   2.81      &   2.67\;    &  0.13       &             &  0.14           & -0.19           &  0.08           & -0.24           & -0.17       &             & -0.27           &  0.88           &                 &                 &                 &                 & $u[uc]_{1^{-}}$                \\
$\Xi_{c}$          &  2.79    &  2.79      &   2.77      &   2.71\;    &  0.07       &  0.51       &                 &                 &  0.09           & -0.05           & -0.20       & -0.78       &                 &                 & -0.06           &  0.07           &  0.09           &  0.25\;           & $s[uc]_{0^{-}}-u[sc]_{0^{-}}$  \\
$\Xi_{c}^{\prime}$ &  2.94    &           &   2.93      &   2.71\;    &  0.15       &             &  0.13           & -0.19           &  0.08           & -0.24           & -0.16       &             & -0.27           &  0.87           &                 &                 &                 &                 & $s[uc]_{1^{-}}+u[sc]_{1^{-}}$  \\
$\Omega_{c}$       &  3.03    &           &   3.04      &   2.74\;   &  0.18       &             &  0.13           & -0.19           &  0.08           & -0.25           & -0.13       &             & -0.27           &  0.87           &                 &                 &                 &                 & $s[sc]_{1^{-}}$                \\
\hline
$\Lambda_{b}$      &  6.03    &  5.91      &            &   5.92\;    &  0          & -0.22       &                 &                 &  0.12           & -0.03           &  0          &  0.15       &                 &                 &  0.14           & -0.72           &  0.44           &  0.43\;           & $b[ud]_{1^{-}}$                \\
$\Sigma_{b}$       &  6.32    &           &            &   5.92\;    &  0.01       &             &  0              &  0              &  -0.05          &  0.25           & -0.01       &             &  0.24           & -0.94           &                 &                 &                 &                 & $u[ub]_{1^{-}}$                \\
$\Xi_{b}$          &  6.15    &           &            &   5.96\;    &  0          & -0.21       &                 &                 &  0.13           & -0.04           &  0          &  0.14       &                 &                 &  0.13           & -0.67           &  0.47           &  0.48\;           & $b[us]_{1^{-}}$                \\
$\Xi_{b}^{\prime}$ &  6.40    &           &            &   5.96\;    &  0.01       &             &  0              &  0              & -0.05           &  0.26           & -0.03       &             &  0.22           & -0.94           &                 &                 &                 &                 & $s[ub]_{1^{-}}+u[sb]_{1^{-}}$  \\
$\Omega_{b}$       &  6.49    &           &            &   5.99\;    &  0.02       &             &  0              &  0              & -0.05           &  0.27           & -0.06       &             &  0.21           & -0.94           &                 &                 &                 &                 & $s[sb]_{1^{-}}$                \\
\hline
$\Xi_{cc}$         &  3.81    &           &   3.93      &   3.79\;    &  0.54       &             &  0.09           & -0.07           & -0.02           &  0.02           & -0.83       &             &  0.03           &  0.10           &                 &                 &                 &                 & $c[uc]_{0^{-}}$                \\
$\Xi_{cb}$         &  7.04    &           &            &   7.04\;    &  0.01       & -0.19       &                 &                 & -0.29           &  0.21           &  0.51       & -0.29       &                 &                 & -0.24           & -0.64           & -0.19           & -0.01\;           & $b[uc]_{1^{-}}-c[ub]_{1^{-}}$  \\
$\Xi_{cb}^{\prime}$&  6.98    &           &            &   7.04\;    &  0.48       &             &  0.11           & -0.03           & -0.05           &  0.02           & -0.75       &             &  0.16           &  0.40           &                 &                 &                 &                 & $b[uc]_{0^{-}}+c[ub]_{0^{-}}$  \\
$\Xi_{bb}$         &  10.22   &           &            &   10.29\;   &  0.62       &             &  0.15           & -0.14           & -0.01           &  0.01           & -0.74       &             &  0.15           & -0.04           &                 &                 &                 &                 & $b[ub]_{0^{-}}$                \\
\hline
$\Omega_{cc}$      &  3.93    &           &   4.04      &   3.83\;    &  0.66       &             &  0.16           & -0.15           & -0.05           &  0.04           & -0.70       &             &  0.17           & -0.02           &                 &                 &                 &                 & $c[sc]_{0^{-}}$                \\
$\Omega_{cb}$      &  7.15    &           &            &   7.08\;    &  0.01       & -0.28       &                 &                 & -0.23           &  0.17           &  0.68       & -0.31       &                 &                 & -0.11           & -0.48           & -0.20           & -0.02\;           & $b[sc]_{0^{-}}-c[sb]_{0^{-}}$  \\
$\Omega_{cb}^{\prime}$ &  7.16    &           &            &   7.08\;    &  0.53       &             &  0.14           & -0.05           & -0.08           &  0.04           & -0.69       &             &  0.21           &  0.40           &                 &                 &                 &                 & $b[sc]_{0^{-}}+c[sb]_{0^{-}}$  \\
$\Omega_{bb}$      &  10.42   &           &            &   10.33\;   &  0          &             &  0              &  0              & -0.01           &  0.01           &  0.01       &             &  0.72           & -0.69           &                 &                 &                 &                 & $b[sb]_{1^{-}}$                \\
\hline
$\Omega_{ccb}$     &  8.10    &           &            &   8.16\;    &  0.24       &             &  0.12           & -0.15           &  0.08           & -0.34           & -0.02       &             & -0.21           &  0.86           &                 &                 &                 &                 & $c[cb]_{1^{-}}$                \\
$\Omega_{cbb}$     &  11.23   &           &            &   11.41\;   &  0.01       &             & -0.09           &  0.33           & -0.07           &  0.14           &  0.54       &             &  0.56           & -0.52           &                 &                 &                 &                 & $b[cb]_{1^{-}}$                \\
\hline\hline
\end{tabular*}
\end{center}
\end{table*}

In corroborating these conclusions, we reaffirm the arguments against treatments of the baryon problem which assume they can be described as effectively two-body in nature, \emph{e.g}.\ as being built from a constituent-quark and static constituent-diquark.  The dynamical character of diquark correlations is paramount because their breakup and reformation are crucial in defining baryon structure.  This is confirmed for light-quark baryons by the fact that lQCD calculations produce a spectrum whose richness cannot be explained by a two-body model \cite{Edwards:2011jj}.  The implications extend to baryons involving one or more heavy quarks, challenging both (\emph{i}) the treatment of singly-heavy baryons as two-body light-diquark+heavy-quark bound-states and (\emph{ii}) analyses of doubly-heavy baryons which assume such systems can be considered as two-body light-quark+heavy-diquark bound-states.  These observations may also be relevant to few-body studies of tetra- and penta-quark problems; in particular, those involving $\{u,d,s\}$ quarks.

\subsubsection{$J^P=3/2^+$}
\label{subsubThreeHalf}
Our results for the masses and amplitudes of flavour-SU$(5)$ $J^P = 3/2^+$ baryons are listed in Table~\ref{ResultsThreeHalf}.
Akin to Eq.\,\eqref{Nrescale}, from each of the values in column~1 we have subtracted
\begin{equation}
{\mathpzc z}_{\Delta^+} = 0.12\,{\rm GeV}
\label{Drescale}
\end{equation}
so as to express an empirically informed estimate of meson-cloud corrections: this value is the average difference between the unmodified CI result and experiment in the first four rows.  Again, it matches expectations described in connection with Eq.\,\eqref{gNgD}.

The computed $J^P=3/2^+$ masses are compared with empirical/lQCD values in Fig.\,\ref{figResultsHalf}B: the mean-abso-lute-relative-difference is $1.0(0.8)$\%.
Furthermore, the results compare well with the three-body calculation described in Ref.\,\cite{Qin:2019hgk}, for which the analogous difference is $2.6(1.6)$\%.
Notwithstanding the fact that we used $g_\Delta$, ${\mathpzc z}_{\Delta^+}$ -- Eqs.\,\eqref{staticexchange}, \eqref{Drescale} -- to adjust the overall scale of the $J^P=3/2^+$ spectrum, this improvement over the results in Ref.\,\cite{Qin:2019hgk} again highlights the utility of our CI formulation.

The CI Faddeev amplitudes of $J=3/2^+$ baryons are also listed in Table~\ref{ResultsThreeHalf}.  Once more, the fully dynamical nature of the diquarks and the character of the Faddeev kernel work together to ensure a continual reshuffling of the dressed-quarks within the diquark correlations.  Consequently, in all cases involving more than one quark flavour, the diquark combination with maximal flavour shuffling is favoured because it is fed by twice as many exchange processes as the less-mixed correlation.
This differs from the result for the $\Sigma_b^\ast$-baryon in Ref.\,\cite{Yin:2019bxe}.  Therein, however, $g_\Delta$ was 26\% larger than here, thereby providing additional enhancement for the single $u$-quark exchange process over the doubly active, but $b-u$ mass-splitting-suppressed, $b$-quark exchange contribution which enables the shuffling.


\begin{figure*}[t]
\vspace*{2ex}

\leftline{\hspace*{0.5em}{\large{\textsf{A}}}}
\vspace*{-3ex}
\includegraphics[width=0.99\textwidth]{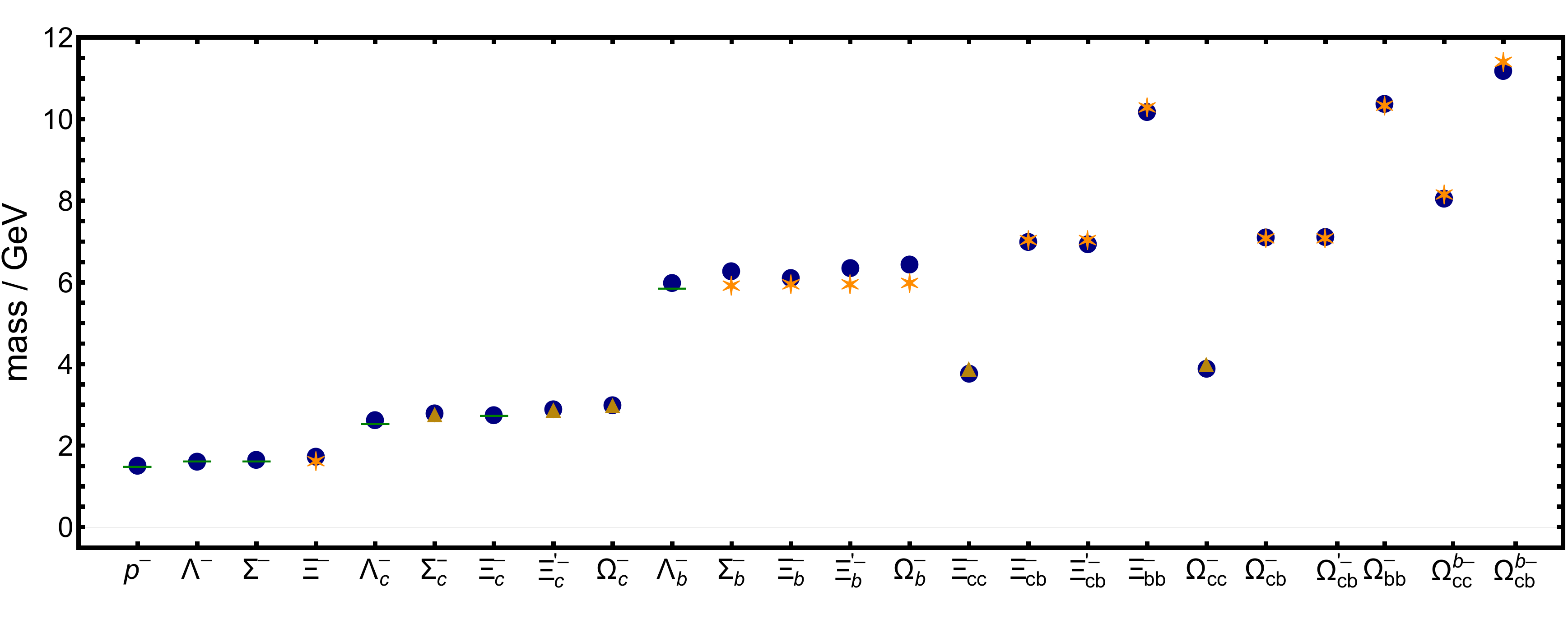}
\vspace*{1ex}

\leftline{\hspace*{0.5em}{\large{\textsf{B}}}}
\vspace*{-3ex}
\includegraphics[width=0.99\textwidth]{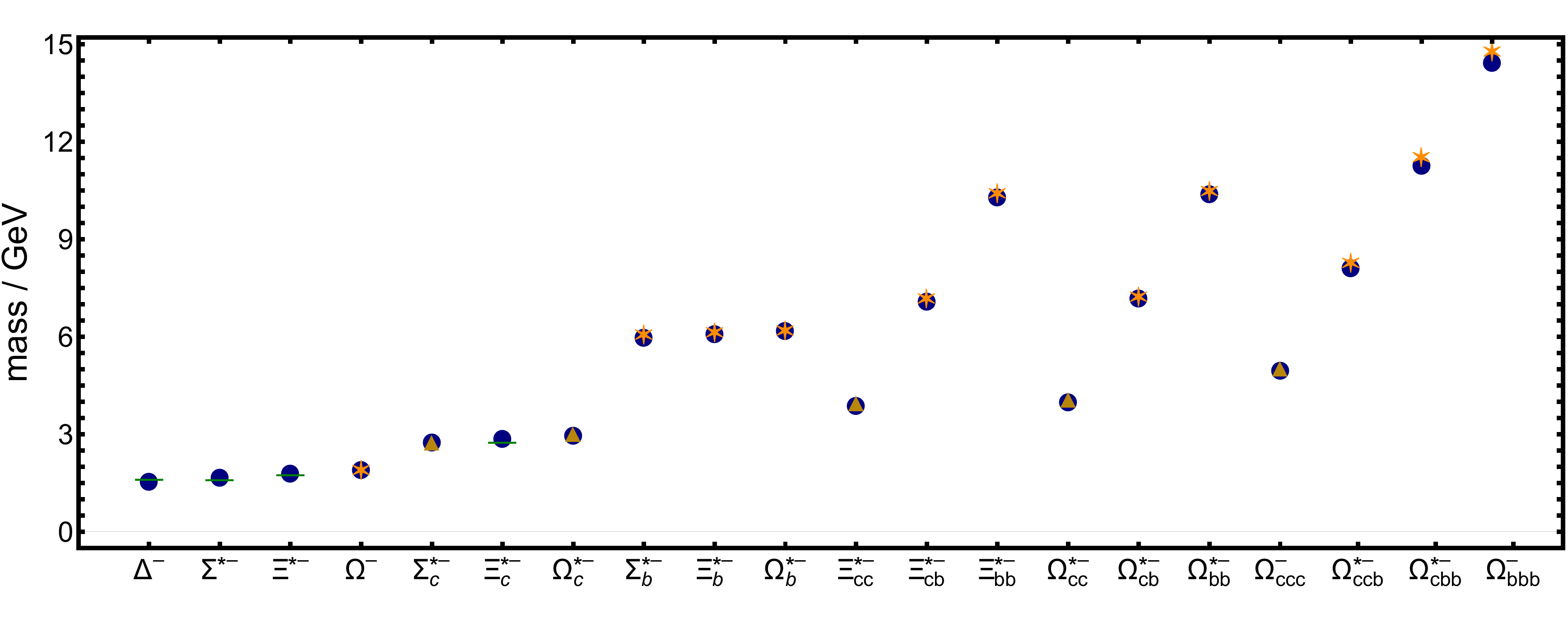}
\caption{\label{figResultsHalfNegative}
\emph{Upper panel}\,--\,{\sf A}.
CI computed masses (in GeV) of ground-state flavour-SU$(5)$ $J^P=1/2^-$ baryons in Table~\ref{ResultsHalfMinus} compared with experiment \cite{Zyla:2020zbs} (green bars), lQCD \cite{Bahtiyar:2020uuj} (gold triangles), or three-body Faddeev equation results \cite{Qin:2019hgk} (orange asterisks).
\emph{Lower panel}\,--\,{\sf B}.
Similar comparison for ground-state flavour-SU$(5)$ $J^P=3/2^-$ baryons, compiled using Table~\ref{ResultsThreeHalfMinus}.
Both panels: the states are labelled as parity partners of the lightest same-$J$ state, e.g. $p^-$ denotes the $N(1535)\,(1/2)^-$ baryon and $\Delta^-$ indicates $\Delta(1700)\,(3/2)^-$.
}
\end{figure*}

\subsubsection{$J^P=1/2^-$}
\label{subsubHalf}
CI results for the masses and amplitudes of flavour-SU$(5)$ $J^P = 1/2^-$ baryons are listed in Table~\ref{ResultsHalfMinus}.
From each of the values in column~1, we subtracted
\begin{equation}
{\mathpzc z}_{N^-} = 0.27\,{\rm GeV}
\label{NrescaleM}
\end{equation}
in order to express an empirically informed estimate of meson cloud corrections: this value is the average difference between the unmodified CI result and experiment in the first three rows.

The computed masses in Table~\ref{ResultsHalfMinus} are also depicted in Fig.\,\ref{figResultsHalfNegative}A.  Compared with empirical values, or lQCD results when empirical values are unavailable, the mean absolute-relative-difference is $1.4(1.1)$\%; and compared with the results in column~4, drawn from Ref.\,\cite{Qin:2019hgk}, the analogous difference is $3.7(3.5)$\%.

The CI Faddeev amplitudes of $J^P=1/2^-$ baryons are also listed in Table~\ref{ResultsHalfMinus}.  They are readily understood because they follow the pattern of the $J=1/2^+$ amplitude in almost all cases: typically, the lightest like-parity $J=1$ diquark correlation dominates.  The exception is $\Xi_{cb}^-$/$\Omega_{cb}^-$, wherein the lightest like-parity diquark correlations dominate, which is a $0^-$ combination in this case.

\begin{table*}[t]
\caption{\label{ResultsThreeHalfMinus}
Computed mass (in GeV) and Faddeev amplitude for each ground-state flavour-$SU(5)$ $J^P=3/2^-$ baryon: the last column identifies the baryon's dominant spin-flavour correlation.  The possibilities are given in Eqs.\,\eqref{ThreeHalfDelta}, \eqref{ThreeHalfqQQ}, \eqref{ThreeHalfOneTwo}, \eqref{ThreeHalfQQQ}.
Empirical mass values, $M^{e}$, are taken from Ref.\,\cite{Zyla:2020zbs}; and lQCD results, $M^{l}$, from Ref.\,\cite{Bahtiyar:2020uuj}.
The results under heading $M^3$ are the three-body Faddeev equation predictions in Ref.\,\cite{Qin:2019hgk}.
}
\begin{center}
\begin{tabular*}
{\hsize}
{l@{\extracolsep{0ptplus1fil}}|
r@{\extracolsep{0ptplus1fil}}
r@{\extracolsep{0ptplus1fil}}
r@{\extracolsep{0ptplus1fil}}
r@{\extracolsep{0ptplus1fil}}|
r@{\extracolsep{0ptplus1fil}}
r@{\extracolsep{0ptplus1fil}}|
c@{\extracolsep{0ptplus1fil}}}
\hline\hline
 & $M^{\rm CI}$ & $M^{e}$ & $M^{l} $ & $M^{3}$\; & $a^{r_{1}}$ & $a^{r_{2}}$\; & dom.\ corr.\ \\
\hline
$\Delta$           &   1.59   &   1.67     &           &   1.73\;    &  1       &            &  $u\{uu\}_{1^{+}}$                  \\
$\Sigma^{*}$       &   1.72   &   1.66     &           &   1.79\;    &  0.72       &   0.69\;      &  $s\{uu\}_{1^{+}}$                  \\
$\Xi^{*}$          &   1.84   &   1.82     &           &   1.84\;    &  0.91       &   0.42\;      &  $s\{us\}_{1^{+}}$                  \\
$\Omega$           &   1.95   &          &           &   1.90\;    &  1       &            &  $s\{ss\}_{1^{+}}$                  \\
\hline
$\Sigma_{c}^{*}$   &   2.80   &          &   2.80      &   2.83\;    &  0.85       &   0.53\;      &  $c\{uu\}_{1^{+}}$                  \\
$\Xi_{c}^{*}$      &   2.91   &   2.82     &   2.80      &   2.89\;    &  0.74       &   0.67\;      &  
$c\{us\}_{1^{+}}$                  \\
$\Omega_{c}^{*}$   &   3.01   &          &   3.07      &   2.94\;    &  0.62       &   0.79\;      &  $s\{sc\}_{1^{+}}$                  \\
\hline
$\Sigma_{b}^{*}$   &   6.03   &          &           &   6.07\;    &  0.84       &   0.54\;      &  $b\{uu\}_{1^{+}}$                  \\
$\Xi_{b}^{*}$      &   6.13   &          &           &   6.13\;    &  0.69       &   0.72\;      &  $s\{ub\}_{1^{+}}+u\{sb\}_{1^{+}}$  \\
$\Omega_{b}^{*}$   &   6.23   &          &           &   6.19\;    &  0.52       &   0.86\;      &  $s\{sb\}_{1^{+}}$                  \\
\hline
$\Xi_{cc}^{*}$     &   3.93   &          &   4.01      &   3.93\;    &  0.98       &   0.18\;      &  $c\{uc\}_{1^{+}}$                  \\
$\Xi_{cb}^{*}$     &   7.13   &          &           &   7.18\;    &  0.96       &   0.28\;      &  $b\{uc\}_{1^{+}}+c\{ub\}_{1^{+}}$  \\
$\Xi_{bb}^{*}$     &   10.35  &          &           &   10.42\;   &  0.99       &   0.08\;      &  $b\{ub\}_{1^{+}}$                  \\
\hline
$\Omega_{cc}^{*}$  &   4.04   &          &   4.12      &   3.99\;    &  0.96       &   0.26\;      &  $c\{sc\}_{1^{+}}$                  \\
$\Omega_{cb}^{*}$  &   7.23   &          &           &   7.23\;    &  0.92       &   0.39\;      &  $b\{sc\}_{1^{+}}+c\{sb\}_{1^{+}}$  \\
$\Omega_{bb}^{*}$  &   10.44  &          &           &   10.48\;   &  0.99       &   0.11\;      &  $b\{sb\}_{1^{+}}$                  \\
\hline
$\Omega_{ccc}$     &   5.01   &          &   5.08      &   5.03\;    &  1       &         &  $c\{cc\}_{1^{+}}$                  \\
$\Omega_{ccb}^{*}$ &   8.17   &          &           &   8.28\;    &  0.63       &   0.77\;      &  $c\{cb\}_{1^{+}}$                  \\
$\Omega_{cbb}^{*}$ &   11.32  &          &           &   11.52\;   &  0.96       &   0.28\;      &  $b\{cb\}_{1^{+}}$                  \\
$\Omega_{bbb}$     &   14.49  &          &           &   14.77\;   &  1       &         &  $b\{bb\}_{1^{+}}$                  \\
\hline\hline
\end{tabular*}
\end{center}
\end{table*}

\subsubsection{$J^P=3/2^-$}
\label{subsubThreeHalf}
These states are somewhat unusual.  Given that they must be completely symmetric under permutations of flavour labels but the only negative-parity diquarks supported by the CI are flavour-$\overline{10}$ (antisymmetric) states, then the $J=3/2^-$ baryons can only contain positive parity pseudovector diquarks, Eq.\,\eqref{ThreeHalfQQQ}.  Owing to this peculiarity, we judged that opposite-parity diquark correlations should not receive additional suppression in these systems;
%
so we solved the associated Faddeev equations with
\begin{equation}
g_{DB}^{\pi_{3/2^-} \pi_{1^+}}=g_{DB}^{\pi_{3/2^+} \pi_{1^+}}=1\,.
\end{equation}

Working with the results thus obtained, the average difference between the CI result and experiment in the first three rows of Table~\ref{ResultsThreeHalfMinus} is
\begin{equation}
z_{\Delta^-} = 0.13\,{\rm GeV}.
\end{equation}
Consequently, in column~1 of Table~\ref{ResultsThreeHalfMinus} we list the result obtained by subtracting $z_{\Delta^-}$ from the directly calculated CI result.  Comparing column~1 with experiment, where known, and available lQCD results otherwise, the mean absolute-relative-difference is $2.2(1.4)$\%, and with the results in column~4, drawn from Ref.\,\cite{Qin:2019hgk}, the analogous difference is $1.4(1.9)$\%.

The computed masses in Table~\ref{ResultsThreeHalfMinus} are also depicted in Fig.\,\ref{figResultsHalfNegative}B alongside, as available, in this order: experiment, lQCD, or three-body Faddeev equation results.

The CI Faddeev amplitudes of $J=3/2^-$ baryons are also listed in Table~\ref{ResultsThreeHalfMinus}.  They are similar to those of the $J=3/2^+$ states in Table~\ref{ResultsThreeHalf}, with the dominant diquark correlation being the same in all cases except $\Sigma^\ast$, $\Sigma^\ast_c$, $\Sigma^\ast_b$.  Considering Eq.\,\eqref{SigmaQ}, one sees that in these cases there is a competition between the lightest diquark correlation and the heavy-light diquark, which is doubly-fed by the exchange kernel.  The former wins in the negative-parity systems, whereas the latter is dominant in the positive-parity systems.  This ordering reverses in the $3/2^-$ states as $g_{DB}^{\pi_{3/2^-} \pi_{1^+}}$ is reduced from unity.  The $\Sigma^\ast$, $\Sigma^\ast_c$, $\Sigma^\ast_b$ Faddeev amplitudes should therefore be considered as somewhat uncertain.


\section{Summary and Outlook}
\label{Sec:Epilogue}
A confining, symmetry-preserving treatment of a vec-tor$\times$vector contact interaction (CI) was used to compute spectra of ground-state $J^P = 0^\pm, 1^\pm$ $(f\bar g)$ mesons and $J^P=1/2^\pm, 3/2^\pm$ $(fgh)$ baryons, where $f,g,h \in \{u,d,s,c,b\}$.  The calculated meson masses agree well with experiment [Sec.\,\ref{Sec:Meson}]: the mean-relative-difference for 33 states is 5(6)\%.  Expressing effects tied to the emergence of hadronic mass (EHM) was crucial to achieving this level of agreement.  Regarding meson leptonic decay constants, empirical (or lattice QCD -- lQCD) values are reproduced with an accuracy of 15(11)\%; and predictions were made for the currently unknown decay constants of 19 $0^+$, $1^+$ states [Table~\ref{Tab:MesonSpectrum}].

A quark+diquark approximation to the baryon Faddeev equation was used herein.  Its formulation required the calculation of masses and correlation strengths for all 38 distinct participating diquarks [Sec.\,\ref{SecDiquarks}].  As in all studies to date, the level ordering of the $J^{P}$ diquarks matches that of their $J^{-P}$ meson partners.  Scalar and pseudovector diquarks are heavier than their partner mesons; but in our CI formulation, this ordering is reversed for pseudoscalar and vector diquark correlations.  Nevertheless, omitting the diquarks partnered with\linebreak would-be Nambu-Goldstone mode mesons, the mass of a diquark's partner meson is a reasonable guide to the diquark's mass: the mean difference in absolute value is $0.08(7)\,$GeV.

A static approximation to the quark exchange kernel was used to solve the Faddeev equation.  It produces momentum-independent Faddeev amplitudes, thereby ensuring a level of consistency with CI two-body bound-state amplitudes, and introduces four parameters.  Correcting also for the omission of meson cloud contributions to the quark-core Faddeev kernels, four additional parameters were necessary to complete the CI definition in the baryon sector.  In total, the CI predicts 88 distinct baryons, \emph{i.e}.\ every possible three-quark $1/2^\pm, 3/2^\pm$ ground-state baryon is realised.  Of this number, 34 states are already known empirically and lQCD-computed masses are available for another 30.  For this collection of 64 states, the mean absolute-relative-diffe-rence between CI prediction and experiment/lQCD mass is $\overline{\rm ard} = 1.4(1.2)$\% [Figs.\,\ref{figResultsHalf}, \ref{figResultsHalfNegative}].  Implementation of EHM-induced effects associated with spin-orbit repulsion in $1/2^-$ baryons was important to achieving this outcome.  The same 88 ground-states are also predicted by a three-body Faddeev equation \cite{Qin:2019hgk}; and in comparison with those results, the analogous difference is 3.4(3.0)\%.

A primary merit of the framework employed herein is its simplicity, enabling all analyses and calculations to be completed algebraically.  In total, there are twelve parameters: four used to define the interaction and its scale dependence via $\pi$, $K$, $\eta_c$, $\eta_b$ properties; and eight introduced to complete the baryon Faddeev equations.
From this foundation, the CI delivers predictions for 164 distinct quantities.  Thus far, 114 of these observables have either been measured or computed using lQCD; and a comparison on this subset yields $\overline{\rm ard} = 4.5(7.1)$\%.

This level of quantitative success suggests that some credibility be given to the qualitative conclusions that follow from our CI analysis.
%
(\emph{I}) Nonpointlike, dynamical diquark correlations play an important role in all baryons; and, typically, the lightest allowed diquark is the most important component of a baryon's Faddeev amplitude.
(\emph{II}) Positive-parity diquarks dominate in positive-parity baryons, with $J=1^+$ diquarks being prominent in all of them.
(\emph{III}) Negative-parity diquarks can be neglected when studying positive-parity baryons; but owing to EHM, they are significant, even dominant, in $J=1/2^-$ baryons.
(\emph{IV}) On the other hand, $J=3/2^-$ baryons are built (almost) exclusively from $J=1^+$ diquark correlations.
Naturally, these conclusions should be checked using more sophisticated Faddeev equations with momentum-dependent exchange interactions, \emph{e.g}.\ extending Refs.\,\cite{Chen:2017pse, Chen:2019fzn}.

An extension of this analysis to radial excitations of the states considered herein is possible.  However, that requires additional intervention because one must, by-hand, force zeros into what would otherwise be momen-tum-independent Faddeev amplitudes \cite{Roberts:2011cf}.
A potentially more worthwhile direction would be to adapt the framework to the challenge of understanding tetra- and penta-quark states \cite{Barabanov:2020jvn, Eichmann:2020oqt}.

A key longer-term goal is extension of the rainbow-ladder truncation three-body Faddeev equation study in Ref.\,\cite{Qin:2019hgk} so that it can: (a) directly handle non-degene-rate valence-quark systems (reducing/eliminating the need to use equal spacing rules); and (b) include beyond-rainbow-ladder contributions and thus both search for signals indicating the appearance of diquark correlations and deliver improved \emph{ab initio} predictions.  In such an effort, the use of high-performance computing resources will be necessary.


\begin{acknowledgements}
We are grateful for constructive comments from C.~Chen, Y.~Lu and S-.S.~Xu.
Work supported by:
National Natural Science Foundation of China (under grant nos.\,11747140, 11805097);
Jiangsu Provincial Natural Science Foundation of China (under grant nos.\,BK20180738, BK20180323);
Nanjing University of Posts and Telecommunications Natural Science Foundation (under grant nos.\,NY218132, NY219063);
Jiangsu Province Innovation Program (under grant no.\,CZ0070619001);
Jiangsu Province \emph{Hundred Talents Plan for Professionals};
Chinese Ministry of Science and Technology \emph{International Expert Involvement Programme};
Ministerio Espa\~nol de Ciencia e Innovaci\'on, grant no. PID2019-107844GB-C22;
and Junta de Andaluc\'ia, contract nos.\ P18-FRJ-1132 and Operativo FEDER Andaluc\'ia 2014-2020 UHU-1264517.
\end{acknowledgements}


\appendix

\section{Contact Interaction}
\label{Sec:Contact}
We employ the symmetry preserving treatment of the vector$\times$vector contact interaction (CI) described in\linebreak Ref.\,\cite{Roberts:2011wy}.
The key element is the quark+antiquark scattering kernel, which in rainbow-ladder (RL) truncation can be written ($k = p_1-p_1^\prime = p_2^\prime -p_2$):
\begin{subequations}
\label{KDinteraction}
\begin{align}
\mathscr{K}_{\alpha_1\alpha_1',\alpha_2\alpha_2'}  & = {\mathpzc G}_{\mu\nu}(k) [i\gamma_\mu]_{\alpha_1\alpha_1'} [i\gamma_\nu]_{\alpha_2\alpha_2'}\,,\\
 {\mathpzc G}_{\mu\nu}(k)  & = \tilde{\mathpzc G}(k^2) T_{\mu\nu}(k)\,,
\end{align}
\end{subequations}
where $k^2 T_{\mu\nu}(k) = k^2 \delta_{\mu\nu} - k_\mu k_\nu$.
(Our Euclidean metric and Dirac-matrix conventions are given in Ref.\,\cite[Appendix~A]{Chen:2012qr}.)

The defining quantity is $\tilde{\mathpzc G}$.  Following two decades of study, much has been learnt about its pointwise behaviour.  The qualitative conclusion is that owing to the emergence of a gluon mass-scale in QCD \cite{Aguilar:2015bud, Huber:2018ned}, $\tilde{\mathpzc G}$ saturates at infrared momenta.  Hence, one may write
\begin{align}
\tilde{\mathpzc G}(k^2) & \stackrel{k^2 \simeq 0}{=} \frac{4\pi \alpha_{IR}}{m_G^2}\,.
\end{align}

In QCD, $m_G \approx 0.5\,$GeV, $\alpha_{\rm IR} \approx \pi$ \cite{Cui:2019dwv}.  Following Ref.\,\cite{Yin:2019bxe}, we retain this value of $m_G$, but reduce $\alpha_{IR}$ to a parameter.  This is needed because the integrals appearing in CI bound-state equations require ultraviolet regularisation and this undermines the link between infrared and ultraviolet scales that is distinctive of QCD.  Furthermore, since a CI cannot support relative momentum between bound-state constituents, it is sensible to simplify the tensor structure in Eqs.\,\eqref{KDinteraction} such that, in operation:
\begin{align}
\label{KCI}
\mathscr{K}_{\alpha_1\alpha_1',\alpha_2\alpha_2'}^{\rm CI}  & = \frac{4\pi \alpha_{IR}}{m_G^2}
 [i\gamma_\mu]_{\alpha_1\alpha_1'} [i\gamma_\mu]_{\alpha_2\alpha_2'}\,.
 \end{align}

As remarked already, any use of Eq.\,\eqref{KCI} in an equation related to the continuum bound-state problem will require ultraviolet regularisation of any integrals involved.  Additionally, the theory is not renormalisable, so the associated mass-scales, $\Lambda_{\rm uv}$, are additional physical parameters.  They may usefully be interpreted as upper bounds on the momentum domains within which the properties of the associated systems are effectively momentum-independent.  For instance, the $\pi$-meson is larger in size than the $\eta_b$; hence, one should expect to use $1/\Lambda_{\rm uv}^\pi > 1/\Lambda_{\rm uv}^{\eta_b}$.

For the purpose of expressing confinement, we also introduce an infrared regularisation scale, $\Lambda_{\rm ir}$, when defining bound-state equations \cite{Ebert:1996vx}.  This has the effect of excising momenta less than $\Lambda_{\rm ir}$, thereby eliminating quark+antiquark production thresholds \cite{Krein:1990sf}.  A typical choice for this scale is $\Lambda_{\rm ir} = 0.24\,$GeV \cite{Roberts:2011wy}.

%
In implementing Eq.\,\eqref{KCI}, Ref.\,\cite{Yin:2019bxe} fixed the parameters via the masses and leptonic decay constants of the $\pi$, $K$, $\eta_c$, $\eta_b$, \emph{i.e}.\ $0^{-+}$ $f \bar f$ mesons, $f\in \{l=u=d, s, c, b\}$, where isospin symmetry is assumed.   The starting point is the dressed-quark gap equation, which in RL truncation, using Eq.\,\eqref{KCI}, takes the following form:
\begin{align}
\label{GapEqn}
S_f^{-1}(p)  & = i\gamma\cdot p +m_f \nonumber \\
& \quad + \frac{16 \pi}{3} \frac{\alpha_{\rm IR}}{m_G^2}
\int \frac{d^4q}{(2\pi)^4} \gamma_\mu S_f(q) \gamma_\mu\,,
\end{align}
where $m_f$ is the quark's current-mass.  Regularising the integral in a Poincar\'e-invariant manner, the solution is
\begin{equation}
\label{genS}
S_f(p)^{-1} = i \gamma\cdot p + M_f\,,
\end{equation}
with $M_f$ obtained as the solution of:
\begin{equation}
M_f = m_f + M_f\frac{4\alpha_{\rm IR}}{3\pi m_G^2}\,\,{\cal C}_0^{\rm iu}(M_f^2)\,,
\label{gapactual}
\end{equation}
where
\begin{align}
\nonumber
{\cal C}_0^{\rm iu}(\sigma) &=
\int_0^\infty\! ds \, s \int_{\tau_{\rm uv}^2}^{\tau_{\rm ir}^2} d\tau\,{\rm e}^{-\tau (s+\sigma)}\\
& =
\sigma \big[\Gamma(-1,\sigma \tau_{\rm uv}^2) - \Gamma(-1,\sigma \tau_{\rm ir}^2)\big].
\label{eq:C0}
\end{align}
$\Gamma(\alpha,y)$ is the incomplete gamma-function.

The following functions arise in solving the bound-state equations considered herein ($\tau_{\rm uv}^2=1/\Lambda_{\textrm{uv}}^{2}$, $\tau_{\rm ir}^2=1/\Lambda_{\textrm{ir}}^{2}$):
\begin{align}
%
%
%
\overline{\cal C}^{\rm iu}_n(\sigma) & = \Gamma(n-1,\sigma \tau_{\textrm{uv}}^{2}) - \Gamma(n-1,\sigma \tau_{\textrm{ir}}^{2})\,,
\end{align}
${\cal C}^{\rm iu}_n(\sigma)=\sigma \overline{\cal C}^{\rm iu}_n(\sigma)$, $n=0,1,2,\ldots$.

The contact interaction Bethe-Salpeter amplitude for a $0^{-+}$ $f\bar g$ meson has the following form \cite{Roberts:2011wy, Chen:2012txa}:
\begin{align}
\Gamma_{0^-}(Q) = \gamma_5 \left[ i E_{0^-} + \frac{1}{2 M_{f g}}\gamma\cdot Q F_{0^-}\right]\,.
\label{PSBSA}
\end{align}
$Q$ is the bound-state's total momentum, $Q^2 = -m_{0^-}^2$, $m_{0^-}$ is the meson's mass; and $M_{f g}= M_f M_{g}/[M_f + M_{g}]$.

The amplitude is obtained by solving $(t_+ = t+Q)$:
\begin{align}
\Gamma_{0^-}(Q)  & =  - \frac{16 \pi}{3} \frac{\alpha_{\rm IR}}{m_G^2} \nonumber \\
&\times
\int \! \frac{d^4t}{(2\pi)^4} \gamma_\mu S_f(t_+) \Gamma_{0^-}(Q)S_g(t) \gamma_\mu \,.
\label{LBSEI}
\end{align}
Using the symmetry-preserving regularisation scheme introduced in Refs.\,\cite{Roberts:2011wy, Chen:2012txa}, which requires
\begin{equation}
0 = \int_0^1d\alpha \,
\big[ {\cal C}_0^{\rm iu}(\omega_{fg}(\alpha,Q^2))
%
+ \, {\cal C}^{\rm iu}_1(\omega_{f g}(\alpha,Q^2))\big], \label{avwtiP}
\end{equation}
where ($\hat \alpha = 1-\alpha$)
\begin{align}
\omega_{f g}(\alpha,Q^2) &= M_f^2 \hat \alpha + \alpha M_{g}^2 + \alpha \hat\alpha Q^2\,,
\label{eq:omega}
\end{align}
one arrives at the following Bethe-Salpeter equation:
\begin{equation}
\label{bsefinalE}
\left[
\begin{array}{c}
E_{0^-}(Q)\\
F_{0^-}(Q)
\end{array}
\right]
= \frac{4 \alpha_{\rm IR}}{3\pi m_G^2}
\left[
\begin{array}{cc}
{\cal K}_{EE}^{0^-} & {\cal K}_{EF}^{0^-} \\
{\cal K}_{FE}^{0^-} & {\cal K}_{FF}^{0^-}
\end{array}\right]
\left[\begin{array}{c}
E_{0^-}(Q)\\
F_{0^-}(Q)
\end{array}
\right],
\end{equation}
with
{\allowdisplaybreaks
\begin{subequations}
\label{fgKernel}
\begin{eqnarray}
\nonumber
{\cal K}_{EE}^{0^-} &=&
\int_0^1d\alpha \bigg\{
{\cal C}_0^{\rm iu}(\omega_{f g}( \alpha, Q^2))  \\
&&+ \bigg[ M_f M_{g}-\alpha \hat\alpha Q^2 - \omega_{f g}( \alpha, Q^2)\bigg]\nonumber \\
&&
\quad \times
\overline{\cal C}^{\rm iu}_1(\omega_{f g}(\alpha, Q^2))\bigg\},\\
\nonumber
{\cal K}_{EF}^{0^-} &=& \frac{Q^2}{2 M_{f g}} \int_0^1d\alpha\, \bigg[\hat \alpha M_f+\alpha M_{g}\bigg]\\
&& \quad \times \overline{\cal C}^{\rm iu}_1(\omega_{f g}(\alpha, Q^2)),\\
{\cal K}_{FE}^{0^-} &=& \frac{2 M_{f g}^2}{Q^2} {\cal K}_{EF}^{0^-} ,\\
\nonumber
{\cal K}_{FF}^{0^-} &=& - \frac{1}{2} \int_0^1d\alpha\, \bigg[ M_f M_{g}+\hat\alpha M_f^2+\alpha M_{g}^2\bigg]\\
&& \quad \times \overline{\cal C}^{\rm iu}_1(\omega_{f g}(\alpha, Q^2))\,.
\end{eqnarray}
\end{subequations}}

Eq.\,\eqref{bsefinalE} is an eigenvalue problem. It has a solution for $Q^2=-m_{0^-}^2$, at which point the eigenvector is the meson's Bethe-Salpeter amplitude.  In the calculation of observables, one must use the canonically normalised amplitude, \emph{viz}.\ the amplitude rescaled such that
\begin{equation}
\label{normcan}
1=\left. \frac{d}{d Q^2}\Pi_{0^-}(Z,Q)\right|_{Z=Q},
\end{equation}
where
\begin{equation}
\Pi_{0^-}(Z,Q) = 6 {\rm tr}_{\rm D} \!\! \int\! \frac{d^4t}{(2\pi)^4} \Gamma_{0^-}(-Z)
 S_f(t_+) \, \Gamma_{0^-}(Z)\, S_g(t)\,.
 \label{normcan2}
\end{equation}

\begin{table}[t]
\caption{\label{Tab:DressedQuarks}
Couplings, ultraviolet cutoffs and current-quark masses that deliver a good description of pseudoscalar meson properties, along with the dressed-quark masses and chosen pseudoscalar meson properties they produce; all obtained with $m_G=0.5\,$GeV, $\Lambda_{\rm ir} = 0.24\,$GeV.
Empirically, at a sensible level of precision \cite{Zyla:2020zbs}:
$m_\pi =0.14$, $f_\pi=0.092$; $m_K=0.50$, $f_K=0.11$; $m_{\eta_c} =2.98$, $f_{\eta_c}=0.24$; $m_{\eta_b}=9.40$.
The value of $f_{\eta_b}$ is discussed in connection with Eq.\,\eqref{fetab}.
(Dimensioned quantities in GeV.)}
\begin{center}
\begin{tabular*}
{\hsize}
{
c@{\extracolsep{0ptplus1fil}}|
c@{\extracolsep{0ptplus1fil}}
|c@{\extracolsep{0ptplus1fil}}
c@{\extracolsep{0ptplus1fil}}
|c@{\extracolsep{0ptplus1fil}}
c@{\extracolsep{0ptplus1fil}}
c@{\extracolsep{0ptplus1fil}}}\hline\hline
quark & $\alpha_{\rm IR}/\pi\ $ & $\Lambda_{\rm uv}$ & $m$ &   $M$ &  $m_{0^-}$ & $f_{0^-}$ \\\hline
$l=u/d\ $  & $0.36\phantom{1}$ & $0.91\ $ & $0.007\ $ & 0.37$\ $ & 0.14 & 0.10  \\\hline
$s$  & $0.36\phantom{1}$ & $0.91\ $ & $0.17\phantom{7}\ $ & 0.53$\ $ & 0.50 & 0.11 \\\hline
$c$  & $0.053$ & $1.89\ $ & $1.23\phantom{7}\ $ & 1.60$\ $ & 2.98 & 0.24 \\\hline
$b$  & $0.012$ & $3.54\ $ & $4.66\phantom{7}\ $ & 4.83$\ $ & 9.40 & 0.41
\\\hline\hline
\end{tabular*}
\end{center}
\end{table}

The pseudoscalar meson's leptonic decay constant is given by:
\begin{align}
f_{0^-} &= \frac{N_c}{4\pi^2}\frac{1}{ M_{f g}}\,
\big[ E_{0^-} {\cal K}_{FE}^{0^-} + F_{0^-}{\cal K}_{FF}^{0^-} \big]_{Q^2=-m_{0^-}^2}\,. \label{ffg}
\end{align}

Following Ref.\,\cite{Yin:2019bxe}, we use the light-quark results from Refs.\,\cite{Roberts:2011cf, Chen:2012txa}, listed in Table~\ref{Tab:DressedQuarks}.  The fitted value of $m_s/m_l = 24$ is compatible with estimates in QCD \cite{Zyla:2020zbs}, even though the individual current-masses are too large by a factor of $\lesssim 2$ because of the CI's deficiencies in connection with ultraviolet quantities.  The result $M_s/M_l =1.4$ is commensurate with the value obtained in efficacious RL studies with momentum-dependent interactions \cite{Roberts:2020udq}: $M_s/M_l =1.25(9)$.

In heavy-quark systems, $\Lambda_{\rm uv}^{0^-}$ is allowed to vary with the meson's mass and the associated coupling is fixed by requiring
\begin{equation}
\alpha_{\rm IR}(\Lambda_{\rm uv}^{0^-}) [\Lambda_{\rm uv}^{0^-}]^2 \ln\frac{\Lambda_{\rm uv}^{0^-}}{\Lambda_{\rm ir}}
=
\alpha_{\rm IR}(\Lambda_{\rm uv}^{\pi}) [\Lambda_{\rm uv}^{\pi}]^2 \ln\frac{\Lambda_{\rm uv}^{\pi}}{\Lambda_{\rm ir}}\,.
\label{alphaLambda}
\end{equation}
This procedure serves to limit the number of parameters, so that in fitting the $\eta_{c,b}$ quantities in Table~\ref{Tab:DressedQuarks} there are only two parameters for each case: $m_{c,b}$, $\Lambda_{\rm uv}^{\eta_{c,b}}$.  Considering the $\eta_b$, a lQCD calculation reports $f_{\eta_b}=0.472(4)$ \cite{McNeile:2012qf}, but this is larger than the result for $f_{\Upsilon}=0.459(22)$ \cite{Colquhoun:2015oha}; hence, it is contrary to the experimental pattern: $f_{\pi}<f_{\rho}$, $f_{\eta_c}<f_{J/\psi}$.  Therefore, Ref.\,\cite{Yin:2019bxe} chose to constrain $m_{b}$, $\Lambda_{\rm uv}^{\eta_{b}}$ via known experimental results \cite{Zyla:2020zbs}:
\begin{equation}
\label{fetab}
f_{\eta_b} = f_{\Upsilon} [ f_{\eta_c}/f_{J/\psi} ] = 0.41(2)\,.
\end{equation}
The fitted values of $m_c$, $m_b-m_c$ are aligned with QCD estimates and the results for $M_{c,b}$ are commensurate with typical values of the heavy-quark pole masses \cite{Zyla:2020zbs}.

Equation~\eqref{alphaLambda} implements another physical constraint, \emph{viz}.\ any increase in the momentum-space extent of a hadron wave function is accompanied by a decrease in the effective coupling between the constituents. This avoids critical over-binding.  One finds
\begin{equation}
\Lambda_{\rm uv}(s=m_{0^-}^2) \stackrel{m_{0^-} \geq m_K}{=} 0.83 \ln[2.79 + s/(4.66 \Lambda_{\rm ir})^2 ]\,;
\end{equation}
hence, via Eq.\,\eqref{alphaLambda}, evolution of the quark+antiquark coupling that is well approximated by
\begin{align}
\label{alphaIRMass}
\alpha_{\rm IR}(s) \stackrel{m_{0^-} \geq m_K}= \frac{0.047\,\alpha_{\rm IR}(m_K^2) }{\ln[1.04 + s/(21.77 \Lambda_{\rm ir})^2]}\,.
\end{align}
The physical origin of these outcomes is clear: the $f_{0^-}$ integral diverges logarithmically with increasing $\Lambda_{\rm uv}$; and the flow of $\alpha_{\rm IR}$ compensates for analogous behaviour in the Bethe-Salpeter kernel, thereby maintaining the given meson's mass.


\section{Faddeev Equations}
\label{AppFaddeev}
\subsection{Dirac structure of the amplitudes}
Adapting the notation in Ref.\,\cite{Cahill:1988dx}, the Faddeev equation depicted in Fig.\,\ref{figFaddeev} can be written practically in the following form:
\begin{align}
\Psi_{(fg)h}(P)  &=
\big\{ g_{DB}^{\pi_\Psi \pi_d} \Gamma_{(gh)}(k_{(gh)}) \nonumber \\
& \quad \times S_{g}^{\textrm{T}}(q_{g})
g_{DB}^{\pi_\Psi \pi_d}\overline{\Gamma}_{(fg)}(-p_{(fg)})\big\} \nonumber \\
& \quad \times S_{f}(k_{f}) \Delta_{(gh)}(k_{(gh)})\Psi_{(gh)f}(P)\,,
\label{AppendixFE}
\end{align}
where:
the first two lines on the right-hand-side, parenthesised, express the exchange kernel;
a sum over all contributing diquark correlations is implicit;
$k_{(gh)}=-k+(2/3)P$, $q_{g}=p_{(fg)}-k_{f}$, $p_{(fg)}=-p+(2/3)P$, $k_{f}=k+ P/3$;
``T'' indicates matrix transpose;
$\bar\Gamma(Q) = C^\dagger \Gamma(Q)^{\rm T} C$; and $g_{DB}^{\pi_\Psi \pi_d}$ is described in connection with Eq.\,\eqref{gDBval}.  The complete amplitude for the bound-state is
\begin{equation}
\Psi =  \Psi_{(gh)f} + \Psi_{(hf)g}+\Psi_{(fg)h} \,.
\end{equation}

Regarding $J=1/2$ solutions of Eq.\,\eqref{AppendixFE}, one has
\begin{equation}
\label{PsiuP}
\Psi(P) = \psi(P) u(P)\,,
\end{equation}
where the positive energy spinor satisfies
\begin{equation}
\bar u(P)(i\gamma\cdot P+M) = 0=(i\gamma\cdot P + M)u(P)
\end{equation}
and is normalised such that $\bar u(P) u(P) = 2 M$.  (See Ref.\,\cite[Appendix~A]{Chen:2012qr} for more details.)  Using this form, then the complete CI solution for $\psi(P)$, the Faddeev amplitude of a $1/2^\pm$ baryon, is a sum of the following Dirac structures:
\begin{subequations}
\label{SAPD}
\begin{align}
\mathcal{S}^{\pm}&= \mathcal{G}^{\pm}\big({\mathpzc s}^{\pm}\,\mathbf{I}_{\rm D}\big),\\
\mathcal{A}_{\mu}^{\pm}&= \mathcal{G}^{\pm}\big({\mathpzc a}_{1}^{\pm}\,i\gamma_{5}\gamma_{\mu}+{\mathpzc a}_{2}^{\pm}\gamma_{5}\hat P_{\mu}\big),
\label{Aa1}\\
\mathcal{P}^{\pm}&= \mathcal{G}^{\pm}\big({\mathpzc p}^{\pm}\,i\gamma_{5}\big),\\
\mathcal{V}_{\mu}^{\pm}&= \mathcal{G}^{\pm}\big({\mathpzc v}_{1}^{\pm}\,i\gamma_{\mu}+{\mathpzc v}_{2}^{\pm}\mathbf{I}_{\rm D}\hat P_{\mu}\big),
\end{align}
\end{subequations}
where $\mathcal{G}^{+(-)}=\mathbf{I}_{\rm D}(\gamma_{5})$.  Faddeev equation dynamics determines the values of the coefficients: $\{{\mathpzc s}^{\pm}, {\mathpzc a}_{1,2}, {\mathpzc p}^{\pm}, {\mathpzc v}_{1,2}\}$, each of which is a vector in flavour space whose structure is determined by the baryon under consideration, as will soon be explicated.

For $J=3/2^\pm$ baryon, Eq.\,\eqref{PsiuP} becomes
\begin{equation}
\label{PsiuPRS}
\Psi_{\mu\nu}^\pm(P) = \psi^\pm_{\mu\nu}(P) u_\nu(P)\,,
\end{equation}
where $u_\nu(P)$ is a Rarita-Schwinger spinor; and
\begin{equation}
\psi_{\mu\nu}^{\pm}= \mathcal{G}^{\pm}\big({\mathpzc f}^{\pm}\,\mathbf{I}_{\rm D}\delta_{\mu\nu}\big)\,,
\end{equation}
where, again, the coefficients ${\mathpzc f}^{\pm}$ carry flavour labels appropriate to the baryon.

The diquark propagators in Eq.\,\eqref{AppendixFE} take the following forms, depending on the quantum numbers of the propagating correlation:
\begin{subequations}
\label{diquarkprops}
\begin{align}
\Delta_{0^{+}}(k) &= \frac{1}{k^{2}+m_{0^{+}}^{2}}, \\
\Delta_{\mu\nu}^{1^{+}}(k)&= \frac{\delta_{\mu\nu} +\frac{k_{\mu}k_{\nu}}{m_{1^{+}}^{2}}}{k^{2}+m_{1^{+}}^{2}},\\
\Delta_{0^{-}}(k)&= \frac{1}{k^{2}+m_{0^{-}}^{2}},\\
\Delta_{\mu\nu}^{1^{-}}(k)&= \frac{\delta_{\mu\nu} +\frac{k_{\mu}k_{\nu}}{m_{1^{-}}^{2}}}{k^{2}+m_{1^{-}}^{2}},
\end{align}
\end{subequations}
where the masses are given in Table~\ref{diquarkspectrum}.

\subsection{Flavour structure of the amplitudes}
Working in the isospin-symmetry limit, the Faddeev equation in Fig.\,\ref{figFaddeev} supports $88$ distinct baryon states, which we list below.  The flavour structure of the amplitudes reflects the quark+diquark approximation.

\subsubsection{Three light quarks}
There are $16 = 2^\pm\times 8$ light-quark baryons (``$2^\pm$'' indicates positive/negative parity), with $8=2^\pm \times 4$ $J=1/2$:
{\allowdisplaybreaks
\begin{subequations}
\label{halfnucleon}
\begin{align}
\Psi_N & =
\left[
\begin{array}{ll}
{\rm r}_1 & u[ud]_{0^{+}}   \\
{\rm r}_2 &  d\{uu\}_{1^{+}}    \\
{\rm r}_3 &  u\{ud\}_{1^{+}}    \\
{\rm r}_4 &  u[ud]_{0^{-}}      \\
{\rm r}_5 &  u[ud]_{1^{-}}      \\
\end{array} \right],  \\
\Psi_\Lambda & =
\frac{1}{\sqrt{2}}\left[
\begin{array}{ll}
{\rm r}_1 & \sqrt{2}s[ud]_{0^{+}}           \\
{\rm r}_2 & d[us]_{0^{+}}-u[ds]_{0^{+}}     \\
{\rm r}_3 & d\{us\}_{1^{+}}-u\{ds\}_{1^{+}} \\
{\rm r}_4 & \sqrt{2}s[ud]_{0^{-}}           \\
{\rm r}_5 & d[us]_{0^{-}}-u[ds]_{0^{-}}     \\
{\rm r}_6 & \sqrt{2}s[ud]_{1^{-}}           \\
{\rm r}_7 & d[us]_{1^{-}}-u[ds]_{1^{-}}     \\
\end{array}
\right],  \\
\rule{0em}{10ex} \Psi_{\Sigma} & =
\left[
\begin{array}{ll}
{\rm r}_1 & u[us]_{0^{+}}   \\
{\rm r}_2 & s\{uu\}_{1^{+}} \\
{\rm r}_3 & u\{us\}_{1^{+}} \\
{\rm r}_4 & u[us]_{0^{-}}   \\
{\rm r}_5 & u[us]_{1^{-}}   \\
\end{array}
\right],\\
\Psi_{\Xi} & =
\left[
\begin{array}{ll}
{\rm r}_1 & s[us]_{0^{+}}   \\
{\rm r}_2 & s\{us\}_{1^{+}} \\
{\rm r}_3 & u\{ss\}_{1^{+}} \\
{\rm r}_4 & s[us]_{0^{-}}   \\
{\rm r}_5 & s[us]_{1^{-}}   \\
\end{array}
\right],
\end{align}
\end{subequations}
}
\hspace*{-0.35\parindent}in which the row number is listed explicitly for future use;
%
and $2^\pm\times 4$ $J=3/2$:
\begin{subequations}
\label{ThreeHalfDelta}
\begin{align}
\Psi_\Delta & =
\left[
\begin{array}{c}
u\{uu\}_{1^{+}} \\
\end{array}
\right], \\
\Psi_{\Sigma^{\ast}} & =
\left[
\begin{array}{c}
s\{uu\}_{1^{+}} \\
u\{us\}_{1^{+}} \\
\end{array}
\right], \\
\Psi_{\Xi^{\ast}} & =
\left[
\begin{array}{c}
s\{us\}_{1^{+}} \\
u\{ss\}_{1^{+}} \\
\end{array}
\right], \\
\Psi_{\Omega} & = \left[
\begin{array}{c}
s\{ss\}_{1^{+}} \\
\end{array}
\right].
\end{align}
\end{subequations}

\subsubsection{Flavour amplitudes: two light quarks and one heavy}
With $Q=c,b$, there are $32 = 2^\pm \times 2_Q\times 8$ distinct states in this case, $20=2^\pm \times 2_Q \times 5$ $J=1/2$:
{\allowdisplaybreaks
\begin{subequations}
\label{half2q1Q}
\begin{align}
\Psi_{\Sigma_{Q}} & =
\left[
\begin{array}{ll}
{\rm r}_1 & u[uQ]_{0^{+}}   \\
{\rm r}_2 & Q\{uu\}_{1^{+}} \\
{\rm r}_3 & u\{uQ\}_{1^{+}} \\
{\rm r}_4 & u[uQ]_{0^{-}}   \\
{\rm r}_5 & u[uQ]_{1^{-}}   \\
\end{array}
\right], \\
\Psi_{\Lambda_{Q}} & =\frac{1}{\sqrt{2}}
\left[
\begin{array}{ll}
{\rm r}_1 & \sqrt{2} Q[ud]_{0^{+}}           \\
{\rm r}_2 &d[uQ]_{0^{+}}-u[dQ]_{0^{+}}     \\
{\rm r}_3 &d\{uQ\}_{1^{+}}-u\{dQ\}_{1^{+}} \\
{\rm r}_4 & \sqrt{2}Q[ud]_{0^{-}}           \\
{\rm r}_5 &d[uQ]_{0^{-}}-u[dQ]_{0^{-}}     \\
{\rm r}_6 &\sqrt{2}Q[ud]_{1^{-}}           \\
{\rm r}_7 &d[uQ]_{1^{-}}-u[dQ]_{1^{-}}     \\
\end{array}
\right], \\
\Psi_{\Xi_{Q}}  & =
\frac{1}{\sqrt{2}} \left[
\begin{array}{ll}
{\rm r}_1 & \sqrt{2}Q[us]_{0^{+}}           \\
{\rm r}_2 & s[uQ]_{0^{+}}-u[sQ]_{0^{+}}     \\
{\rm r}_3 & s\{uQ\}_{1^{+}}-u\{sQ\}_{1^{+}} \\
{\rm r}_4 & \sqrt{2} Q[us]_{0^{-}}           \\
{\rm r}_5 & s[uQ]_{0^{-}}-u[sQ]_{0^{-}}     \\
{\rm r}_6 & \sqrt{2}Q[us]_{1^{-}}           \\
{\rm r}_7 & s[uQ]_{1^{-}}-u[sQ]_{1^{-}}     \\
\end{array}
\right], \\
\Psi_{\Xi_{Q}^{\prime}} & =
\frac{1}{\sqrt{2}} \left[
\begin{array}{ll}
{\rm r}_1 & s[uQ]_{0^{+}}+u[sQ]_{0^{+}}     \\
{\rm r}_2 & \sqrt{2}Q\{us\}_{1^{+}}         \\
{\rm r}_3 & s\{uQ\}_{1^{+}}+u\{sQ\}_{1^{+}} \\
{\rm r}_4 & s[uQ]_{0^{-}}+u[sQ]_{0^{-}}     \\
{\rm r}_5 & s[uQ]_{1^{-}}+u[sQ]_{1^{-}}     \\
\end{array}
\right], \\
\Psi_{\Omega_{Q}} & =
\left[
\begin{array}{ll}
{\rm r}_1 & s[sQ]_{0^{+}}   \\
{\rm r}_2 & Q\{ss\}_{1^{+}} \\
{\rm r}_3 & s\{sQ\}_{1^{+}} \\
{\rm r}_4 & s[sQ]_{0^{-}}   \\
{\rm r}_5 & s[sQ]_{1^{-}}   \\
\end{array}
\right],
\end{align}
\end{subequations}}
\hspace*{-0.4\parindent}where $\Xi_{Q}$/$\Xi_{Q}^{\prime}$ are antisymmetric/symmetric under $u\leftrightarrow s$; and $12=2^\pm \times 2_Q\times 3$ $J=3/2$:
{\allowdisplaybreaks
\begin{subequations}
\label{ThreeHalfqQQ}
\begin{align}
\Psi_{\Sigma_{Q}^{\ast}} & =
\left[
\begin{array}{c}
Q\{uu\}_{1^{+}} \\
u\{uQ\}_{1^{+}} \\
\end{array}
\right], \label{SigmaQ}\\
\Psi_{\Xi_{Q}^{\ast}} & =
\frac{1}{\sqrt{2}} \left[
\begin{array}{ll}
\sqrt{2} Q\{us\}_{1^{+}}         \\
s\{uQ\}_{1^{+}}+u\{sQ\}_{1^{+}} \\
\end{array}
\right], \\
\Psi_{\Omega_{Q}^{\ast}}^{\pm} & =
\left[
\begin{array}{c}
Q\{ss\}_{1^{+}} \\
s\{sQ\}_{1^{+}} \\
\end{array}
\right].
\end{align}
\end{subequations}}

\subsubsection{Flavour amplitudes: one light quark and two heavy}
Writing $q=u,s$ and remembering that isospin-symmetry is assumed, there are $28 = 2^\pm \times 2_q \times 7$ distinct states in this case, $16 = 2^\pm \times 2_q\times 4$ $J=1/2$:
{\allowdisplaybreaks
\begin{subequations}
\label{HalfOneTwo}
\begin{align}
\Psi^{\Xi_{cc}^{q=u}}_{\Omega_{cc}^{q=s}} & =
\left[
\begin{array}{ll}
{\rm r}_1 & c[qc]_{0^{+}}   \\
{\rm r}_2 & c\{qc\}_{1^{+}} \\
{\rm r}_3 & q\{cc\}_{1^{+}} \\
{\rm r}_4 & c[qc]_{0^{-}}   \\
{\rm r}_5 & c[qc]_{1^{-}}   \\
\end{array}
\right],\\
\Psi^{\Xi_{cb}^{q=u}}_{\Omega_{cb}^{q=s}} & =
\frac{1}{\sqrt{2}}
\left[
\begin{array}{ll}
{\rm r}_1 & b[qc]_{0^{+}}-c[qb]_{0^{+}}     \\
{\rm r}_2 & \sqrt{2}q[cb]_{0^{+}}           \\
{\rm r}_3 & b\{qc\}_{1^{+}}-c\{qb\}_{1^{+}} \\
{\rm r}_4 & b[qc]_{0^{-}}-c[qb]_{0^{-}}     \\
{\rm r}_5 & \sqrt{2}q[cb]_{0^{-}}           \\
{\rm r}_6 & b[qc]_{1^{-}}-c[qb]_{1^{-}}     \\
{\rm r}_7 & \sqrt{2}q[cb]_{1^{-}}           \\
\end{array}
\right], \\
\Psi^{\Xi_{cb}^{\prime q=u }}_{\Omega_{cb}^{\prime q=s }}& =
\frac{1}{\sqrt{2}}
\left[
\begin{array}{ll}
{\rm r}_1 & b[qc]_{0^{+}}+c[qb]_{0^{+}}     \\
{\rm r}_2 & b\{qc\}_{1^{+}}+c\{qb\}_{1^{+}} \\
{\rm r}_3 & \sqrt{2}q\{cb\}_{1^{+}}         \\
{\rm r}_4 & b[qc]_{0^{-}}+c[qb]_{0^{-}}     \\
{\rm r}_5 & b[qc]_{1^{-}}+c[qb]_{1^{-}}     \\
\end{array}
\right], \\
%
\Psi^{\Xi_{bb}^{q=u}}_{\Omega_{bb}^{q=s}} & =
\left[
\begin{array}{ll}
{\rm r}_1 & b[qb]_{0^{+}}   \\
{\rm r}_2 & b\{qb\}_{1^{+}} \\
{\rm r}_3 & q\{bb\}_{1^{+}} \\
{\rm r}_4 & b[qb]_{0^{-}}   \\
{\rm r}_5 & b[qb]_{1^{-}}   \\
\end{array}
\right],
\end{align}
\end{subequations}}
\hspace*{-0.4\parindent}where $\Xi_{cb}$/$\Xi_{cb}^{\prime}$ and $\Omega_{cb}$/$\Omega_{cb}^{\prime}$ are antisymmetric/symmetric under $c\leftrightarrow b$; and $12=2^\pm \times 2_q \times 3$ $J=3/2$:
{\allowdisplaybreaks
\begin{subequations}
\label{ThreeHalfOneTwo}
\begin{align}
\Psi^{\Xi_{cc}^{\ast q=u}}_{\Omega_{cc}^{\ast q=s}} & =
\left[
\begin{array}{l}
c\{qc\}_{1^{+}} \\
q\{cc\}_{1^{+}} \\
\end{array}
\right], \\
\Psi^{\Xi_{cb}^{\ast q=u}}_{\Omega_{cb}^{\ast q=s}} & =
\frac{1}{\sqrt{2}}
\left[
\begin{array}{l}
b\{qc\}_{1^{+}}+c\{qb\}_{1^{+}} \\
\sqrt{2}q\{cb\}_{1^{+}}         \\
\end{array}
\right], \\
\Psi^{\Xi_{bb}^{\ast} q=u}_{\Omega_{bb}^{\ast} q=s} & =
\left[
\begin{array}{c}
b\{qb\}_{1^{+}} \\
q\{bb\}_{1^{+}} \\
\end{array}
\right].
\end{align}
\end{subequations}}

\subsubsection{Flavour amplitudes: three heavy quarks}
Finally, there are $12 = 2^\pm \times 6$ triply-heavy baryons, $4=2^\pm \times 2$ $J=1/2$:
{\allowdisplaybreaks
\begin{subequations}
\label{OneHalfQQQ}
\begin{align}
\Psi_{\Omega_{ccb}} & =
\left[
\begin{array}{ll}
{\rm r}_1 & c[cb]_{0^{+}}   \\
{\rm r}_2 & b\{cc\}_{1^{+}} \\
{\rm r}_3 & c\{cb\}_{1^{+}} \\
{\rm r}_4 & c[cb]_{0^{-}}   \\
{\rm r}_5 & c[cb]_{1^{-}}   \\
\end{array}
\right], \\
\Psi_{\Omega_{cbb}} & =
\left[
\begin{array}{ll}
{\rm r}_1 & b[cb]_{0^{+}}   \\
{\rm r}_2 & b\{cb\}_{1^{+}} \\
{\rm r}_3 & c\{bb\}_{1^{+}} \\
{\rm r}_4 & b[cb]_{0^{-}}   \\
{\rm r}_5 & b[cb]_{1^{-}}   \\
\end{array}
\right];
\end{align}
\end{subequations}}
\hspace*{-0.4ex}and $8=2^\pm \times 4$ $J=3/2$:
{\allowdisplaybreaks
\begin{subequations}
\label{ThreeHalfQQQ}
\begin{align}
\Psi_{\Omega_{ccc}^{\ast}} & =
\left[
\begin{array}{c}
c\{cc\}_{1^{+}} \\
\end{array}
\right], \\
\Psi_{\Omega_{ccb}^{\ast}} & =
\left[
\begin{array}{c}
b\{cc\}_{1^{+}} \\
c\{cb\}_{1^{+}} \\
\end{array}
\right], \\
\Psi_{\Omega_{cbb}^{\ast}} & =
\left[
\begin{array}{c}
b\{cb\}_{1^{+}} \\
c\{bb\}_{1^{+}} \\
\end{array}
\right], \\
\Psi_{\Omega_{bbb}^{\ast}} & =
\left[
\begin{array}{c}
b\{bb\}_{1^{+}} \\
\end{array}
\right].
\end{align}
\end{subequations}}



\begin{thebibliography}{97}
\providecommand{\natexlab}[1]{#1}
\providecommand{\url}[1]{\texttt{#1}}
\providecommand{\urlprefix}{URL }
\expandafter\ifx\csname urlstyle\endcsname\relax
  \providecommand{\doi}[1]{doi:\discretionary{}{}{}#1}\else
  \providecommand{\doi}[1]{doi:\discretionary{}{}{}\begingroup
  \urlstyle{rm}\url{#1}\endgroup}\fi
\providecommand{\bibinfo}[2]{#2}

\bibitem[{Zyla et~al.(2020)}]{Zyla:2020zbs}
\bibinfo{author}{P.~Zyla}, et~al., \bibinfo{title}{{Review of Particle
  Physics}}, \bibinfo{journal}{PTEP} \bibinfo{volume}{2020}
  (\bibinfo{year}{2020}) \bibinfo{pages}{083C01}.

\bibitem[{Richard(2016)}]{Richard:2016eis}
\bibinfo{author}{J.-M. Richard}, \bibinfo{title}{{Exotic hadrons: review and
  perspectives}}, \bibinfo{journal}{Few Body Syst.} \bibinfo{volume}{57}
  (\bibinfo{year}{2016}) \bibinfo{pages}{1185--1212}.

\bibitem[{Esposito et~al.(2017)Esposito, Pilloni, and
  Polosa}]{Esposito:2016noz}
\bibinfo{author}{A.~Esposito}, \bibinfo{author}{A.~Pilloni},
  \bibinfo{author}{A.~D. Polosa}, \bibinfo{title}{{Multiquark Resonances}},
  \bibinfo{journal}{Phys. Rept.} \bibinfo{volume}{668} (\bibinfo{year}{2017})
  \bibinfo{pages}{1--97}.

\bibitem[{Liu et~al.(2019)Liu, Chen, Chen, Liu, and Zhu}]{Liu:2019zoy}
\bibinfo{author}{Y.-R. Liu}, \bibinfo{author}{H.-X. Chen},
  \bibinfo{author}{W.~Chen}, \bibinfo{author}{X.~Liu}, \bibinfo{author}{S.-L.
  Zhu}, \bibinfo{title}{{Pentaquark and Tetraquark states}},
  \bibinfo{journal}{Prog. Part. Nucl. Phys.} \bibinfo{volume}{107}
  (\bibinfo{year}{2019}) \bibinfo{pages}{237--320}.

\bibitem[{Barabanov et~al.(2021)}]{Barabanov:2020jvn}
\bibinfo{author}{M.~Y. Barabanov}, et~al., \bibinfo{title}{{Diquark
  Correlations in Hadron Physics: Origin, Impact and Evidence}},
  \bibinfo{journal}{Prog. Part. Nucl. Phys.} \bibinfo{volume}{116}
  (\bibinfo{year}{2021}) \bibinfo{pages}{103835}.

\bibitem[{Eichmann et~al.(2020)Eichmann, Fischer, Heupel, Santowsky, and
  Wallbott}]{Eichmann:2020oqt}
\bibinfo{author}{G.~Eichmann}, \bibinfo{author}{C.~S. Fischer},
  \bibinfo{author}{W.~Heupel}, \bibinfo{author}{N.~Santowsky},
  \bibinfo{author}{P.~C. Wallbott}, \bibinfo{title}{{Four-quark states from
  functional methods}}, \bibinfo{journal}{Few Body Syst.} \bibinfo{volume}{61}
  (\bibinfo{year}{2020}) \bibinfo{pages}{38}.

\bibitem[{Galanti(2018)}]{Galanti:2018bnp}
\bibinfo{author}{M.~Galanti}, \bibinfo{title}{{Production, spectroscopy and
  properties of heavy hadrons}}, \bibinfo{journal}{PoS}
  \bibinfo{volume}{LHCP2018} (\bibinfo{year}{2018}) \bibinfo{pages}{132}.

\bibitem[{Ketzer et~al.(2020)Ketzer, Grube, and Ryabchikov}]{KETZER2020103755}
\bibinfo{author}{B.~Ketzer}, \bibinfo{author}{B.~Grube},
  \bibinfo{author}{D.~Ryabchikov}, \bibinfo{title}{Light-meson spectroscopy
  with COMPASS}, \bibinfo{journal}{Prog. Part. Nucl. Phys.}
  \bibinfo{volume}{113} (\bibinfo{year}{2020}) \bibinfo{pages}{103755}.

\bibitem[{Carman et~al.(2020)Carman, Joo, and Mokeev}]{Carman:2020qmb}
\bibinfo{author}{D.~Carman}, \bibinfo{author}{K.~Joo},
  \bibinfo{author}{V.~Mokeev}, \bibinfo{title}{{Strong QCD Insights from
  Excited Nucleon Structure Studies with CLAS and CLAS12}},
  \bibinfo{journal}{Few Body Syst.} \bibinfo{volume}{61} (\bibinfo{year}{2020})
  \bibinfo{pages}{29}.

\bibitem[{Brodsky et~al.(2020)}]{Brodsky:2020vco}
\bibinfo{author}{S.~J. Brodsky}, et~al., \bibinfo{title}{{Strong QCD from
  Hadron Structure Experiments}}, \bibinfo{journal}{Intern. J. Mod. Phys. E}
  \bibinfo{volume}{124} (\bibinfo{year}{2020}) \bibinfo{pages}{2030006}.

\bibitem[{{R. Beck, A. Thiel, U. Thoma, Y. Wunderlich
  (Eds.)}(2020)}]{NStar2020}
\bibinfo{author}{{R. Beck, A. Thiel, U. Thoma, Y. Wunderlich (Eds.)}},
  \bibinfo{title}{{Proceedings of The 12th International Workshop on the
  Physics of Excited Nucleons (NSTAR 2019)}}, \bibinfo{journal}{EPJ Web Conf.}
  \bibinfo{volume}{241} (\bibinfo{year}{2020}) \bibinfo{pages}{01000}.

\bibitem[{Bazavov et~al.(2010)}]{Bazavov:2009bb}
\bibinfo{author}{A.~Bazavov}, et~al., \bibinfo{title}{{Full nonperturbative QCD
  simulations with 2+1 flavors of improved staggered quarks}},
  \bibinfo{journal}{Rev. Mod. Phys.} \bibinfo{volume}{82}
  (\bibinfo{year}{2010}) \bibinfo{pages}{1349--1417}.

\bibitem[{Fodor and Hoelbling(2012)}]{Fodor:2012gf}
\bibinfo{author}{Z.~Fodor}, \bibinfo{author}{C.~Hoelbling},
  \bibinfo{title}{{Light Hadron Masses from Lattice QCD}},
  \bibinfo{journal}{Rev. Mod. Phys.} \bibinfo{volume}{84}
  (\bibinfo{year}{2012}) \bibinfo{pages}{449}.

\bibitem[{Brice{\~{n}}o et~al.(2018)Brice{\~{n}}o, Dudek, and
  Young}]{Briceno:2017max}
\bibinfo{author}{R.~A. Brice{\~{n}}o}, \bibinfo{author}{J.~J. Dudek},
  \bibinfo{author}{R.~D. Young}, \bibinfo{title}{{Scattering processes and
  resonances from lattice QCD}}, \bibinfo{journal}{Rev. Mod. Phys.}
  \bibinfo{volume}{90} (\bibinfo{year}{2018}) \bibinfo{pages}{025001}.

\bibitem[{Hansen and Sharpe(2019)}]{Hansen:2019nir}
\bibinfo{author}{M.~T. Hansen}, \bibinfo{author}{S.~R. Sharpe},
  \bibinfo{title}{{Lattice QCD and Three-particle Decays of Resonances}},
  \bibinfo{journal}{Ann. Rev. Nucl. Part. Sci.} \bibinfo{volume}{69}
  (\bibinfo{year}{2019}) \bibinfo{pages}{65--107}.

\bibitem[{Durr et~al.(2008)}]{Durr:2008zz}
\bibinfo{author}{S.~Durr}, et~al., \bibinfo{title}{{Ab-Initio Determination of
  Light Hadron Masses}}, \bibinfo{journal}{Science} \bibinfo{volume}{322}
  (\bibinfo{year}{2008}) \bibinfo{pages}{1224--1227}.

\bibitem[{Edwards et~al.(2011)Edwards, Dudek, Richards, and
  Wallace}]{Edwards:2011jj}
\bibinfo{author}{R.~G. Edwards}, \bibinfo{author}{J.~J. Dudek},
  \bibinfo{author}{D.~G. Richards}, \bibinfo{author}{S.~J. Wallace},
  \bibinfo{title}{{Excited state baryon spectroscopy from lattice QCD}},
  \bibinfo{journal}{Phys. Rev. D} \bibinfo{volume}{84} (\bibinfo{year}{2011})
  \bibinfo{pages}{074508}.

\bibitem[{Brown et~al.(2014)Brown, Detmold, Meinel, and
  Orginos}]{Brown:2014ena}
\bibinfo{author}{Z.~S. Brown}, \bibinfo{author}{W.~Detmold},
  \bibinfo{author}{S.~Meinel}, \bibinfo{author}{K.~Orginos},
  \bibinfo{title}{{Charmed bottom baryon spectroscopy from lattice QCD}},
  \bibinfo{journal}{Phys. Rev. D} \bibinfo{volume}{90} (\bibinfo{year}{2014})
  \bibinfo{pages}{094507}.

\bibitem[{Lubicz et~al.(2017)Lubicz, Melis, and Simula}]{Lubicz:2017asp}
\bibinfo{author}{V.~Lubicz}, \bibinfo{author}{A.~Melis},
  \bibinfo{author}{S.~Simula}, \bibinfo{title}{{Masses and decay constants of
  D*$_{(s)}$ and B*$_{(s)}$ mesons with N$_f = 2 + 1 + 1$ twisted mass
  fermions}}, \bibinfo{journal}{Phys. Rev. D} \bibinfo{volume}{96}
  (\bibinfo{year}{2017}) \bibinfo{pages}{034524}.

\bibitem[{Bali et~al.(2017)Bali, Collins, Cox, and Sch{\"a}fer}]{Bali:2017pdv}
\bibinfo{author}{G.~S. Bali}, \bibinfo{author}{S.~Collins},
  \bibinfo{author}{A.~Cox}, \bibinfo{author}{A.~Sch{\"a}fer},
  \bibinfo{title}{{Masses and decay constants of the $D_{s0}^*(2317)$ and
  $D_{s1}(2460)$ from $N_f=2$ lattice QCD close to the physical point}},
  \bibinfo{journal}{Phys. Rev. D} \bibinfo{volume}{96} (\bibinfo{year}{2017})
  \bibinfo{pages}{074501}.

\bibitem[{Mathur et~al.(2018)Mathur, Padmanath, and Mondal}]{Mathur:2018epb}
\bibinfo{author}{N.~Mathur}, \bibinfo{author}{M.~Padmanath},
  \bibinfo{author}{S.~Mondal}, \bibinfo{title}{{Precise predictions of
  charmed-bottom hadrons from lattice QCD}}, \bibinfo{journal}{Phys. Rev.
  Lett.} \bibinfo{volume}{121}~(\bibinfo{number}{20}) (\bibinfo{year}{2018})
  \bibinfo{pages}{202002}.

\bibitem[{Bahtiyar et~al.(2020)Bahtiyar, Can, Erkol, Gubler, Oka, and
  Takahashi}]{Bahtiyar:2020uuj}
\bibinfo{author}{H.~Bahtiyar}, \bibinfo{author}{K.~U. Can},
  \bibinfo{author}{G.~Erkol}, \bibinfo{author}{P.~Gubler},
  \bibinfo{author}{M.~Oka}, \bibinfo{author}{T.~T. Takahashi},
  \bibinfo{title}{{Charmed baryon spectrum from lattice QCD near the physical
  point}}, \bibinfo{journal}{Phys. Rev. D}
  \bibinfo{volume}{102}~(\bibinfo{number}{5}) (\bibinfo{year}{2020})
  \bibinfo{pages}{054513}.

\bibitem[{Weinberg(1967)}]{Weinberg:1967kj}
\bibinfo{author}{S.~Weinberg}, \bibinfo{title}{{Precise relations between the
  spectra of vector and axial vector mesons}}, \bibinfo{journal}{Phys. Rev.
  Lett.} \bibinfo{volume}{18} (\bibinfo{year}{1967}) \bibinfo{pages}{507--509}.

\bibitem[{Roberts and Schmidt(2020)}]{Roberts:2020udq}
\bibinfo{author}{C.~D. Roberts}, \bibinfo{author}{S.~M. Schmidt},
  \bibinfo{title}{{Reflections upon the Emergence of Hadronic Mass}},
  \bibinfo{journal}{Eur. Phys. J. ST}
  \bibinfo{volume}{229}~(\bibinfo{number}{22-23}) (\bibinfo{year}{2020})
  \bibinfo{pages}{3319--3340}.

\bibitem[{Roberts(2020)}]{Roberts:2020hiw}
\bibinfo{author}{C.~D. Roberts}, \bibinfo{title}{{Empirical Consequences of
  Emergent Mass}}, \bibinfo{journal}{Symmetry} \bibinfo{volume}{12}
  (\bibinfo{year}{2020}) \bibinfo{pages}{1468}.

\bibitem[{Roberts(2021)}]{Roberts:2021xnz}
\bibinfo{author}{C.~D. Roberts}, \bibinfo{title}{{On Mass and Matter --
  arXiv:2101.08340 [hep-ph]}}, \bibinfo{journal}{AAPPS Bulletin}
  (\bibinfo{year}{2021}) \bibinfo{pages}{\emph{in press}}.

\bibitem[{Roberts et~al.(2021)Roberts, Richards, Horn, and
  Chang}]{Roberts:2021nhw}
\bibinfo{author}{C.~D. Roberts}, \bibinfo{author}{D.~G. Richards},
  \bibinfo{author}{T.~Horn}, \bibinfo{author}{L.~Chang},
  \bibinfo{title}{{\emph{Insights into the Emergence of Mass from Studies of
  Pion and Kaon Structure} -- arXiv:2102.01765 [hep-ph]}} .

\bibitem[{Aznauryan et~al.(2013)Aznauryan, Bashir, Braun, Brodsky, Burkert
  et~al.}]{Aznauryan:2012ba}
\bibinfo{author}{I.~Aznauryan}, \bibinfo{author}{A.~Bashir},
  \bibinfo{author}{V.~Braun}, \bibinfo{author}{S.~Brodsky},
  \bibinfo{author}{V.~Burkert}, et~al., \bibinfo{title}{{Studies of Nucleon
  Resonance Structure in Exclusive Meson Electroproduction}},
  \bibinfo{journal}{Int. J. Mod. Phys. E} \bibinfo{volume}{22}
  (\bibinfo{year}{2013}) \bibinfo{pages}{1330015}.

\bibitem[{Crede and Roberts(2013)}]{Crede:2013sze}
\bibinfo{author}{V.~Crede}, \bibinfo{author}{W.~Roberts},
  \bibinfo{title}{{Progress towards understanding baryon resonances}},
  \bibinfo{journal}{Rept. Prog. Phys.} \bibinfo{volume}{76}
  (\bibinfo{year}{2013}) \bibinfo{pages}{076301}.

\bibitem[{Segovia et~al.(2013)Segovia, Entem, Fernandez, and
  Hernandez}]{Segovia:2013wma}
\bibinfo{author}{J.~Segovia}, \bibinfo{author}{D.~R. Entem},
  \bibinfo{author}{F.~Fernandez}, \bibinfo{author}{E.~Hernandez},
  \bibinfo{title}{{Constituent quark model description of charmonium
  phenomenology}}, \bibinfo{journal}{Int. J. Mod. Phys. E} \bibinfo{volume}{22}
  (\bibinfo{year}{2013}) \bibinfo{pages}{1330026}.

\bibitem[{Giannini and Santopinto(2015)}]{Giannini:2015zia}
\bibinfo{author}{M.~M. Giannini}, \bibinfo{author}{E.~Santopinto},
  \bibinfo{title}{{The hypercentral Constituent Quark Model and its application
  to baryon properties}}, \bibinfo{journal}{Chin. J. Phys.}
  \bibinfo{volume}{53} (\bibinfo{year}{2015}) \bibinfo{pages}{020301}.

\bibitem[{Fern\'andez and Segovia(2021)}]{Fernandez:2021zjq}
\bibinfo{author}{F.~Fern\'andez}, \bibinfo{author}{J.~Segovia},
  \bibinfo{title}{{Historical Introduction to Chiral Quark Models}},
  \bibinfo{journal}{Symmetry} \bibinfo{volume}{13}~(\bibinfo{number}{2})
  (\bibinfo{year}{2021}) \bibinfo{pages}{252}.

\bibitem[{Horn and Roberts(2016)}]{Horn:2016rip}
\bibinfo{author}{T.~Horn}, \bibinfo{author}{C.~D. Roberts},
  \bibinfo{title}{{The pion: an enigma within the Standard Model}},
  \bibinfo{journal}{J. Phys. G.} \bibinfo{volume}{43} (\bibinfo{year}{2016})
  \bibinfo{pages}{073001}.

\bibitem[{Eichmann et~al.(2016{\natexlab{a}})Eichmann, Sanchis-Alepuz,
  Williams, Alkofer, and Fischer}]{Eichmann:2016yit}
\bibinfo{author}{G.~Eichmann}, \bibinfo{author}{H.~Sanchis-Alepuz},
  \bibinfo{author}{R.~Williams}, \bibinfo{author}{R.~Alkofer},
  \bibinfo{author}{C.~S. Fischer}, \bibinfo{title}{{Baryons as relativistic
  three-quark bound states}}, \bibinfo{journal}{Prog. Part. Nucl. Phys.}
  \bibinfo{volume}{91} (\bibinfo{year}{2016}{\natexlab{a}})
  \bibinfo{pages}{1--100}.

\bibitem[{Burkert and Roberts(2019)}]{Burkert:2017djo}
\bibinfo{author}{V.~D. Burkert}, \bibinfo{author}{C.~D. Roberts},
  \bibinfo{title}{{Roper resonance: Toward a solution to the fifty-year
  puzzle}}, \bibinfo{journal}{Rev. Mod. Phys.} \bibinfo{volume}{91}
  (\bibinfo{year}{2019}) \bibinfo{pages}{011003}.

\bibitem[{Fischer(2019)}]{Fischer:2018sdj}
\bibinfo{author}{C.~S. Fischer}, \bibinfo{title}{{QCD at finite temperature and
  chemical potential from Dyson--Schwinger equations}}, \bibinfo{journal}{Prog.
  Part. Nucl. Phys.} \bibinfo{volume}{105} (\bibinfo{year}{2019})
  \bibinfo{pages}{1--60}.

\bibitem[{Qin and Roberts(2020{\natexlab{a}})}]{Qin:2020rad}
\bibinfo{author}{S.-X. Qin}, \bibinfo{author}{C.~D. Roberts},
  \bibinfo{title}{{Impressions of the Continuum Bound State Problem in QCD}},
  \bibinfo{journal}{Chin. Phys. Lett.}
  \bibinfo{volume}{37}~(\bibinfo{number}{12})
  (\bibinfo{year}{2020}{\natexlab{a}}) \bibinfo{pages}{121201}.

\bibitem[{Qin et~al.(2018)Qin, Roberts, and Schmidt}]{Qin:2018dqp}
\bibinfo{author}{S.-X. Qin}, \bibinfo{author}{C.~D. Roberts},
  \bibinfo{author}{S.~M. Schmidt}, \bibinfo{title}{{Poincar{\'e}-covariant
  analysis of heavy-quark baryons}}, \bibinfo{journal}{Phys. Rev. D}
  \bibinfo{volume}{97} (\bibinfo{year}{2018}) \bibinfo{pages}{114017}.

\bibitem[{Qin et~al.(2019)Qin, Roberts, and Schmidt}]{Qin:2019hgk}
\bibinfo{author}{S.-X. Qin}, \bibinfo{author}{C.~D. Roberts},
  \bibinfo{author}{S.~M. Schmidt}, \bibinfo{title}{{Spectrum of light- and
  heavy-baryons}}, \bibinfo{journal}{Few Body Syst.} \bibinfo{volume}{60}
  (\bibinfo{year}{2019}) \bibinfo{pages}{26}.

\bibitem[{Binosi et~al.(2016)Binosi, Chang, Qin, Papavassiliou, and
  Roberts}]{Binosi:2016rxz}
\bibinfo{author}{D.~Binosi}, \bibinfo{author}{L.~Chang}, \bibinfo{author}{S.-X.
  Qin}, \bibinfo{author}{J.~Papavassiliou}, \bibinfo{author}{C.~D. Roberts},
  \bibinfo{title}{{Symmetry preserving truncations of the gap and
  Bethe-Salpeter equations}}, \bibinfo{journal}{Phys. Rev. D}
  \bibinfo{volume}{93} (\bibinfo{year}{2016}) \bibinfo{pages}{096010}.

\bibitem[{El-Bennich et~al.(2016)El-Bennich, Krein, Rojas, and
  Serna}]{El-Bennich:2016qmb}
\bibinfo{author}{B.~El-Bennich}, \bibinfo{author}{G.~Krein},
  \bibinfo{author}{E.~Rojas}, \bibinfo{author}{F.~E. Serna},
  \bibinfo{title}{{Excited hadrons and the analytical structure of bound-state
  interaction kernels}}, \bibinfo{journal}{Few Body Syst.}
  \bibinfo{volume}{57}~(\bibinfo{number}{10}) (\bibinfo{year}{2016})
  \bibinfo{pages}{955--963}.

\bibitem[{Binosi et~al.(2017)Binosi, Chang, Papavassiliou, Qin, and
  Roberts}]{Binosi:2016wcx}
\bibinfo{author}{D.~Binosi}, \bibinfo{author}{L.~Chang},
  \bibinfo{author}{J.~Papavassiliou}, \bibinfo{author}{S.-X. Qin},
  \bibinfo{author}{C.~D. Roberts}, \bibinfo{title}{{Natural constraints on the
  gluon-quark vertex}}, \bibinfo{journal}{Phys. Rev. D} \bibinfo{volume}{95}
  (\bibinfo{year}{2017}) \bibinfo{pages}{031501(R)}.

\bibitem[{Fu et~al.(2017)Fu, Wang, and Jiang}]{Fu:2017azw}
\bibinfo{author}{H.-F. Fu}, \bibinfo{author}{Q.~Wang},
  \bibinfo{author}{L.~Jiang}, \bibinfo{title}{{Derivation of the gap and
  Bethe-Salpeter equations at large $N_c$ limit and symmetry preserving
  truncations}}, \bibinfo{journal}{Phys. Rev. D}
  \bibinfo{volume}{96}~(\bibinfo{number}{9}) (\bibinfo{year}{2017})
  \bibinfo{pages}{094023}.

\bibitem[{Williams(2019)}]{Williams:2018adr}
\bibinfo{author}{R.~Williams}, \bibinfo{title}{{Vector mesons as dynamical
  resonances in the Bethe--Salpeter framework}}, \bibinfo{journal}{Phys. Lett.
  B} \bibinfo{volume}{798} (\bibinfo{year}{2019}) \bibinfo{pages}{134943}.

\bibitem[{Xu et~al.(2019)}]{Xu:2018cor}
\bibinfo{author}{S.-S. Xu}, et~al., \bibinfo{title}{{New perspective on hybrid
  mesons}}, \bibinfo{journal}{Eur. Phys. J. A (Lett.)} \bibinfo{volume}{55}
  (\bibinfo{year}{2019}) \bibinfo{pages}{113}.

\bibitem[{Oliveira et~al.(2018)Oliveira, Frederico, de~Paula, and
  de~Melo}]{Oliveira:2018fkj}
\bibinfo{author}{O.~Oliveira}, \bibinfo{author}{T.~Frederico},
  \bibinfo{author}{W.~de~Paula}, \bibinfo{author}{J.~P. B.~C. de~Melo},
  \bibinfo{title}{{Exploring the Quark-Gluon Vertex with Slavnov-Taylor
  Identities and Lattice Simulations}}, \bibinfo{journal}{Eur. Phys. J. C}
  \bibinfo{volume}{78} (\bibinfo{year}{2018}) \bibinfo{pages}{553}.

\bibitem[{Qin and Roberts(2020{\natexlab{b}})}]{Qin:2020jig}
\bibinfo{author}{S.-X. Qin}, \bibinfo{author}{C.~D. Roberts},
  \bibinfo{title}{{\emph{Resolving the Bethe-Salpeter kernel} --
  arXiv:2009.13637 [hep-ph]}} .

\bibitem[{Yin et~al.(2019)Yin, Chen, Krein, Roberts, Segovia, and
  Xu}]{Yin:2019bxe}
\bibinfo{author}{P.-L. Yin}, \bibinfo{author}{C.~Chen},
  \bibinfo{author}{G.~Krein}, \bibinfo{author}{C.~D. Roberts},
  \bibinfo{author}{J.~Segovia}, \bibinfo{author}{S.-S. Xu},
  \bibinfo{title}{{Masses of ground-state mesons and baryons, including those
  with heavy quarks}}, \bibinfo{journal}{Phys. Rev. D}
  \bibinfo{volume}{100}~(\bibinfo{number}{3}) (\bibinfo{year}{2019})
  \bibinfo{pages}{034008}.

\bibitem[{Guti{\'e}rrez-Guerrero et~al.(2019)Guti{\'e}rrez-Guerrero, Bashir,
  Bedolla, and Santopinto}]{Gutierrez-Guerrero:2019uwa}
\bibinfo{author}{L.~Guti{\'e}rrez-Guerrero}, \bibinfo{author}{A.~Bashir},
  \bibinfo{author}{M.~A. Bedolla}, \bibinfo{author}{E.~Santopinto},
  \bibinfo{title}{{Masses of Light and Heavy Mesons and Baryons: A Unified
  Picture}}, \bibinfo{journal}{Phys. Rev. D} \bibinfo{volume}{100}
  (\bibinfo{year}{2019}) \bibinfo{pages}{114032}.

\bibitem[{Roberts et~al.(2011{\natexlab{a}})Roberts, Bashir,
  Guti{\'e}rrez-Guerrero, Roberts, and Wilson}]{Roberts:2011wy}
\bibinfo{author}{H.~L.~L. Roberts}, \bibinfo{author}{A.~Bashir},
  \bibinfo{author}{L.~X. Guti{\'e}rrez-Guerrero}, \bibinfo{author}{C.~D.
  Roberts}, \bibinfo{author}{D.~J. Wilson}, \bibinfo{title}{{$\pi$- and
  $\rho$-mesons, and their diquark partners, from a contact interaction}},
  \bibinfo{journal}{Phys. Rev. C} \bibinfo{volume}{83}
  (\bibinfo{year}{2011}{\natexlab{a}}) \bibinfo{pages}{065206}.

\bibitem[{Cahill et~al.(1989)Cahill, Roberts, and Praschifka}]{Cahill:1988dx}
\bibinfo{author}{R.~T. Cahill}, \bibinfo{author}{C.~D. Roberts},
  \bibinfo{author}{J.~Praschifka}, \bibinfo{title}{{Baryon structure and QCD}},
  \bibinfo{journal}{Austral. J. Phys.} \bibinfo{volume}{42}
  (\bibinfo{year}{1989}) \bibinfo{pages}{129--145}.

\bibitem[{Burden et~al.(1989)Burden, Cahill, and Praschifka}]{Burden:1988dt}
\bibinfo{author}{C.~J. Burden}, \bibinfo{author}{R.~T. Cahill},
  \bibinfo{author}{J.~Praschifka}, \bibinfo{title}{{Baryon Structure and {QCD}:
  Nucleon Calculations}}, \bibinfo{journal}{Austral. J. Phys.}
  \bibinfo{volume}{42} (\bibinfo{year}{1989}) \bibinfo{pages}{147--159}.

\bibitem[{Reinhardt(1990)}]{Reinhardt:1989rw}
\bibinfo{author}{H.~Reinhardt}, \bibinfo{title}{{Hadronization of Quark Flavor
  Dynamics}}, \bibinfo{journal}{Phys. Lett. B} \bibinfo{volume}{244}
  (\bibinfo{year}{1990}) \bibinfo{pages}{316--326}.

\bibitem[{Efimov et~al.(1990)Efimov, Ivanov, and Lyubovitskij}]{Efimov:1990uz}
\bibinfo{author}{G.~V. Efimov}, \bibinfo{author}{M.~A. Ivanov},
  \bibinfo{author}{V.~E. Lyubovitskij}, \bibinfo{title}{{Quark - diquark
  approximation of the three quark structure of baryons in the quark
  confinement model}}, \bibinfo{journal}{Z. Phys. C} \bibinfo{volume}{47}
  (\bibinfo{year}{1990}) \bibinfo{pages}{583--594}.

\bibitem[{Roberts et~al.(2011{\natexlab{b}})Roberts, Chang, Cloet, and
  Roberts}]{Roberts:2011cf}
\bibinfo{author}{H.~L.~L. Roberts}, \bibinfo{author}{L.~Chang},
  \bibinfo{author}{I.~C. Cloet}, \bibinfo{author}{C.~D. Roberts},
  \bibinfo{title}{{Masses of ground and excited-state hadrons}},
  \bibinfo{journal}{Few Body Syst.} \bibinfo{volume}{51}
  (\bibinfo{year}{2011}{\natexlab{b}}) \bibinfo{pages}{1--25}.

\bibitem[{Eichmann et~al.(2016{\natexlab{b}})Eichmann, Fischer, and
  Sanchis-Alepuz}]{Eichmann:2016hgl}
\bibinfo{author}{G.~Eichmann}, \bibinfo{author}{C.~S. Fischer},
  \bibinfo{author}{H.~Sanchis-Alepuz}, \bibinfo{title}{{Light baryons and their
  excitations}}, \bibinfo{journal}{Phys. Rev. D} \bibinfo{volume}{94}
  (\bibinfo{year}{2016}{\natexlab{b}}) \bibinfo{pages}{094033}.

\bibitem[{Lu et~al.(2017)Lu, Chen, Roberts, Segovia, Xu, and Zong}]{Lu:2017cln}
\bibinfo{author}{Y.~Lu}, \bibinfo{author}{C.~Chen}, \bibinfo{author}{C.~D.
  Roberts}, \bibinfo{author}{J.~Segovia}, \bibinfo{author}{S.-S. Xu},
  \bibinfo{author}{H.-S. Zong}, \bibinfo{title}{{Parity partners in the baryon
  resonance spectrum}}, \bibinfo{journal}{Phys. Rev. C} \bibinfo{volume}{96}
  (\bibinfo{year}{2017}) \bibinfo{pages}{015208}.

\bibitem[{Llewellyn-Smith(1969)}]{LlewellynSmith:1969az}
\bibinfo{author}{C.~H. Llewellyn-Smith}, \bibinfo{title}{{A relativistic
  formulation for the quark model for mesons}}, \bibinfo{journal}{Annals Phys.}
  \bibinfo{volume}{53} (\bibinfo{year}{1969}) \bibinfo{pages}{521--558}.

\bibitem[{Krassnigg(2009)}]{Krassnigg:2009zh}
\bibinfo{author}{A.~Krassnigg}, \bibinfo{title}{{Survey of J=0,1 mesons in a
  Bethe-Salpeter approach}}, \bibinfo{journal}{Phys. Rev. D}
  \bibinfo{volume}{80} (\bibinfo{year}{2009}) \bibinfo{pages}{114010}.

\bibitem[{Maris et~al.(2001)Maris, Roberts, Schmidt, and Tandy}]{Maris:2000ig}
\bibinfo{author}{P.~Maris}, \bibinfo{author}{C.~D. Roberts},
  \bibinfo{author}{S.~M. Schmidt}, \bibinfo{author}{P.~C. Tandy},
  \bibinfo{title}{{T-dependence of pseudoscalar and scalar correlations}},
  \bibinfo{journal}{Phys. Rev. C} \bibinfo{volume}{63} (\bibinfo{year}{2001})
  \bibinfo{pages}{025202}.

\bibitem[{Chang and Roberts(2012)}]{Chang:2011ei}
\bibinfo{author}{L.~Chang}, \bibinfo{author}{C.~D. Roberts},
  \bibinfo{title}{{Tracing masses of ground-state light-quark mesons}},
  \bibinfo{journal}{Phys. Rev. C} \bibinfo{volume}{85} (\bibinfo{year}{2012})
  \bibinfo{pages}{052201(R)}.

\bibitem[{Qin et~al.(2012)Qin, Chang, Liu, Roberts, and Wilson}]{Qin:2011xq}
\bibinfo{author}{S.-X. Qin}, \bibinfo{author}{L.~Chang}, \bibinfo{author}{Y.-x.
  Liu}, \bibinfo{author}{C.~D. Roberts}, \bibinfo{author}{D.~J. Wilson},
  \bibinfo{title}{{Investigation of rainbow-ladder truncation for excited and
  exotic mesons}}, \bibinfo{journal}{Phys. Rev. C} \bibinfo{volume}{85}
  (\bibinfo{year}{2012}) \bibinfo{pages}{035202}.

\bibitem[{Bhagwat et~al.(2007)Bhagwat, H{\"o}ll, Krassnigg, and
  Roberts}]{Bhagwat:2006py}
\bibinfo{author}{M.~S. Bhagwat}, \bibinfo{author}{A.~H{\"o}ll},
  \bibinfo{author}{A.~Krassnigg}, \bibinfo{author}{C.~D. Roberts},
  \bibinfo{title}{{Theory and phenomenology of hadrons}},
  \bibinfo{journal}{Nucl. Phys. A} \bibinfo{volume}{790} (\bibinfo{year}{2007})
  \bibinfo{pages}{10--16}.

\bibitem[{Becirevic et~al.(1999)Becirevic, Boucaud, Leroy, Lubicz, Martinelli,
  Mescia, and Rapuano}]{Becirevic:1998ua}
\bibinfo{author}{D.~Becirevic}, \bibinfo{author}{P.~Boucaud},
  \bibinfo{author}{J.~P. Leroy}, \bibinfo{author}{V.~Lubicz},
  \bibinfo{author}{G.~Martinelli}, \bibinfo{author}{F.~Mescia},
  \bibinfo{author}{F.~Rapuano}, \bibinfo{title}{{Nonperturbatively improved
  heavy - light mesons: Masses and decay constants}}, \bibinfo{journal}{Phys.
  Rev. D} \bibinfo{volume}{60} (\bibinfo{year}{1999}) \bibinfo{pages}{074501}.

\bibitem[{Chiu and Hsieh(2007)}]{Chiu:2007bc}
\bibinfo{author}{T.-W. Chiu}, \bibinfo{author}{T.-H. Hsieh},
  \bibinfo{title}{{B(s) and B(c) mesons in lattice QCD with exact chiral
  symmetry}}, \bibinfo{journal}{PoS} \bibinfo{volume}{LAT2006}
  (\bibinfo{year}{2007}) \bibinfo{pages}{180}.

\bibitem[{Davies et~al.(2010)}]{Davies:2010ip}
\bibinfo{author}{C.~T.~H. Davies}, et~al., \bibinfo{title}{{Update: Precision
  $D_s$ decay constant from full lattice QCD using very fine lattices}},
  \bibinfo{journal}{Phys. Rev. D} \bibinfo{volume}{82} (\bibinfo{year}{2010})
  \bibinfo{pages}{114504}.

\bibitem[{McNeile et~al.(2012)McNeile, Davies, Follana, Hornbostel, and
  Lepage}]{McNeile:2012qf}
\bibinfo{author}{C.~McNeile}, \bibinfo{author}{C.~T.~H. Davies},
  \bibinfo{author}{E.~Follana}, \bibinfo{author}{K.~Hornbostel},
  \bibinfo{author}{G.~P. Lepage}, \bibinfo{title}{{Heavy meson masses and decay
  constants from relativistic heavy quarks in full lattice QCD}},
  \bibinfo{journal}{Phys. Rev. D} \bibinfo{volume}{86} (\bibinfo{year}{2012})
  \bibinfo{pages}{074503}.

\bibitem[{Donald et~al.(2012)}]{Donald:2012ga}
\bibinfo{author}{G.~C. Donald}, et~al., \bibinfo{title}{{Precision tests of the
  $J/{\psi}$ from full lattice QCD: mass, leptonic width and radiative decay
  rate to ${\eta}_c$}}, \bibinfo{journal}{Phys. Rev. D} \bibinfo{volume}{86}
  (\bibinfo{year}{2012}) \bibinfo{pages}{094501}.

\bibitem[{Colquhoun et~al.(2015)Colquhoun, Davies, Dowdall, Kettle, Koponen,
  Lepage, and Lytle}]{Colquhoun:2015oha}
\bibinfo{author}{B.~Colquhoun}, \bibinfo{author}{C.~T.~H. Davies},
  \bibinfo{author}{R.~J. Dowdall}, \bibinfo{author}{J.~Kettle},
  \bibinfo{author}{J.~Koponen}, \bibinfo{author}{G.~P. Lepage},
  \bibinfo{author}{A.~T. Lytle}, \bibinfo{title}{{B-meson decay constants: a
  more complete picture from full lattice QCD}}, \bibinfo{journal}{Phys. Rev.
  D} \bibinfo{volume}{91} (\bibinfo{year}{2015}) \bibinfo{pages}{114509}.

\bibitem[{Chen et~al.(2012)Chen, Chang, Roberts, Wan, and Wilson}]{Chen:2012qr}
\bibinfo{author}{C.~Chen}, \bibinfo{author}{L.~Chang}, \bibinfo{author}{C.~D.
  Roberts}, \bibinfo{author}{S.-L. Wan}, \bibinfo{author}{D.~J. Wilson},
  \bibinfo{title}{{Spectrum of hadrons with strangeness}},
  \bibinfo{journal}{Few Body Syst.} \bibinfo{volume}{53} (\bibinfo{year}{2012})
  \bibinfo{pages}{293--326}.

\bibitem[{Chang and Roberts(2009)}]{Chang:2009zb}
\bibinfo{author}{L.~Chang}, \bibinfo{author}{C.~D. Roberts},
  \bibinfo{title}{{Sketching the Bethe-Salpeter kernel}},
  \bibinfo{journal}{Phys. Rev. Lett.} \bibinfo{volume}{103}
  (\bibinfo{year}{2009}) \bibinfo{pages}{081601}.

\bibitem[{Fischer and Williams(2009)}]{Fischer:2009jm}
\bibinfo{author}{C.~S. Fischer}, \bibinfo{author}{R.~Williams},
  \bibinfo{title}{{Probing the gluon self-interaction in light mesons}},
  \bibinfo{journal}{Phys. Rev. Lett.} \bibinfo{volume}{103}
  (\bibinfo{year}{2009}) \bibinfo{pages}{122001}.

\bibitem[{Chang et~al.(2011)Chang, Liu, and Roberts}]{Chang:2010hb}
\bibinfo{author}{L.~Chang}, \bibinfo{author}{Y.-X. Liu}, \bibinfo{author}{C.~D.
  Roberts}, \bibinfo{title}{{Dressed-quark anomalous magnetic moments}},
  \bibinfo{journal}{Phys. Rev. Lett.} \bibinfo{volume}{106}
  (\bibinfo{year}{2011}) \bibinfo{pages}{072001}.

\bibitem[{Roberts et~al.(2010)Roberts, Roberts, Bashir, Guti{\'e}rrez-Guerrero,
  and Tandy}]{Roberts:2010rn}
\bibinfo{author}{H.~L.~L. Roberts}, \bibinfo{author}{C.~D. Roberts},
  \bibinfo{author}{A.~Bashir}, \bibinfo{author}{L.~X. Guti{\'e}rrez-Guerrero},
  \bibinfo{author}{P.~C. Tandy}, \bibinfo{title}{{Abelian anomaly and neutral
  pion production}}, \bibinfo{journal}{Phys. Rev. C} \bibinfo{volume}{82}
  (\bibinfo{year}{2010}) \bibinfo{pages}{{\mbox{065202}}}.

\bibitem[{Okubo(1962)}]{Okubo:1961jc}
\bibinfo{author}{S.~Okubo}, \bibinfo{title}{{Note on unitary symmetry in strong
  interactions}}, \bibinfo{journal}{Prog. Theor. Phys.} \bibinfo{volume}{27}
  (\bibinfo{year}{1962}) \bibinfo{pages}{949--966}.

\bibitem[{Gell-Mann(1962)}]{GellMann:1962xb}
\bibinfo{author}{M.~Gell-Mann}, \bibinfo{title}{{Symmetries of baryons and
  mesons}}, \bibinfo{journal}{Phys. Rev.} \bibinfo{volume}{125}
  (\bibinfo{year}{1962}) \bibinfo{pages}{1067--1084}, \bibinfo{note}{see also
  {``\emph{The Eightfold Way: A Theory of Strong Interaction Symmetry},''} DOE
  Technical Report TID-12608, 1961}.

\bibitem[{Chen et~al.(2019)Chen, Krein, Roberts, Schmidt, and
  Segovia}]{Chen:2019fzn}
\bibinfo{author}{C.~Chen}, \bibinfo{author}{G.~I. Krein},
  \bibinfo{author}{C.~D. Roberts}, \bibinfo{author}{S.~M. Schmidt},
  \bibinfo{author}{J.~Segovia}, \bibinfo{title}{{Spectrum and structure of
  octet and decuplet baryons and their positive-parity excitations}},
  \bibinfo{journal}{Phys. Rev. D} \bibinfo{volume}{100} (\bibinfo{year}{2019})
  \bibinfo{pages}{054009}.

\bibitem[{Maris et~al.(1998)Maris, Roberts, and Tandy}]{Maris:1997hd}
\bibinfo{author}{P.~Maris}, \bibinfo{author}{C.~D. Roberts},
  \bibinfo{author}{P.~C. Tandy}, \bibinfo{title}{{Pion mass and decay
  constant}}, \bibinfo{journal}{Phys. Lett. B} \bibinfo{volume}{420}
  (\bibinfo{year}{1998}) \bibinfo{pages}{267--273}.

\bibitem[{Ivanov et~al.(1999)Ivanov, Kalinovsky, and Roberts}]{Ivanov:1998ms}
\bibinfo{author}{M.~A. Ivanov}, \bibinfo{author}{{\mbox{Yu}}.~L. Kalinovsky},
  \bibinfo{author}{C.~D. Roberts}, \bibinfo{title}{{Survey of heavy-meson
  observables}}, \bibinfo{journal}{Phys. Rev. D} \bibinfo{volume}{60}
  (\bibinfo{year}{1999}) \bibinfo{pages}{034018}.

\bibitem[{Bender et~al.(1996)Bender, Blaschke, Kalinovsky, and
  Roberts}]{Bender:1996bm}
\bibinfo{author}{A.~Bender}, \bibinfo{author}{D.~Blaschke},
  \bibinfo{author}{Y.~Kalinovsky}, \bibinfo{author}{C.~D. Roberts},
  \bibinfo{title}{{Continuum study of deconfinement at finite temperature}},
  \bibinfo{journal}{Phys. Rev. Lett.} \bibinfo{volume}{77}
  (\bibinfo{year}{1996}) \bibinfo{pages}{3724--3727}.

\bibitem[{Anselmino et~al.(1993)Anselmino, Predazzi, Ekelin, Fredriksson, and
  Lichtenberg}]{Anselmino:1992vg}
\bibinfo{author}{M.~Anselmino}, \bibinfo{author}{E.~Predazzi},
  \bibinfo{author}{S.~Ekelin}, \bibinfo{author}{S.~Fredriksson},
  \bibinfo{author}{D.~B. Lichtenberg}, \bibinfo{title}{{Diquarks}},
  \bibinfo{journal}{Rev. Mod. Phys.} \bibinfo{volume}{65}
  (\bibinfo{year}{1993}) \bibinfo{pages}{1199--1234}.

\bibitem[{Chen et~al.(2018)Chen, El-Bennich, Roberts, Schmidt, Segovia, and
  Wan}]{Chen:2017pse}
\bibinfo{author}{C.~Chen}, \bibinfo{author}{B.~El-Bennich},
  \bibinfo{author}{C.~D. Roberts}, \bibinfo{author}{S.~M. Schmidt},
  \bibinfo{author}{J.~Segovia}, \bibinfo{author}{S.~Wan},
  \bibinfo{title}{{Structure of the nucleon's low-lying excitations}},
  \bibinfo{journal}{Phys. Rev. D} \bibinfo{volume}{97} (\bibinfo{year}{2018})
  \bibinfo{pages}{034016}.

\bibitem[{Cahill et~al.(1987)Cahill, Roberts, and Praschifka}]{Cahill:1987qr}
\bibinfo{author}{R.~T. Cahill}, \bibinfo{author}{C.~D. Roberts},
  \bibinfo{author}{J.~Praschifka}, \bibinfo{title}{{Calculation of diquark
  masses in QCD}}, \bibinfo{journal}{Phys. Rev. D} \bibinfo{volume}{36}
  (\bibinfo{year}{1987}) \bibinfo{pages}{2804}.

\bibitem[{Maris(2002)}]{Maris:2002yu}
\bibinfo{author}{P.~Maris}, \bibinfo{title}{{Effective masses of diquarks}},
  \bibinfo{journal}{Few Body Syst.} \bibinfo{volume}{32} (\bibinfo{year}{2002})
  \bibinfo{pages}{41--52}.

\bibitem[{Maris(2004)}]{Maris:2004bp}
\bibinfo{author}{P.~Maris}, \bibinfo{title}{{Electromagnetic properties of
  diquarks}}, \bibinfo{journal}{Few Body Syst.} \bibinfo{volume}{35}
  (\bibinfo{year}{2004}) \bibinfo{pages}{117--127}.

\bibitem[{Munczek and Nemirovsky(1983)}]{Munczek:1983dx}
\bibinfo{author}{H.~J. Munczek}, \bibinfo{author}{A.~M. Nemirovsky},
  \bibinfo{title}{{The Ground State $q\bar q$ Mass Spectrum in QCD}},
  \bibinfo{journal}{Phys. Rev. D} \bibinfo{volume}{28} (\bibinfo{year}{1983})
  \bibinfo{pages}{181--186}.

\bibitem[{Bhagwat et~al.(2004)Bhagwat, H{\"o}ll, Krassnigg, Roberts, and
  Tandy}]{Bhagwat:2004hn}
\bibinfo{author}{M.~S. Bhagwat}, \bibinfo{author}{A.~H{\"o}ll},
  \bibinfo{author}{A.~Krassnigg}, \bibinfo{author}{C.~D. Roberts},
  \bibinfo{author}{P.~C. Tandy}, \bibinfo{title}{{Aspects and consequences of a
  dressed-quark-gluon vertex}}, \bibinfo{journal}{Phys. Rev. C}
  \bibinfo{volume}{70} (\bibinfo{year}{2004}) \bibinfo{pages}{035205}.

\bibitem[{Bloch et~al.(1999)Bloch, Roberts, and Schmidt}]{Bloch:1999vk}
\bibinfo{author}{J.~C.~R. Bloch}, \bibinfo{author}{C.~D. Roberts},
  \bibinfo{author}{S.~M. Schmidt}, \bibinfo{title}{{Diquark condensation and
  the quark quark interaction}}, \bibinfo{journal}{Phys. Rev. C}
  \bibinfo{volume}{60} (\bibinfo{year}{1999}) \bibinfo{pages}{065208}.

\bibitem[{Buck et~al.(1992)Buck, Alkofer, and Reinhardt}]{Buck:1992wz}
\bibinfo{author}{A.~Buck}, \bibinfo{author}{R.~Alkofer},
  \bibinfo{author}{H.~Reinhardt}, \bibinfo{title}{{Baryons as bound states of
  diquarks and quarks in the Nambu-Jona-Lasinio model}},
  \bibinfo{journal}{Phys. Lett. B} \bibinfo{volume}{286} (\bibinfo{year}{1992})
  \bibinfo{pages}{29--35}.

\bibitem[{Xu et~al.(2015)Xu, Chen, Cloet, Roberts, Segovia, and
  Zong}]{Xu:2015kta}
\bibinfo{author}{S.-S. Xu}, \bibinfo{author}{C.~Chen}, \bibinfo{author}{I.~C.
  Cloet}, \bibinfo{author}{C.~D. Roberts}, \bibinfo{author}{J.~Segovia},
  \bibinfo{author}{H.-S. Zong}, \bibinfo{title}{{Contact-interaction Faddeev
  equation and, \emph{inter alia}, proton tensor charges}},
  \bibinfo{journal}{Phys. Rev. D} \bibinfo{volume}{92} (\bibinfo{year}{2015})
  \bibinfo{pages}{114034}.

\bibitem[{Hecht et~al.(2002)Hecht, Oettel, Roberts, Schmidt, Tandy, and
  Thomas}]{Hecht:2002ej}
\bibinfo{author}{M.~B. Hecht}, \bibinfo{author}{M.~Oettel},
  \bibinfo{author}{C.~D. Roberts}, \bibinfo{author}{S.~M. Schmidt},
  \bibinfo{author}{P.~C. Tandy}, \bibinfo{author}{A.~W. Thomas},
  \bibinfo{title}{{Nucleon mass and pion loops}}, \bibinfo{journal}{Phys. Rev.
  C} \bibinfo{volume}{65} (\bibinfo{year}{2002}) \bibinfo{pages}{055204}.

\bibitem[{Aguilar et~al.(2016)Aguilar, Binosi, and
  Papavassiliou}]{Aguilar:2015bud}
\bibinfo{author}{A.~C. Aguilar}, \bibinfo{author}{D.~Binosi},
  \bibinfo{author}{J.~Papavassiliou}, \bibinfo{title}{{The Gluon Mass
  Generation Mechanism: A Concise Primer}}, \bibinfo{journal}{Front. Phys.
  China} \bibinfo{volume}{11} (\bibinfo{year}{2016}) \bibinfo{pages}{111203}.

\bibitem[{Huber(2020)}]{Huber:2018ned}
\bibinfo{author}{M.~Q. Huber}, \bibinfo{title}{{Nonperturbative properties of
  Yang-Mills theories}}, \bibinfo{journal}{Phys. Rept.} \bibinfo{volume}{879}
  (\bibinfo{year}{2020}) \bibinfo{pages}{1 -- 92}.

\bibitem[{Cui et~al.(2020)Cui, Zhang, Binosi, de~Soto, Mezrag, Papavassiliou,
  Roberts, Rodr{\'{\i}}guez-Quintero, Segovia, and Zafeiropoulos}]{Cui:2019dwv}
\bibinfo{author}{Z.-F. Cui}, \bibinfo{author}{J.-L. Zhang},
  \bibinfo{author}{D.~Binosi}, \bibinfo{author}{F.~de~Soto},
  \bibinfo{author}{C.~Mezrag}, \bibinfo{author}{J.~Papavassiliou},
  \bibinfo{author}{C.~D. Roberts},
  \bibinfo{author}{J.~Rodr{\'{\i}}guez-Quintero}, \bibinfo{author}{J.~Segovia},
  \bibinfo{author}{S.~Zafeiropoulos}, \bibinfo{title}{{Effective charge from
  lattice QCD}}, \bibinfo{journal}{Chin. Phys. C} \bibinfo{volume}{44}
  (\bibinfo{year}{2020}) \bibinfo{pages}{083102}.

\bibitem[{Ebert et~al.(1996)Ebert, Feldmann, and Reinhardt}]{Ebert:1996vx}
\bibinfo{author}{D.~Ebert}, \bibinfo{author}{T.~Feldmann},
  \bibinfo{author}{H.~Reinhardt}, \bibinfo{title}{{Extended NJL model for light
  and heavy mesons without $q \bar q$ thresholds}}, \bibinfo{journal}{Phys.
  Lett. B} \bibinfo{volume}{388} (\bibinfo{year}{1996})
  \bibinfo{pages}{154--160}.

\bibitem[{Roberts et~al.(1992)Roberts, Williams, and Krein}]{Krein:1990sf}
\bibinfo{author}{C.~D. Roberts}, \bibinfo{author}{A.~G. Williams},
  \bibinfo{author}{G.~Krein}, \bibinfo{title}{{On the implications of
  confinement}}, \bibinfo{journal}{Int. J. Mod. Phys. A} \bibinfo{volume}{7}
  (\bibinfo{year}{1992}) \bibinfo{pages}{5607--5624}.

\bibitem[{Chen et~al.(2013)Chen, Chang, Roberts, Wan, Schmidt, and
  Wilson}]{Chen:2012txa}
\bibinfo{author}{C.~Chen}, \bibinfo{author}{L.~Chang}, \bibinfo{author}{C.~D.
  Roberts}, \bibinfo{author}{S.-L. Wan}, \bibinfo{author}{S.~M. Schmidt},
  \bibinfo{author}{D.~J. Wilson}, \bibinfo{title}{{Features and flaws of a
  contact interaction treatment of the kaon}}, \bibinfo{journal}{Phys. Rev. C}
  \bibinfo{volume}{87} (\bibinfo{year}{2013}) \bibinfo{pages}{045207}.

\end{thebibliography}

INSPIRE-00557473
0000-0003-3890-0242
0000-0002-2937-1361
0000-0001-5838-7103

\end{document}